\documentclass[aps,floats,prb,superscriptaddress,
showpacs,
twocolumn]{revtex4}

\usepackage{amsfonts,amsmath,mathrsfs} \usepackage{bm} \usepackage{bbm} \usepackage{dcolumn}
\usepackage{epsfig} \usepackage{latexsym} \usepackage{multirow}

\newcommand{\tp}{{\tilde \psi}}

\begin{document}

\title{Emergence of integer quantum Hall effect from chaos}

\author{Chushun Tian}
\affiliation{Institute for Advanced Study, Tsinghua University, Beijing, 100084, China}
\author{Yu Chen}
\affiliation{Institute for Advanced Study, Tsinghua University, Beijing, 100084, China}
\affiliation{Department of Physics and Center for Theoretical Physics, Capital Normal University, Beijing, 100048, China}
\author{Jiao Wang}
\affiliation{Department of Physics and Institute of Theoretical
Physics and Astrophysics,
Xiamen University, Xiamen, 361005, China}


\date{\today}

\begin{abstract}

We present an analytic microscopic theory showing that in a large class of spin-$\frac{1}{2}$ quasiperiodic quantum kicked rotors, a dynamical analog of the integer quantum Hall effect (IQHE) emerges from an intrinsic chaotic structure. Specifically, the inverse of the Planck's quantum ($h_e$) and the rotor's energy growth rate mimic the `filling fraction' and the `longitudinal conductivity' in conventional IQHE, respectively, and a hidden quantum number is found to mimic the `quantized Hall conductivity'. We show that for an infinite discrete set of critical values of $h_e$, the long-time energy growth rate is universal and of order of unity (`metallic' phase), but
otherwise vanishes (`insulating' phase).
Moreover, the rotor insulating phases are topological, each of which is characterized by a hidden quantum number. This number exhibits universal behavior 
for small $h_e
$, i.e., it jumps by unity whenever
$h_e$ decreases, passing through each critical value. This intriguing phenomenon is not triggered by the like of Landau band filling, well-known to be the mechanism for conventional IQHE, and far beyond the canonical Thouless-Kohmoto-Nightingale-Nijs paradigm for quantum Hall transitions. Instead, this dynamical phenomenon is of strong chaos origin; it does not occur when the dynamics is (partially) regular. More precisely, we find that, for the first time, a topological object, similar to the topological theta angle in quantum chromodynamics, emerges from strongly chaotic motion at microscopic scales, and its renormalization gives the hidden quantum number.
Our analytic results are confirmed by numerical simulations.
Our findings indicate that rich topological quantum phenomena can emerge from chaos and might point to a new direction of study in the interdisciplinary area straddling chaotic dynamics and condensed matter physics. This work is a substantial extension of a short paper published earlier by two of us [Y. Chen and C. Tian, Phys. Rev. Lett. \textbf{113}, 216802 (2014)].

\end{abstract}
\maketitle

\section{Introduction}
\label{sec:introduction}

Chaos is ubiquitous in Nature. In quantum chaotic systems a wealth of phenomena arise from the interplay between chaotic motion and quantum interference \cite{Gutzwiller90,Haake}. A dimensionless parameter governing this interplay is the so-called Planck's quantum, $h_e$, which is the ratio of Planck's
constant $\hbar$ to the system's characteristic action
(see Refs.~\onlinecite{Zaslavsky81,Izrailev90,Larkin96,Tian04,Wimberger,Zurek03} and references therein).
A `standard model' in studies of such interplay is the quantum
kicked rotor (QKR) \cite{Izrailev90,QKR79,Chirikov79,Fishman10,Fishman84,Altland10} -- a
particle moving on a ring under the influence of a sequential driving
force (`kicking'). This kicking, making the
particle rapidly lose the memory about its angular position, leads to strong chaos.
The realization of QKR in atom optics in the
mid-nineties \cite{Raizen95} has boosted interests in the study of quantum chaos
\cite{Zoller97,Ammann98,Hensinger01,Raizen01,Deland08,Chaudhury09,Altland11}, opening a
route to experimental studies of the interplay between chaos and interference \cite{Phillips06,Raizen00,Arcy01,Steinberg07}.
In particular, realization of the time modulation of the angular profile of kicking potential \cite{Deland08} affords opportunities to explore this interplay in higher dimensions. Indeed, $(d-1)$ modulated phase parameters introduce a virtual $(d-1)$-dimensional space. When the modulation frequencies are incommensurate with each other as well as the kicking frequency, the system, so-called quasiperiodic QKR, effectively
simulates a $d$-dimensional disordered system \cite{Deland08,Altland11,Casati89}. In the present work, we focus on the case of $d=2$.

\begin{figure}[h]
\includegraphics[width=8.6cm]{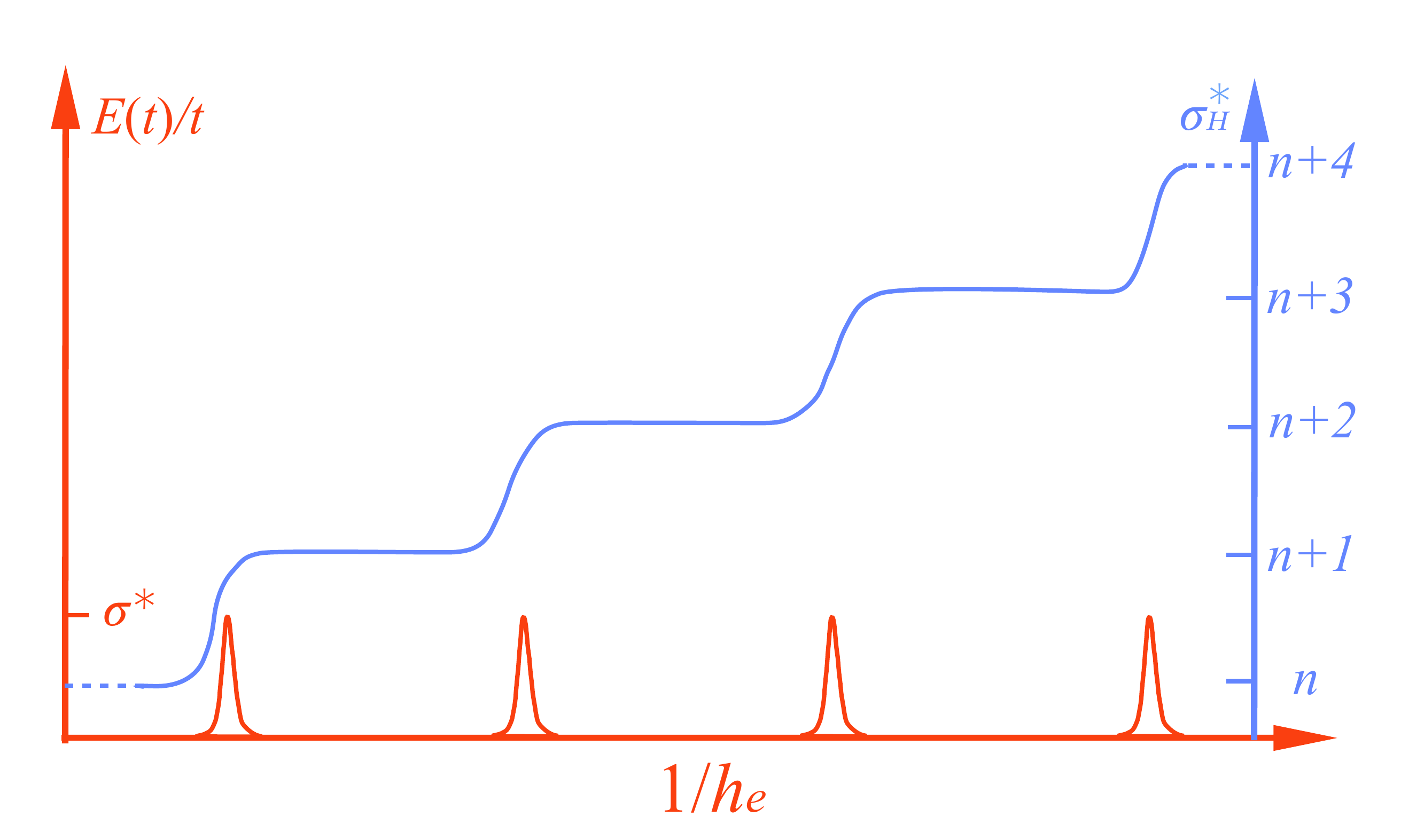}
\caption{Schematic representation of Planck-IQHE 
for small $h_e
$ in spin-$\frac{1}{2}$ quasiperiodic QKR.
Red line: for an infinite discrete set of critical values of $h_e$
the rotor's energy increases linearly at long times, i.e.,
$\lim_{t\rightarrow \infty} \frac{E(t)}{t}=\sigma^*={\cal O}(1)$, which is a characteristic of metals; for other values the rotor's energy
saturates at long times, i.e., $\lim_{t\rightarrow \infty} \frac{E(t)}{t}=0$, which is a characteristic of insulators.
Blue line: the insulator is characterized by a hidden quantum number $\sigma_{\rm H}^*$;
this number jumps by unity whenever $h_e^{-1}$ increases passing through a critical
value. (The sharp transitions are smeared at finite times.)}
\label{fig:5}
\end{figure}

For QKR, the Planck's quantum $h_e=\tau\hbar/I$, where $I$ is the particle's moment of inertia and $\tau$ the kicking period \cite{Izrailev90}. The system's behavior turns out to be extremely sensitive to the number-theoretic properties of this parameter \cite{Izrailev90,Fishman10,Altland10,Wimberger,Altland11}. That is, depending on whether the value of $h_e/(4\pi)$ is (i) irrational or (ii) rational, qualitatively different quantum phenomena occur. (i) For
(generic) irrational values of $h_e/(4\pi)$, the rotor's energy growth is bounded for periodic QKR, i.e., when the driving force is strictly periodic. This is the celebrated dynamical localization \cite{QKR79} -- an analog of Anderson localization in quasi one-dimensional ($1$D) disordered systems \cite{Fishman84,Altland10}. For quasiperiodic QKR, richer phenomena arise. Notably, the system can exhibit a transition from bounded to unbounded growth as the kicking strength increases. This is an analog of Anderson transition \cite{Deland08,Altland11}. (ii) For rational $h_e/(4\pi)$, the energy of a periodic QKR grows quadratically at long times, for quasiperiodic QKR increasing the kicking strength results in a transition from quadratic to linear growth, simulating a supermetal-metal transition \cite{Altland11}.

The Planck's quantum-driven phenomena in QKR have been well documented (see Refs.~\onlinecite{Altland10,Altland11,Izrailev90,Wimberger,Casati00,Fishman03a,Wang11} and references therein). The abundance of these phenomena notwithstanding, they can all be attributed to the translation symmetry or its breaking in angular momentum space. Indeed, when $h_e/(4\pi)$ is irrational, the system, or more precisely, the one-step evolution operator governing the quantum dynamics within a single time period, does not exhibit translational invariance. As a result, the QKR behaves as a genuine disordered system. As opposed to this, when $h_e/(4\pi)$ is rational, the system possesses the translation symmetry and therefore behaves as a perfect crystal.

Most theoretical and experimental studies of QKR pay no attention to the spin degree of freedom of the rotating particle. The subject of the impact of spin on the dynamics of QKR was pioneered by Scharf \cite{Scharf89} and subsequently studied in several works \cite{Kravtsov04,Bluemel94,Beenakker11}. The proposal of using spinful QKR to simulate a topological quantum phenomenon in paramagnetic semiconductors \cite{Zhang06} was made in Ref.~\onlinecite{Beenakker11}. These works, however, do not address the sensitivity of system's behavior to $h_e$, which, as mentioned above, is of fundamental importance to quantum chaos. It was not until very recently that this task was undertaken by two of us \cite{Tian14}. It is found that the spin affects profoundly the interplay -- controlled by $h_e$ -- between chaos and quantum interference.

In this work, we substantially extend this earlier investigation \cite{Tian14}. We present an analytic microscopic theory for a large class of spinful quasiperiodic QKR. In essence, these systems differ from spinless quasiperiodic QKR \cite{Altland11,Deland08} in that the particle has spin and the kicking potential couples the particle's angular and spin degrees of freedom, i.e., upon kicking the particle undergoes an abrupt change in the angular momentum and a flip of spin simultaneously. Based on the microscopic theory developed, we show analytically a striking dynamical phenomenon driven by the Planck's quantum, which bears a close resemblance to the integer quantum Hall effect (IQHE) \cite{Klitzing80} in condensed matter physics. (We make a clear distinction between IQHE and the quantum anomalous Hall effect \cite{Haldane88}. In the present work we are concerned in the former only.) This phenomenon, dubbed `the Planck's quantum-driven integer quantum Hall effect (Planck-IQHE)', is found to be rooted in the strong chaos brought about by the coupling between angular and spin degrees of freedom. Our analytic predictions are confirmed by numerical simulations.

At first glance, there is no reason to expect any relationship between dynamical phenomena in simple chaotic systems, such as QKR, and IQHE. Indeed, the IQHE was originally found in electronic systems, such as the metal-oxide field effect transistor (MOSFET), which are totally different from QKR. It arises from the integer filling of the Landau bands \cite{Pruisken84a}. The formation of these bands requires an external magnetic field, while the integer filling of these bands is a profound consequence of the Pauli principle for many-electron systems. These two essential ingredients, however, are both absent for the present QKR systems. In particular, because of the single-particle nature of QKR, the concept of `integer filling' is meaningless. Besides, the QKR is a chaotic system. The basic characteristic of chaos, namely, the extreme sensitivity of system's behavior to disturbances, is seemingly opposite to that of IQHE, namely, the robustness of the Hall conductivity quantization.

Contrary to the intuitive reasonings above, we report here that in a large class of spin-$\frac{1}{2}$ quasiperiodic QKR the Planck-IQHE (cf. Fig.~\ref{fig:5}) occurs. Specifically, we find that the inverse Planck's quantum $h_e^{-1}$ mimics the `filling fraction' in conventional IQHE. We also find that the asymptotic growth rate of (rescaled) energy $E(t)$, i.e., $\lim_{t\rightarrow \infty}\frac{E(t)}{t}$, mimics the longitudinal or diagonal conductivity in conventional IQHE. For almost all values of $h_e$, the `longitudinal conductivity' $\lim_{t\rightarrow \infty}\frac{E(t)}{t}=0$,
and
the system simulates an insulator. Surprisingly, there is an infinite discrete set of critical values of $h_e$, for which the `longitudinal conductivity' $\lim_{t\rightarrow \infty}\frac{E(t)}{t}$ has a universal value $\sigma^*={\cal O}(1)$. Correspondingly, the system simulates a quantum metal. The insulating phases, however, are conceptually different from conventional rotor insulators \cite{QKR79,Izrailev90,Chirikov79,Fishman10,Raizen95,Altland10,Fishman84}. That is, we find that each of them is characterized by
a hidden quantum number, denoted as $\sigma_{\rm H}^*$,
which is of topological nature. 
For small $h_e
$, this number jumps by unity whenever
$h_e$ decreases, passing through a critical value. As such, $\sigma_{\rm H}^*$ mimics the `quantized Hall conductivity' in conventional IQHE and its jump at the critical $h_e$-value simulates a `plateau transition'.

The Planck-IQHE is totally beyond the common wisdom of the translation symmetry-based mechanism for various $h_e$-driven phenomena in QKR. Rather, as we will show below, it is of strong chaos origin. To be specific, the rotor's energy $E(t)$ can be {\it exactly} expressed in terms of a functional integral. The corresponding field configuration induces a mapping from the phase space, whose coordinates are the position and velocity characterizing the coherent propagation of quantum amplitudes, onto a certain target space, and carries complete information about the propagation. When the coherent propagation is strongly chaotic at short (microscopic) scales, the phase space corresponding to the propagation at large scales is reduced effectively. The ensuing mappings fall into different homotopic classes, which form a group $\simeq \mathbb{Z}$. This is the topological structure hidden behind discriminating the insulating phases by the quantum number $\sigma_{\rm H}^*$, see Fig.~\ref{fig:5}.

More precisely, we show that the exact functional integral formalism is reduced to a Pruisken-type effective field theory (see Refs.~\onlinecite{Pruisken10,Pruisken84a} for
reviews) at large scales. The key feature of this field theory is that in the effective action a topological theta term emerges from strong chaos at short scales. We emphasize that this term is {\it not} added by hand. The theta term has many far-reaching consequences. Most importantly, the coefficient of this term, the bare (unrenormalized) topological theta angle, is strongly renormalized at large scales and quantized. The quantization value is essentially the plateau value $\sigma_{\rm H}^*$ (up to an irrelevant factor $2\pi$). The insulating phases, distinct from each other by this value, are thus topological in origin, and so is the metal-insulator transition accompanying a plateau transition. A manifestation of the topological nature of this metal-insulator transition is the universality of the critical growth rate $\sigma^*$. Therefore, this transition is conceptually different from the metal-insulator transition in spinless quasiperiodic QKR \cite{Casati89,Altland11,Deland08}.

We emphasize that the emergence of a theta term does not necessarily lead to the Planck-IQHE. An additional key ingredient is the coupling between the rotor's angular and spin degrees of freedom. We find that this coupling plays certain roles of the magnetic translation in conventional IQHE \cite{Thouless82,Bellisard94}, but physical reasons for this similarity remain unclear. Specifically, combined with the mathematical structure of $\frac{1}{2}$-spin, i.e.,
\begin{equation}\label{eq:317}
    \sigma^i\sigma^j+\sigma^j\sigma^i=2\delta_{ij},\quad [\sigma^i,\sigma^j]=2i\varepsilon^{ijk}\sigma^k,
\end{equation}
with $\sigma^i, i=1,2,3$ being the Pauli matrices and $\varepsilon^{ijk}$ the totally antisymmetric tensor, this coupling gives
\begin{eqnarray}
\label{eq:318}
  {\rm unrenormalized\,\, theta\,\, angle}=\mathscr{F}_{ijk}{\rm tr}(\sigma^i\sigma^j\sigma^k).
\end{eqnarray}
Here the coefficient $\mathscr{F}_{ijk}$ depends on the coupling. The Einstein summation convention is implied throughout. Detailed analysis shows that Eq.~(\ref{eq:318}) bears a close resemblance to a classical Hall conductivity in electronic systems, both formally and physically. In particular,
when $h_e^{-1}$ is sufficiently large, this unrenormalized angle linearly increases with $h_e^{-1}$,
and thereby simulates a classical Hall conductivity in strong magnetic field \cite{Pruisken84a}, thanks to the analogy between $h_e^{-1}$ and the filling fraction. It is the renormalization of this linear scaling law that gives the stair-like pattern in Fig.~\ref{fig:5}.


To the best of our knowledge, this is the first time that a topological theta angle, which leads to remarkable results, is discovered in chaotic systems.
A similar topological theta angle was originally proposed in studies of quantum chromodynamics \cite{Polyakov75,tHooft76,Gross76,Jakiw76}. In a series of works \cite{Pruisken84b,Pruisken84d,Pruisken84c,Pruisken84}, Pruisken and co-workers brought this concept to the condensed matter field in treating the discovery of Klitzing and co-workers and established the profound relation between the renormalization of theta angle and quantized Hall conductivity. However, unlike the situation in conventional IQHE, whether the theta angle here can be directly measured by certain `transport' experiments (real or numerical) remains unclear to us at present.

The remainder of the paper is organized as follows.
In the next section we describe in details the model and summarize the main results.
In addition, we discuss qualitatively the topological nature of these results and,
in particular, sketch how the topological structure emerges from strongly chaotic motion at microscopic scales.
In Sec.~\ref{sec:field_theory} we develop an analytic microscopic theory for
the spin-$\frac{1}{2}$ quasiperiodic QKR, for which the potential profile is modulated in time and
the modulation frequency is incommensurate with the kicking frequency.
In Sec.~\ref{sec:transport_parameter} we introduce two transport parameters and
use the developed microscopic theory to calculate their perturbative and
nonperturbative instanton contributions. The explicit results enable us to construct a two-parameter scaling theory and further
analytically predict the Planck-IQHE, which is the subject of Sec.~\ref{sec:microscopic_theory}.
In Sec.~\ref{sec:numerical_test_incommensurate} we confirm numerically
the predicted Planck-IQHE. In Sec.~\ref{sec:absence_IQHE} we show analytically
that the Planck-IQHE disappears when the modulating
frequency is commensurate with the kicking frequency and confirm this prediction numerically.
The corresponding microscopic mechanism is discussed. Conclusions are made in Sec.~\ref{sec:conclusion}. For the convenience of readers we present many technical details in Appendixes \ref{sec:massive}--\ref{sec:finite_time_effects}.

\section{Main physical results and discussions}
\label{sec:summary_main_results}

In this section we summarize the main physical results. Moreover, because a transparent picture for these results is currently absent, we discuss physical implications covered by the analytic microscopic theory. In particular, because our finding of Planck-IQHE is beyond the canonical Thouless-Kohmoto-Nightingale-Nijs (TKNN) paradigm for general quantum Hall systems \cite{Thouless82,Thouless85}, we will sketch -- leaving the complete microscopic theory in later sections -- how the topological structure emerges from the intrinsic strong chaoticity of dynamics of spin-$\frac{1}{2}$ quasiperiodic QKR and further leads to the Planck-IQHE. In doing so, we hope that the readers who are not interested in technical details could skip the next two sections and move to Sec.~\ref{sec:microscopic_theory} directly.

\begin{figure}[h]
\includegraphics[width=8.6cm]{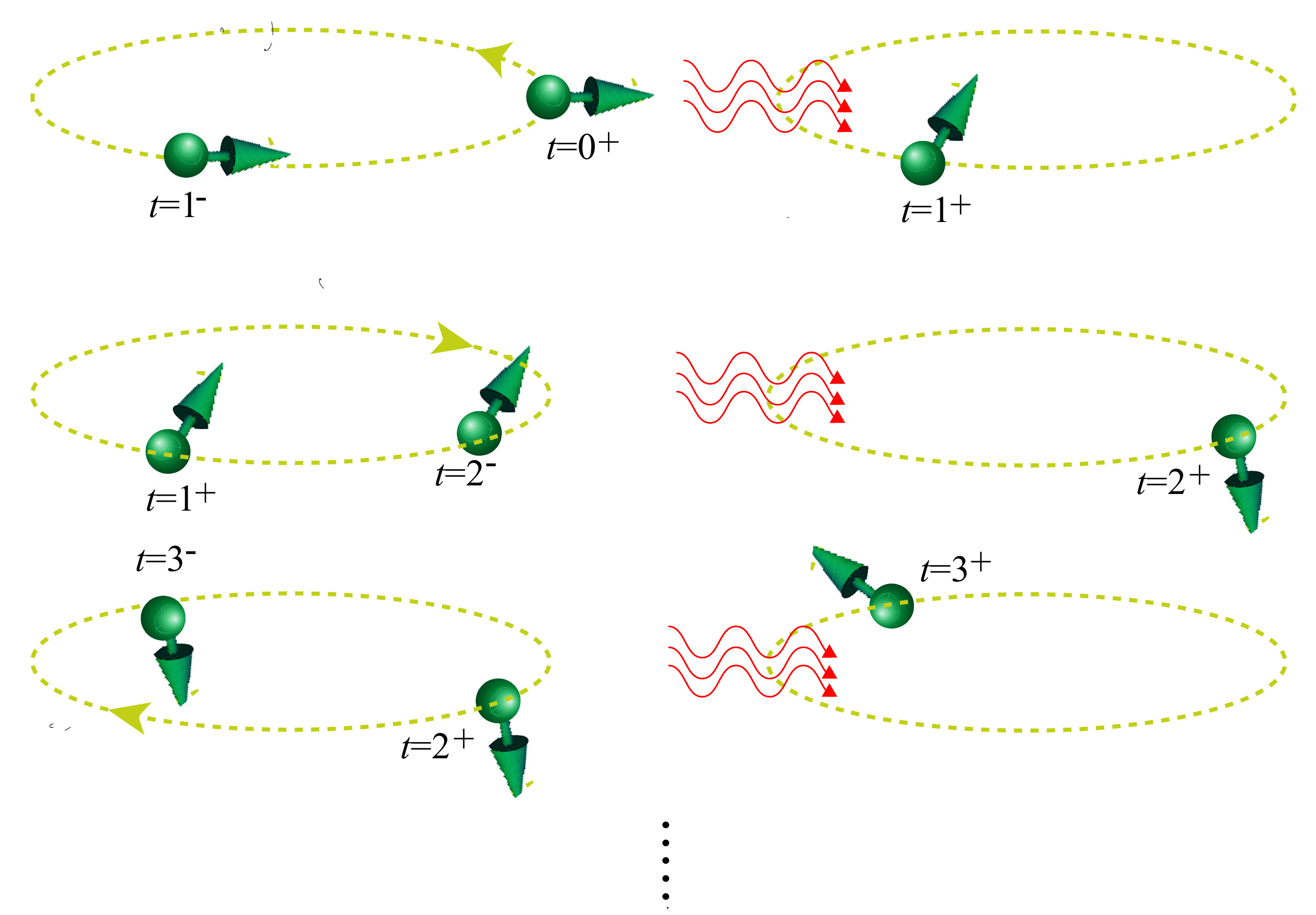}
\caption{A spin-$\frac{1}{2}$ particle moving on a ring (yellow dashed lines) subjected to a train of pulsed external potential -- `kicking' -- switched on at integer times $t=s=0,1,2,\cdots$ (red wavy lines), with the time rescaled by the kicking period. The potential depends on both the angular position and the spin of the particle, and its profile is modulated in
time with a single modulation frequency $\tilde \omega$. Between two successive potential pulses, i.e., $s^+\leq t \leq (s+1)^-$, both the angular momentum and the spin polarization of the particle are conserved (left column, with the arrow of the dashed line representing the direction of rotation); upon kicking, i.e., at time $t=(s+1)^+$, the particle undergoes an abrupt change in the angular momentum and a flip of spin simultaneously (right column). The superscripts `$\pm$' stand for the time right before (after) kicking.}
\label{fig:4}
\end{figure}

\subsection{Description of the model}
\label{sec:description_model}

In the present work, we consider a spin-$\frac{1}{2}$ particle moving on a ring (Fig.~\ref{fig:4}) whose angular position is denoted as $\theta_1$. When the external potential is switched off, the particle has a conserved angular momentum, i.e., moves on the ring with a constant angular speed (Fig.~\ref{fig:4}, left column). This is a completely integrable motion, corresponding to a Hamiltonian, $H_0(-i\hbar\partial_{\theta_1})$ which is a function of the angular momentum $-i\hbar\partial_{\theta_1}$. Here, $H_0$ needs not to be quadratic and takes a very general form, whose details determine how the angular speed is related to the angular momentum. Importantly, this Hamiltonian does not recognize the particle's spin degree of freedom. As such, the spin polarization is also conserved between two successive kickings. At the instant of $t=s\tau$ ($s=0,1,2,\cdots$), when the external potential is switched on, the particle undergoes an abrupt change in the angular momentum and, simultaneously, a spin flip (Fig.~\ref{fig:4}, right column). The angular ($\theta_1$) profile of the potential, denoted as $V(\theta_1,\theta_2+\tilde \omega t)$, is modulated in time with a modulation frequency $\tilde \omega$, where $\theta_{2}$ is an arbitrarily prescribed phase parameter.

The potential is a function of two angular variables, with a general form,
\begin{equation}\label{eq:237}
    V(\theta_1,\theta_2)\equiv V_i(\theta_1,\theta_2)\sigma^i,
\end{equation}
where
none of the coefficients $V_i$ identically vanishes. The parity properties of $V_i$ with respect to the transformation: $\theta_{1,2}\rightarrow -\theta_{1,2}$ are listed in Table~\ref{Table2}. Note that for Eq.~(\ref{eq:237}) the second variable includes the time modulation. $V_i(\theta_{1},\theta_{2})$ is periodic in both variables, i.e.,
\begin{eqnarray}
V_i(\theta_{1}+2\pi,\theta_{2})=V_i(\theta_{1},\theta_{2}+2\pi)=V_i(\theta_{1},\theta_{2}).
\label{eq:312}
\end{eqnarray}

\begin{table}[htbp]
\caption{\label{Table2} Symmetry properties of $V_i$.}
\begin{tabular}{c|c|c|c}
  \hline\hline
   & $V_1(\theta_1,\theta_2)$
   & $V_2(\theta_1,\theta_2)$
   & $V_3(\theta_1,\theta_2)$ \\
   \hline
  $\theta_1$ & odd & even & even \\
  \hline
  $\theta_2$ & even & odd & even \\
  \hline\hline
\end{tabular}
\end{table}

The motion of the moving particle is described by a two-component spinor, $\tp_t$. With the rotor's angular momentum and $\hbar$ rescaled by $I/\tau$ and the time by $\tau$, the quantum dynamics is described by
\begin{eqnarray}
ih_e \partial_t \tp_t = {\hat H}(t) \tp_t\,,
\qquad\qquad\qquad \label{eq:6}\\
{\hat H}(t) =
H_0(-ih_e\partial_{\theta_1})+
V(\theta_1,\theta_2+\tilde \omega t)\sum_s\delta
(t-s).\nonumber
\end{eqnarray}
This is a $1$D motion. The rotor's energy is defined as
\begin{equation}\label{eq:26}
    E(t) \equiv -\frac{1}{2}
\langle\!\langle \tilde \psi_t|\partial_{\theta_1}^2|\tilde \psi_t\rangle\!\rangle_{\theta_2},
\end{equation}
with $\langle\cdot\rangle_{\theta_2}
$ being the average over the prescribed phase
$\theta_2$.
For simplicity we assume that the initial state is uniform in $\theta_1$ throughout.
For (non)vanishing $\lim_{t\rightarrow\infty} \frac{E(t)}{t}$,
the rotor exhibits (un)bounded motion in
angular momentum space and simulates an insulator (a metal) in disordered electronic systems. Note that, exactly speaking, the definition (\ref{eq:26}) has the meaning of rotor's rotation energy only for quadratic $H_0$; in general,
it has the meaning of angular momentum variance instead. Here we follow the convention in most studies of QKR.

\subsection{Topology structure from strong chaos}
\label{sec:origin_IQHE}

\subsubsection{Equivalent time-periodic quantum system}
\label{sec:2D_system}

We notice that for $\hat H(t)$ in Eq.~(\ref{eq:6}), for each unit time interval the increment in the external parameter $(\theta_2+\tilde \omega t)$ is the same, i.e., the modulation frequency $\tilde \omega$. Therefore, one may expect that Eq.~(\ref{eq:6}) could be traded for a two-dimensional ($2$D) strictly periodic system by interpreting the parameter as a `virtual' dynamical variable. This indeed can be achieved by performing the transformation \cite{Casati89,Altland11},
\begin{equation}\label{eq:32}
    \hat H \rightarrow e^{\tilde \omega t \partial_{\theta_2}}\hat H
e^{-\tilde \omega t\partial_{\theta_2}},  \tilde\psi_t
\rightarrow e^{-\tilde \omega t\partial_{\theta_2}} \tilde\psi_t \equiv \psi_t,
\end{equation}
for Eq.~(\ref{eq:6}), which gives
\begin{eqnarray}
ih_e \partial_t \psi_t = (H_0(h_e \hat n_1)+h_e\tilde \omega {\hat n}_2 +V(\Theta)\sum_s\delta (t-s)) \psi_t.
\label{eq:4}
\end{eqnarray}
Here $\Theta\equiv (\theta_{1},\theta_{2})$ is canonically conjugate to angular momenta $\hat N\equiv (\hat n_{1},\hat n_{2})$ and $\theta_2$ now is a virtual dynamical variable. Equation (\ref{eq:4}) describes a generalized QKR. It is very important that this equivalent system is strictly time-periodic and $2$D. For integer times Eq.~(\ref{eq:4}) is reduced to autonomous stroboscopic dynamics,
\begin{eqnarray}
\psi_t &=&
{\hat U}^t \psi_0,\nonumber\\
\hat U &\equiv& e^{-\frac{i}{h_e}\left(H_0(h_e {\hat n}_1)+h_e{\tilde \omega} {\hat n}_2\right)
}e^{-\frac{i}{h_e}V(\Theta)},
\label{eq:5}
\end{eqnarray}
governed by the
Floquet operator, $\hat U$. The initial state, $\psi_0$, corresponding to this $2$D dynamics is uniform in
$\Theta
$-representation. For this $2$D equivalent the (effective) time-reversal symmetry is broken. That is, Eq.~(\ref{eq:4}) is not invariant under the operation $
i\sigma^2 K'$,
where $K'$ is the combination of complex conjugation
and the operation: $t\rightarrow -t,
\Theta\rightarrow -\Theta,\hat N\rightarrow \hat N$.

\begin{figure}[h]
\includegraphics[width=8.6cm]{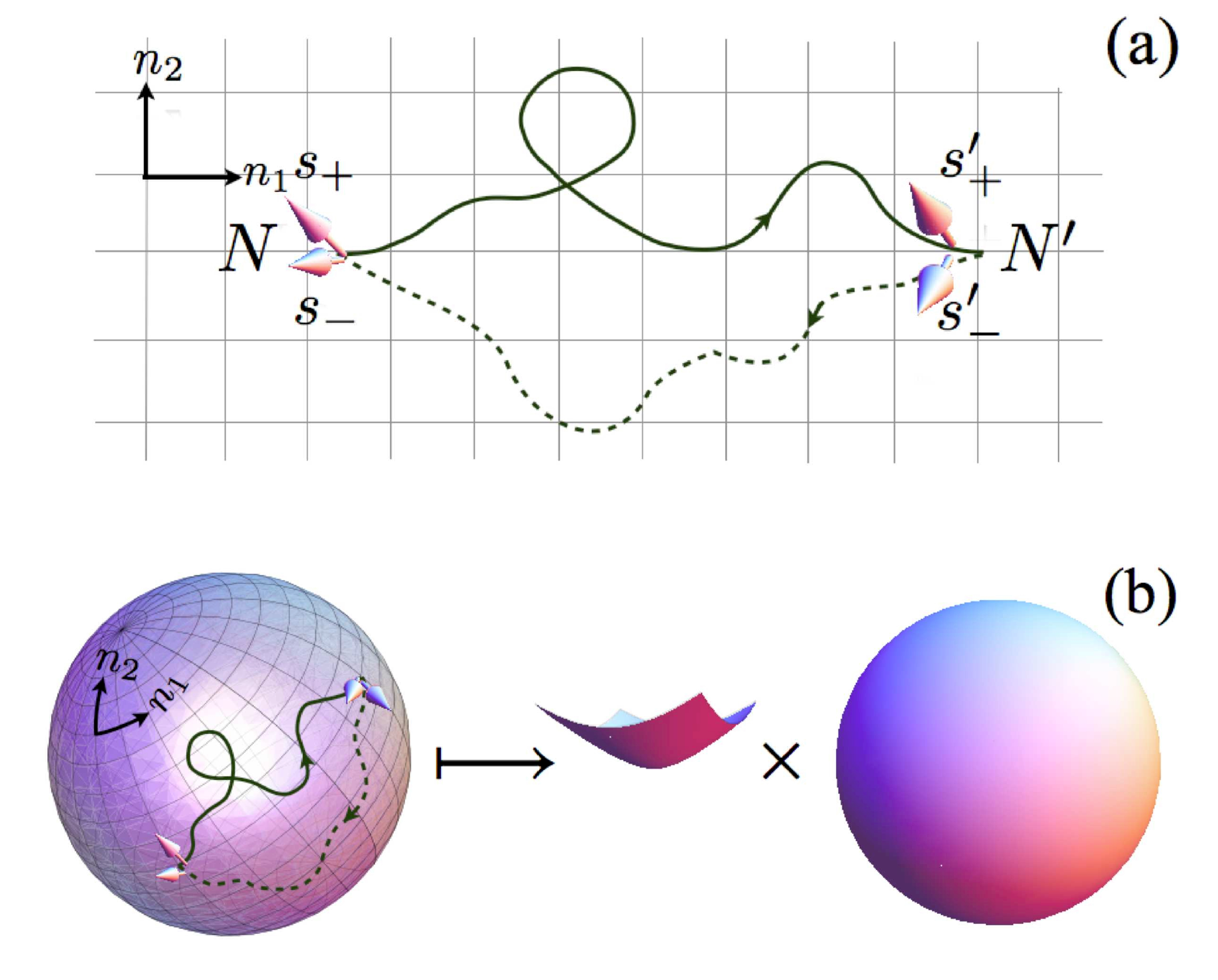}
\caption{
(a) A quasiperiodic QKR with one modulation frequency is equivalent to a $2$D {\it strictly} periodic QKR. The latter corresponds to a quantum motion in the $N\equiv (n_1,n_2)$ space under the influence of a periodic driving force. $n_1$ is the genuine angular momentum and $n_2$ virtual, canonically conjugate to the modulation parameter. The properties of localization in the $N$ space are governed by interference between
the advanced (solid line) and the retarded (dashed line) quantum amplitudes.
(b) When the coherent propagation of the advanced and the retarded quantum amplitudes is chaotic in both $n_1$- and $n_2$-directions, mappings from the $N$ space (with its boundary identified as the same point)
to the supersymmetry $\sigma$-model space of unitary symmetry $\simeq H^2\times S^2$ arise. These mappings are classified by the homotopy
group $\mathbb{Z}$.}
\label{fig:1}
\end{figure}

Starting from the $2$D dynamics described by Eq.~(\ref{eq:5}) we can express the rotor's energy as
\begin{eqnarray}\label{eq:287}
    E(t)&=&\frac{1}{2}\int\frac{d\omega}{2\pi}e^{-i\omega t}\sum_{NN'}\sum_{s_\pm,s'_\pm} \delta_{N'0}\delta_{s_+s_-}\nonumber\\
    &&\times n_1^2 K_\omega(Ns_+s_-,N's'_+s'_-)\psi_{0s'_+}\psi^*_{0s'_-}.
\end{eqnarray}
The function $K_\omega$ may be considered as the correlation between the bilinear $|N's_+'\rangle\langle N's_-'|$ and $|Ns_+\rangle\langle Ns_-|$. Physically, it describes the interference between
the advanced and the retarded quantum
amplitudes for the motion in the angular momentum ($N$) space (Fig.~\ref{fig:1}(a)). This interference governs the localization physics of this $2$D quantum system. The exact definition of $K_\omega$ is not important for present discussions and will be given later (see Eq.~(\ref{eq:107})).

\subsubsection{Emergence of topology structure at irrational $\tilde \omega/(2\pi)$}
\label{sec:topologically_nontrivial_mappings}

It turns out that $K_\omega$ can be {\it exactly} expressed in terms of a functional integral. For discussions in this section it is sufficient to give its symbolic expression,
\begin{equation}\label{eq:288}
    K_\omega\sim \int D(Z,\tilde Z)e^{-S[Z,\tilde Z]}(...),
\end{equation}
and refer to Eq.~(\ref{eq:33}) below for its complete form. In this expression $Z_{N_1,N_2},\tilde Z_{N_1,N_2}$ depend on two angular momenta $N_{1,2}$ and take supermatrix value. Physically, $Z_{N_1,N_2}$ (or $\tilde Z_{N_1,N_2}$) is the representative of the coherent propagation of the advanced and the retarded quantum amplitudes. Specifically, passing to the Wigner representation, $Z_{N_1,N_2}\rightarrow Z(N,\Theta)\equiv \sum_{\Delta N}e^{-i\Delta N \Theta}Z_{N_1,N_2}$ with $\Delta N\equiv N_1-N_2$, $N\equiv (N_1+N_2)/2$ is the `center-of-mass' coordinate and $\Theta$ (more precisely, $(\sin \theta_1,\sin\theta_2)$) the `velocity' of the coherent propagation. Moreover, the action $S[Z,\tilde Z]$ carries the complete information of this propagation. From the Wigner representation, we see that the velocity relaxation is encoded by the components of $Z$ which are off-diagonal in angular momentum space.

\begin{table*}
\newcommand{\tabincell}[2]
{\begin{tabular}{@{}#1@{}}#2
\end{tabular}}
\centering
\caption{\label{Table4} Properties of various quantum phases.}
\begin{tabular}{c|c|c|c|c|c|c}
\hline\hline
\multirow{2}{*}{phase} & \multicolumn{2}{c}{fixed point of RG flow} &
\multicolumn{2}{|c|}{stability of RG flow}& \multirow{2}{*}{energy profile}& \multirow{2}{*}{classification}\\
\cline{2-5}
& $\tilde \sigma_{\rm H}$ & $\tilde \sigma$ & $\tilde \sigma_{\rm H}$-direction & $\tilde \sigma$-direction & &\\
\hline
insulating & \,\,\,\,$\sigma_{\rm H}^*\in \mathbb{Z}$\,\,\,\,\,\,\,\,& $0$ & stable & stable & $E\stackrel{t\rightarrow \infty}{\longrightarrow}const.$&$\mathbb{Z}$\\
\hline
metallic & $\sigma_{\rm H}^*+\frac{1}{2}$&$\sigma^*$ & unstable& stable & $E\stackrel{t\rightarrow \infty}{\longrightarrow}\sigma^* t$&$-$ \\
\hline
\hline
\end{tabular}
\end{table*}

Thanks to its supermatrix structure, $Z(N,\Theta)$ induces a mapping from the `phase space' comprised of coordinates $(N,\Theta)$ to a target space, the so-called supersymmetry $\sigma$-model space of unitary symmetry. Loosely speaking, the latter space may be identified as a product of two-hyperboloid, $H^2$, and a two-sphere, $S^2$. This structure is intrinsic to the broken time-reversal symmetry. Detailed discussions of this structure will be made in Sec.~\ref{sec:physical_meanings_topological_term}.

When the system exhibits strong chaos so that the memory on the velocity is quickly lost, the components $Z_{N,N'}$ with $N\neq N'$ are massive. In other words, the propagation modes represented by these components damp rapidly in time. Since we are interested in dynamics taking place at much longer times, these components are effectively suppressed from the functional integral (\ref{eq:288}). The ensuing Wigner representation, $Z(N)=Z_{N,N}$, thus has no $\Theta$-dependence. In this way, a given field configuration $Z(N)$ entails a mapping from the $2$D $N$ space (with its boundary identified as the same point) onto the target space $H^2\times S^2$ (Fig.~\ref{fig:1}(b)). All such mappings are classified by a nontrivial homotopy group,
\begin{eqnarray}\label{eq:2}
    \pi_2\left(
    H^2\times S^2
    \right)=\pi_2(H^2)\times\pi_2(S^2)=\pi_2(S^2)=\mathbb{Z}.
\end{eqnarray}
Note that $H^2$ has no topological consequences due to its non-compact nature.

This nontrivial homotopy group (\ref{eq:2}) lays down a mathematical foundation for searching topologically insulating phases in simple spin-$\frac{1}{2}$ quasiperiodic QKR with one modulation frequency and possible topological transitions between them. However, we emphasize that this topological structure is not sufficient for the occurrence of Planck-IQHE. As we shall see shortly, an additional key factor responsible for the Planck-IQHE is a universal scaling law, which is rooted in the coupling between the rotor's spin and angular degrees of freedom, but insensitive to the details of the coupling.

We have seen that to establish the topological structure (\ref{eq:2}) the $2$D dynamics is required to be strongly chaotic. That is, $\Theta$ is a fast variable while $N$ a slow variable. To achieve this, it is necessary that the modulation frequency $\tilde \omega$ is incommensurate with $2\pi$, i.e., $\tilde \omega/(2\pi)$ is a (generic) irrational number so that the system is a quasiperiodic QKR \cite{note_nongeneric_omega}.

\subsubsection{Absence of topology structure at rational $\tilde \omega/(2\pi)$}
\label{sec:absence_topologically_nontrivial_mappings}

For rational $\tilde \omega/(2\pi)$, the system (more precisely, the Floquet operator, $\hat U$) is translationally invariant in $n_2$-direction. Associated with this translation symmetry the $2$D coherent propagation is ballistic in $n_2$-direction at long times, while memory on the velocity component in $n_1$-direction is quickly lost. The ballistic motion in $n_2$-direction implies partial restoration of regular dynamics.
In this case, the topological structure shown above is washed out.
Indeed, the $2$D system is decomposed into a family of decoupled (quasi) $1$D subsystems, each of which is governed by a good quantum number, namely, the Bloch momentum. The ensuing $Z$-field configurations entail
mappings from the $n_1$ space
into the same target space, i.e., $H^2\times S^2$, as that for irrational $\tilde \omega/(2\pi)$. These mappings are all topologically trivial, since the corresponding homotopy group is
\begin{eqnarray}\label{eq:31}
    \pi_1\left(H^2\times S^2
    \right)=\pi_1(H^2)\times\pi_1(S^2)=0.
\end{eqnarray}
Because of this -- a result of the restoration of dynamics regularity, no topologically insulating phases arise, and therefore the Planck-IQHE does not occur.

\begin{table*}
\newcommand{\tabincell}[2]
{\begin{tabular}{@{}#1@{}}#2
\end{tabular}}
\centering
\caption{\label{Table1} The analogy between Planck- and conventional IQHE.}
\begin{tabular}{c|c|c}
  \hline\hline
  &Planck-IQHE& conventional IQHE\\
  \hline
  system & spin-$\frac{1}{2}$ quasiperiodic QKR & $2$D electron gas (e.g., MOSFET) \\
  \hline
  driving parameter & Planck's quantum $h_e$ & magnetic field (or inverse filling fraction) \\
  \hline
  characteristic of dissipation& energy growth rate $\lim_{t\rightarrow \infty}\frac{E(t)}{t}$ & longitudinal conductivity \\
  \hline
  characteristic of topology & hidden quantum number $\sigma_{\rm H}^*$ & quantized Hall conductivity\\
  \hline
  characteristic of insulator & $\lim_{t\rightarrow \infty}\frac{E(t)}{t}=0$ & vanishing longitudinal conductivity\\
  \hline
  characteristic of metal & $\lim_{t\rightarrow \infty}\frac{E(t)}{t}=\sigma^*$ & finite longitudinal conductivity\\
  \hline\hline
\end{tabular}
\end{table*}

\subsection{Summary of main physical results}
\label{sec:main_results}

\subsubsection{Irrational $\tilde \omega/(2\pi)$}
\label{sec:irrational_omega}

In this case we find that the action $S[Z,\tilde Z]$ in Eq.~(\ref{eq:288}) is reduced to a $2$D effective action (cf. Eq.~(\ref{eq:111})) at large scales. Most importantly, this effective action includes a term which is purely topological in nature (see the second term in Eq.~(\ref{eq:111})). This is the very topological theta term that does not show up in all previous effective field theories for various QKR systems \cite{Altland11,Altland10}. In addition, the effective action is governed by two (unrenormalized) parameters, $\sigma(h_e)$ and $\sigma_{\rm H}(h_e)$. 
For sufficiently small $h_e
$, they exhibit universal scaling behavior, i.e.,
\begin{eqnarray}\label{eq:291}
    \sigma \sim h_e^{-2}
\end{eqnarray}
and
\begin{equation}\label{eq:9}
    \sigma_{\rm H} \sim h_e^{-1}.
\end{equation}
These two scaling laws are independent of the details of the coupling between the angular and spin degrees of freedom, i.e., $V_i(\Theta)$. The parameter $2\pi \sigma_{\rm H}$ gives the (unrenormalized) topological theta angle, while $\sigma$ is found to be the energy growth rate at short times. Interestingly, if we interpret $h_e^{-1}$ as the filling fraction, Eq.~(\ref{eq:9}) corresponds to the classical Hall conductivity in conventional Hall systems with a strong magnetic field \cite{Pruisken84a}. As shown below, the $h_e^{-1}$--filling fraction analogy persists even at long times.

At long times the scaling laws (\ref{eq:291}) and (\ref{eq:9}) break down. Instead, $\sigma$ and $\sigma_{\rm H}$ are strongly renormalized. In this work we explicitly show that their renormalized values, respectively denoted as $\tilde \sigma(\tilde \lambda)$ and $\tilde\sigma_{\rm H}(\tilde \lambda)$ with $\tilde \lambda$ being the scaling parameter, follow Gell-Mann--Low equations,
\begin{equation}\label{eq:292}
    \frac{d\tilde \sigma}{d\ln \tilde\lambda}=-\frac{1}{
    8\pi^2\tilde \sigma}-\frac{
    32\pi}{e} \tilde \sigma^2 e^{-4\pi\tilde \sigma}\cos 2\pi \tilde \sigma_{\rm H},
\end{equation}
and
\begin{equation}\label{eq:293}
    \frac{d\tilde \sigma_{\rm H}}{d\ln \tilde\lambda}=-\frac{
    64\pi}{e} \tilde \sigma^2 e^{-4\pi\tilde\sigma}\sin 2\pi \tilde \sigma_{\rm H},
\end{equation}
in the weak coupling regime. This renormalization group (RG) flow leads to profound results, which we summarize below. The results also capture the system's behavior in the strong coupling regime, even quantitatively. They are robust against the modification of $H_0$ and $V_i$.

First of all, the fixed points of this RG flow give the realizable quantum phases in considered quasiperiodic QKR. The main properties of these phases are summarized in Table \ref{Table4}. The insulating phases correspond to the plateau regimes in Fig.~\ref{fig:5}. They have a vanishing energy growth rate at $t\rightarrow \infty$. Namely, $E(t)$ saturates at $t\rightarrow \infty$. These phases are distinguished by the plateau value $\sigma_{\rm H}^*$. The formation of plateaus is a result of the renormalization of topological theta angle. As such, the insulating phases are endowed with topological nature and, therefore, are conceptually different from usual rotor insulators \cite{Raizen95,QKR79,Izrailev90,Chirikov79,Fishman10,Casati89,Altland11,Deland08,Fishman84}. The metallic phase corresponds to the peak in Fig.~\ref{fig:5}. It appears only at the plateau transition, i.e., is a critical phase. At this critical phase, $E(t)$ grows linearly at long times, with a small growth rate $\sigma^*={\cal O}(1)$. Strikingly, this growth rate is universal, independent of system's details such as specific forms of $H_0$, $V_i$, and quantum critical points.
This suggests that this rotor metal is of quantum nature.

Next, combined with the universal scaling law (\ref{eq:9}), the RG flow gives rise to the Planck-IQHE. Indeed, following from Eq.~(\ref{eq:9}) the unrenormalized parameter $\sigma_{\rm H}(h_e)$ increases unboundedly with $h_e^{-1}$. As a result, there is an infinite discrete set of critical $h_e$-values namely quantum critical points, at which $\sigma_{\rm H}(h_e)$ is a half-integer,
\begin{equation}\label{eq:60}
    \sigma_{\rm H}(
    h_e)=n+\frac{1}{2}, \quad n\in \mathbb{Z},
\end{equation}
i.e., the zero of the right-hand side of Eq.~(\ref{eq:293}). When $h_e$ decreases, the system successively passes through these critical points, at each of which $\sigma_{\rm H}^*$ jumps by unity (the plateau transition in Fig.~\ref{fig:5}), accompanied by a topological metal-insulator transition (the peak in Fig.~\ref{fig:5}). In addition, the scaling law (\ref{eq:9}) implies that the quantum critical points are evenly spaced along the $h_e^{-1}$-axis.

Comparing the results summarized above with the discovery of Klitzing and co-workers \cite{Klitzing80}, we find that $h_e^{-1}$ mimics the driving parameter, namely the filling fraction in conventional IQHE, as discussed above. Furthermore,
$\lim_{t\rightarrow \infty}\frac{E(t)}{t}$ and $\sigma_{\rm H}^*$ simulate two transport parameters, namely, the longitudinal conductivity and the quantized Hall conductivity, respectively (Table~\ref{Table1}). In this paper, this analogy will be put on a firm ground, both analytically and numerically.

\subsubsection{Rational $\tilde \omega/(2\pi)$}
\label{sec:rational_omega}

Although the Planck-IQHE is found to be very robust against the modification of $H_0$ and $V_i$, we find that it is extremely sensitive to the number-theoretic property of the modulation frequency $\tilde \omega$. Specifically, when $\tilde \omega$ becomes commensurate with $2\pi$ (but $H_0$ and $V_i$ do not change), we find that the action $S[Z,\tilde Z]$ in Eq.~(\ref{eq:288}) is reduced to a $1$D effective action (cf. Eq.~(\ref{eq:45})) at large scales. Most importantly, this effective action does not include any topological term and is essentially the same as that for the conventional QKR \cite{Altland10}. Following from this effective field
theory the system behaves as conventional QKR \cite{QKR79,Fishman84,Izrailev90,Altland10}, i.e., the rotor's energy saturates at long times irrespective of the value of Planck's quantum \cite{footnote_1}. Therefore, the phenomenon of Planck-IQHE is washed out. It is likely that this occurs also for nongeneric irrational values of $\tilde \omega/(2\pi)$.

\subsection{Discussions on critical metallic phase}
\label{sec:quantum_stochastic_web}

\begin{figure}[h]
\includegraphics[width=8.6cm]{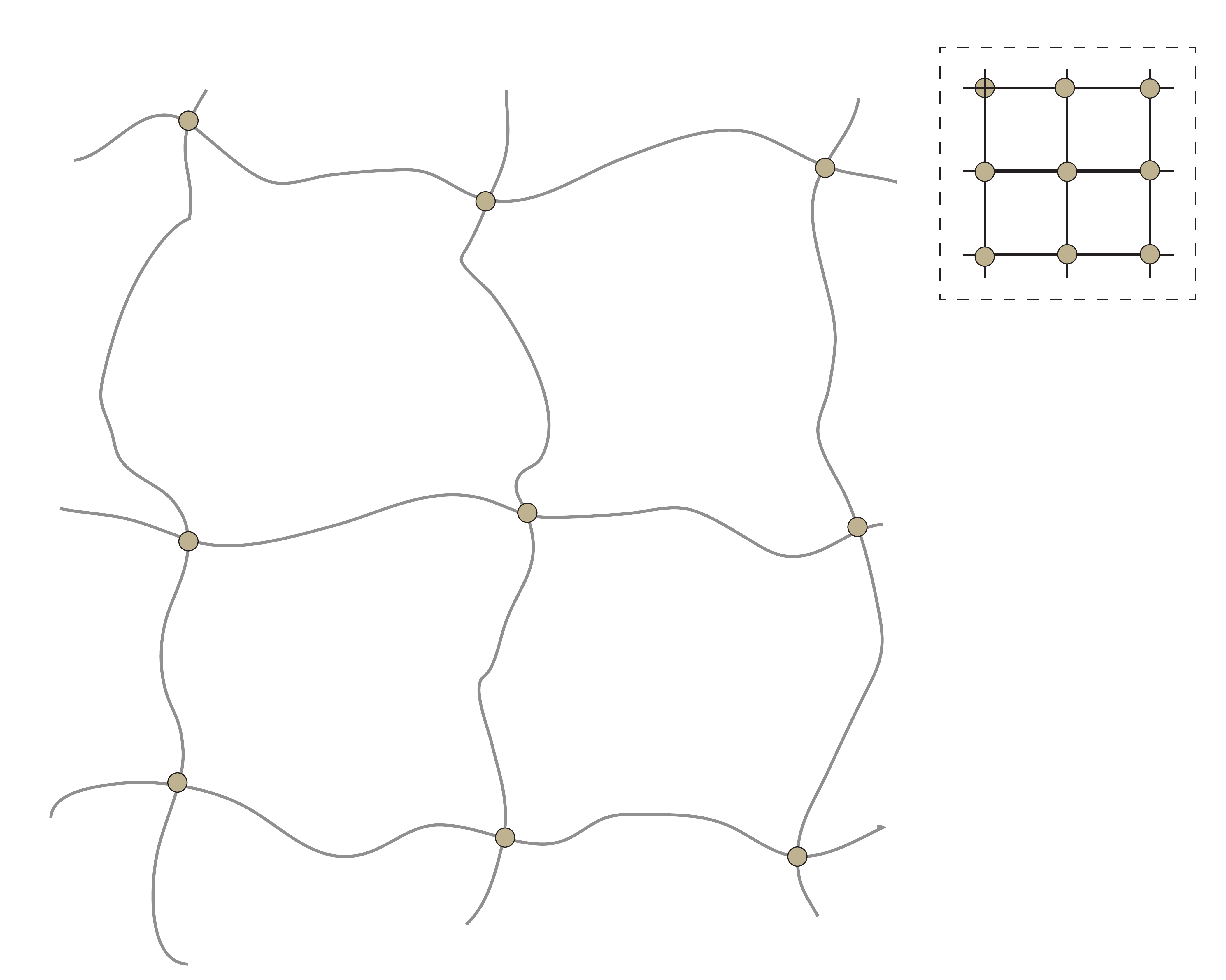}
\caption{The quantum stochastic web in $N$ space as a possible physical picture for the critical metallic phase. The link supports ballistic motion while the node is the origin of quantum stochasticity. Inset: this web is topologically equivalent to a graph made up of links and nodes.}
\label{fig:12}
\end{figure}

The linear energy growth, $E(t\rightarrow \infty)\sim t$, for classical kicked rotors has been well understood \cite{Chirikov79a}. It finds its origin at stochastic diffusion (Brownian motion) in angular momentum space. Although the linear growth also displays at the critical metallic phase, it exhibits considerable `anomalies'. Most strikingly, the growth rate $\sigma^*$ is small, which is order of unity, and universal. This indicates the quantum nature of the critical metal. It also indicates that the canonical physical picture for linear energy growth of classical kicked rotors must break down here, since the picture leads to a growth rate sensitive to system's details such as the kicking strength and potential which is not the case here. Below we discuss a possible picture -- the quantum stochastic web (Fig.~\ref{fig:12}) -- for the critical metal.

First of all, the smallness of $\sigma^*$ cannot be attributed to a small mean free path, since the latter is governed by the system's details. Rather, it is a signature of certain non-ergodic but unbounded motion in $2$D angular momentum ($N$) space. More precisely, a large portion of the $N$ space is `blocked', and the system has to find narrow channels in order to arrive at a remote point. Figure \ref{fig:12} represents a heuristic example for such channel structure. It is an extended web, topologically equivalent to a graph made up of nodes and links -- stretched channels (the inset). The quantum stochastic diffusion in $N$ space, manifesting itself in $E(t\rightarrow \infty)\rightarrow \sigma^*t$, finds its origin at the node. Then, $\sigma^*$ is the total conductance of this web, essentially given by the ballistic conductance of the link (narrow channel), which is of order unity and universal. Provided the skeleton (topological structure) of this web is universal, independent of system's details such as the potential, the critical point, etc., the universality of $\sigma^*$ then follows.

We note that the classical stochastic web was discovered in many dynamical systems long time ago \cite{Zaslavsky86,Arnold64,Chirikov79a}. It gives rise to intriguing transport phenomena, notably the Arnold diffusion \cite{Arnold64}. However, results for quantum stochastic webs are extremely rare. Among them is the non-ergodic metal in many-body localized systems, which is believed to be a manifestation of Arnold diffusion \cite{Basko06}.

\section{Effective field theory}
\label{sec:field_theory}

Having summarized the main results, we present the analytic derivation. The microscopic theory starts from the $2$D autonomous stroboscopic dynamics (\ref{eq:5}).

\subsection{Functional integral formalism}
\label{sec:formalism}

To calculate the rotor's energy at integer times
we introduce the `two-particle Green function', defined as
\begin{eqnarray}
&&K_\omega(Ns_+s_-,N's'_+s'_-)\equiv \label{eq:107}\\
&&\langle\!\langle Ns_+|\frac{1}{1-e^{i\omega_+}{\hat U}}| N's'_+\rangle
\langle N's'_-|\frac{1}{1-e^{-i\omega_-}{\hat U}^\dagger}|Ns_-\rangle\!\rangle_{\omega_0}.
\nonumber
\end{eqnarray}
Here $s_\pm$ and $s'_\pm$ are spin indices. $\omega_\pm\equiv \omega_0\pm\frac{\omega}{2}$ and
$\omega$ is understood as $\omega+\frac{i{\eta}}{2}$ with ${\eta}$ being an infinitesimal positive. The average,
$\langle\cdot\rangle_{\omega_0}\equiv \int_0^{2\pi}\frac{d\omega_0}{2\pi}(\cdot)$, forces the two
paths Fig.~\ref{fig:1}(a) that interfere with each other to last for the same time. The energy is related to $K_\omega$ via
\begin{equation}\label{eq:7}
    E(t)=\frac{1}{2}\int\frac{d\omega}{2\pi}e^{-i\omega t}{\rm Tr} (\hat n_1^2 K_\omega {\psi}_0\otimes {\psi}_0^\dagger).
\end{equation}
Throughout the capital trace
`Tr'
includes the angular momentum
but the small trace `tr' does not.

To proceed, we combine the methods of deriving effective field theories
for conventional QKR \cite{Altland10,Altland11}
and for graphene with long-range disorder \cite{Ostrovskii07}.
Noticing the broken time-reversal symmetry we introduce a superfield
$\psi=\{\psi_{Ns\alpha \lambda}\}$, where $\alpha$
discriminates between the commuting ($\alpha=b$)
and the anticommuting ($\alpha=f$) component, and the index $\lambda$
between the advanced ($\lambda=+$) and the
retarded ($\lambda=-$) space of the theory.
With this introduction Eq.~(\ref{eq:107}) is cast into a Gaussian functional integral,
\begin{eqnarray}
&&K_\omega(Ns_+s_-,N's'_+s'_-) \nonumber\\
&=&\int D({\bar \psi},\psi)
\left\langle \exp\left(-{\bar \psi}
G^{-1}\psi\right) \right\rangle_{\omega_0}\nonumber\\
&&\quad\quad\quad\quad \times{\bar \psi}_{N's'_+b+}\psi_{Ns_+b+}
{\bar\psi}_{Ns_-b-}\psi_{N's'_-b-},
\label{eq:8}
\end{eqnarray}
where
\begin{eqnarray}
G^{-1}=\left(
         \begin{array}{cc}
           (1-e^{i\omega_+}{\hat U})^{-1} & 0 \\
           0 & (1-e^{-i\omega_-}{\hat U}^\dagger)^{-1} \\
         \end{array}
       \right)_{ar}
\label{eq:299}
\end{eqnarray}
is block-diagonal in the advanced-retarded ({\it ar}) space.

Next, we apply a rigorous identity, the color-flavor transformation
for the circular unitary ensemble~\cite{Zirnbauer96}.
The adjective `circular' accounts for the $\omega_0$-average and
`unitary' for the broken time-reversal symmetry.
This transforms the $\psi$-integral and the $\omega_0$-average
into the integral over a supermatrix field, $Z\equiv \{Z_{Ns\alpha,N's'\alpha'}\}$, in the bosonic-fermonic ({\it bf}) space,
as
\begin{eqnarray}
&&K_\omega(Ns_+s_-,N's'_+s'_-) = \int D(Z,\tilde Z)e^{-S[Z,\tilde Z]}\nonumber\\
&\times& ((1-Z \tilde Z)^{-1}Z)_{Ns_+b,Ns_-b}
((1-\tilde Z Z)^{-1}\tilde Z)_{N's'_-b,N's'_+b}.\nonumber\\
\label{eq:33}
\end{eqnarray}
Here $D(Z,\tilde Z)$ is the flat Berezin measure,
$Z$ is subjected to the constraints: $\tilde Z_{b,b}=Z_{b,b}^\dagger$,
$\tilde Z_{f,f}=-Z_{f,f}^\dagger$
and all the eigenvalues of $Z_{b,b} Z_{b,b}^\dagger$ be less than unity.
The action is given by
\begin{eqnarray}
  \label{eq:12}
  S[Z,\tilde Z] = -\mathrm{Str}\ln(1- Z \tilde Z)+\mathrm{Str}\ln(1-e^{i\omega} {\hat U} Z
  {\hat U}^\dagger \tilde Z).\,\,
\end{eqnarray}
The supertrace ${\rm Str}\equiv {\rm Tr}_{bb}-{\rm Tr}_{ff}$
with ${\rm Tr}_{bb,ff}$ being the normal trace of the bosonic-bosonic (fermionic-fermionic) block. Like the definition of normal trace,
throughout the capital supertrace
`Str' includes the angular momentum
but the small one `str' not. Equations (\ref{eq:33}) and (\ref{eq:12}) constitute an exact supersymmetric functional integral formalism for the two-particle Green function.

Let us make the following observation. Expanding the action in $Z$ up to the second order, we find that the kernel of the ensuing Gaussian action is $(1-e^{i\omega}{\rm Ad}_{\hat U})^{-1}$. Here the notation: ${\rm Ad}_{\hat {\cal O}}(\cdot)\equiv \hat {\cal O} (\cdot)\hat {\cal O}^\dagger$ for a general
operator $\hat {\cal O}$. We see that this kernel, namely the propagator corresponding to the $Z$ field, describes the coherent propagation of the advanced and the retarded quantum amplitudes, as discussed in Sec.~\ref{sec:topologically_nontrivial_mappings}.

\subsection{Low-energy effective action}
\label{sec:low_energy_action}

We proceed to derive the effective field theory, which describes the $2$D motion at large scales, from this formalism. To this end we note that in the zero-frequency
limit, $\omega\rightarrow 0$, the action (\ref{eq:12}) vanishes
provided that the field is constrained by $[{\hat U}, Z]=0$.
To solve this constraint we substitute the decomposition,
$Z\equiv Z_0 \sigma^0 + \boldsymbol{Z}\cdot \boldsymbol{\sigma}$, into it, where $\sigma^0$ is the unit matrix in spin space.
We find that the constraint is satisfied provided that $\boldsymbol{Z}=0$ and $Z_0$ is homogeneous in $N$ space (zero mode).
In other words, the $\boldsymbol{Z}$-components are massive
and thereby negligible. Physically, this reflects that
the particle number is conserved but the spin polarization not and, as a result, the latter is
irrelevant to physics at large scales.
We are thereby left with a low-energy effective action of $Z_0$-component only. Therefore, we use the symbols $Z,\tilde Z$ for $Z_0,\tilde Z_0$ hereafter.

The simplified zero-frequency action can be rewritten in a rotationally invariant form,
\begin{equation}\label{eq:1}
    S[Z,\tilde Z]|_{\omega=0}={\rm Str}\ln\left(\frac{1+{\hat U}}{1-{\hat U}}+Q\sigma^0\right)\equiv S[Q]|_{\omega=0},
\end{equation}
where $Q$ is a $4\times 4$ supermatrix field, defined as
\begin{eqnarray}
\label{eq:284}
Q\equiv T^{-1}\Lambda T
\end{eqnarray}
with
\begin{eqnarray}
T=\left(
                                  \begin{array}{cc}
                                    1 & Z \\
                                    \tilde Z & 1 \\
                                  \end{array}
                                \right)_{ar}\equiv 1+iW,\quad
                                \Lambda=\left(\begin{array}{cc}
                                    1 & 0 \\
                                    0 & -1 \\
                                  \end{array}
                                \right)_{ar}.
\label{eq:36}
\end{eqnarray}
In this definition the spin index is excluded.

\subsubsection{Separation of fast and slow modes}
\label{sec:separation_fast_slow_modes}

The action (\ref{eq:1}) is invariant under global rotation,
\begin{equation}\label{eq:13}
    S[Q]|_{\omega=0} = S[T'QT'^{-1}]|_{\omega=0},
\end{equation}
where $T'\in G=U(1,1|2)$ is homogeneous in $N$ space.
As a result, for homogeneous $Q$ configurations
one may always rotate $Q$ back to $\Lambda$ with the action invariant, i.e.,
$S[Q]|_{\omega=0}=S[\Lambda]|_{\omega=0}$. The latter obviously vanishes.
Therefore, a finite but small action $S[Q]$ must result from either shallow variations of $Q$ in $N$ space
or small $\omega$, and they contribute separately (because
their coupling is of higher order.)

By their definitions, $Z,\tilde Z$ and thereby $Q$ are off-diagonal in $N$ space in general.
Accordingly, we divide the components of $Z,\tilde Z$ into two groups. For one group, the components are diagonal in $N$ space and vary smoothly in $N$;
for the other, they are either diagonal, but varies rapidly in $N$, or off-diagonal. Their definitions are essentially the same as those in conventional QKR \cite{Altland10}. Since they are unimportant for present discussions, we refer to Ref.~\onlinecite{Altland10}.
The former group of $(Z,\tilde Z)$ components -- the slow mode -- carries information about the $2$D motion at
large scales, and the latter -- the fast mode -- at short scales. In particular, the latter carries the information about the velocity relaxation. If the chaoticity associated with the motion at short scales is sufficiently strong, the scales are well separated, and so are the fast and slow modes. In this case, as shown in Appendix~\ref{sec:massive}, the fast mode only introduces unimportant corrections to the
bare (unrenormalized) parameters governing motion at large scales. For this reason we shall not discuss the fast mode further and hereafter
the fields, $Z,\tilde Z$, are composed of slow modes only, i.e.,
\begin{eqnarray}\label{eq:300}
    Z= \{Z_{N\alpha,N\alpha'}\}\equiv \{Z_{\alpha,\alpha'}(N)\},\nonumber\\
    \tilde Z= \{\tilde Z_{N\alpha,N\alpha'}\}\equiv \{\tilde Z_{\alpha,\alpha'}(N)\}.
\end{eqnarray}
Moreover, these fields exhibit shallow variations in $N$.

For $Q$ made up of $Z,\tilde Z$ given above, Eq.~(\ref{eq:284}) is simplified to
\begin{eqnarray}
\label{eq:301}
Q(N)&=&T(N)^{-1}\Lambda T(N),\\
T(N)&=&\left(
                                  \begin{array}{cc}
                                    1 & Z(N) \\
                                    \tilde Z(N) & 1 \\
                                  \end{array}
                                \right)_{ar}\equiv 1+iW(N).\nonumber
\end{eqnarray}
Taking this into account, the first exponent of $\hat U$ (i.e., $e^{-\frac{i}{h_e}\left(H_0(h_e {\hat n}_1)+h_e{\tilde \omega} {\hat n}_2\right)}$) is canceled out in the exact action (\ref{eq:12}). As a result, the zero-frequency action (\ref{eq:1}) is substantially simplified, which reads
\begin{eqnarray}
S[Q]|_{\omega=0}={\rm Str}\ln\left(\epsilon+iQ\right),
\label{eq:70}
\end{eqnarray}
with (recall that the index $i$ runs over $1,2,3$.)
\begin{eqnarray}
\epsilon\equiv\epsilon_i\sigma^i,\quad \epsilon_i=\cot\frac{|V|}{2h_e}\frac{V_i}{|V|}.
\label{eq:238}
\end{eqnarray}
This is a key step, which makes subsequent derivation completely different from that of the effective field theory for the conventional QKR \cite{Altland10,Altland11}. In deriving Eqs.~(\ref{eq:70}) and (\ref{eq:238}) we use the relations (\ref{eq:317}). To make the formula
compact we suppress the argument $\Theta$ of $V_i$, $\epsilon_i$, etc., and drop out the unit matrix $\sigma^0$. Note that this action is insensitive to the explicit form of $H_0$ and the value of irrational $\tilde \omega/(2\pi)$. Physically, this is due to the ignorance of all fast modes, namely the immediate loss of the memory on $\Theta$ upon kicking.

The action (\ref{eq:70}) can be rewritten as
\begin{eqnarray}
S[Q]|_{\omega=0}={\rm Str}\ln\left(\epsilon+i\Lambda+T[\epsilon,T^{-1}]\right).
\label{eq:S14}
\end{eqnarray}
Then, we formally express $[\epsilon,T^{-1}]$ as the summation over
the terms of the following form,
$$
\sim\underbrace{[\theta_1,\cdots [\theta_1}_{k},\underbrace{[\theta_2,\cdots [\theta_2}_{j},T^{-1}]\cdots ]\sigma^i,
$$
where $k,j$ are arbitrary (non-negative) integers. This may be considered as a hydrodynamic expansion, since $T(N)$ varies smoothly in $N$ and the operator $-i[\theta_\alpha, \,]$ can be identified as the usual derivative, $\nabla_\alpha$, with respect to $n_\alpha$. Keeping this expansion up
to the second order and then substituting it into Eq.~(\ref{eq:S14}), we obtain (see Appendix~\ref{sec:derivation_1} for the derivation)
\begin{eqnarray}
S[Q]|_{\omega=0}=\qquad\qquad\qquad\qquad\qquad\qquad\qquad\qquad\qquad\label{eq:S30}\\
{\rm Str}\ln\left( {\cal G}_+^{-1}
+i\partial_{\theta_\alpha}\epsilon\, u_\alpha
-\frac{1}{2}\partial_{\theta_\alpha \theta_\beta}^2\epsilon\,
(u_\alpha u_\beta-\nabla_\alpha u_\beta)\right),\nonumber
\end{eqnarray}
where we have introduced the notations,
\begin{equation}\label{eq:10}
    u_\alpha \equiv T\nabla_\alpha T^{-1}
\end{equation}
and
\begin{equation}\label{eq:275}
    {\cal G}_\pm\equiv(\epsilon \pm i\Lambda)^{-1}.
\end{equation}
Throughout the Greek indices $\alpha,\beta=1,2$, and the Einstein summation convention applies to these indices also.

\subsubsection{Fluctuation action}
\label{sec:fluctuation_action}

To simplify technical discussions, below we consider potentials such that
\begin{eqnarray}
V_2(\theta_1,\theta_2)=V_1(\theta_2,\theta_1),\quad V_3(\theta_1,\theta_2)=V_3(\theta_2,\theta_1).
\label{eq:274}
\end{eqnarray}
(This simplification is inessential. Its only effect is to make the ensuing effective field theory namely Eq.~(\ref{eq:111}) isotropic.)
The action (\ref{eq:S30}) can be decomposed as
\begin{equation}\label{eq:S31}
    S[Q]|_{\omega=0}=S_1+S_2,
\end{equation}
with $S_{1}$ being real and $S_{2}$ purely imaginary.
The real part is given by
\begin{eqnarray}
S_1&=&\frac{1}{2}{\rm Str}
\ln\Big( {\cal G}_+^{-1}
+i\partial_{\theta_\alpha}\!\epsilon\, u_\alpha \nonumber\\
&&-
\frac{1}{2}\partial_{\theta_\alpha \theta_\beta}^2\!\epsilon\,
(u_\alpha u_\beta-\nabla_\alpha u_\beta)\Big)+c.c.,
\label{eq:S112}
\end{eqnarray}
where `{\it c.c.}' is the abbreviation of complex conjugate.
Keeping its hydrodynamic expansion up to the second order, we obtain
\begin{eqnarray}
\label{eq:S39}
S_1&=&-\frac{1}{4}{\rm Str}\left(({\cal G}^0_R+{\cal G}^0_A)
\partial_{\theta_\alpha \theta_\beta}^2\!\epsilon\, u_\alpha u_\beta\right)\nonumber\\
&+&\frac{1}{8}{\rm Str}\left(({\cal G}^0_R+{\cal G}^0_A)
\partial_{\theta_\alpha}\!\epsilon
({\cal G}^0_R+{\cal G}^0_A)
\partial_{\theta_\beta}\!\epsilon\, u_\alpha u_\beta\right)\nonumber\\
&+&\frac{1}{8}{\rm Str}\left(({\cal G}^0_R-{\cal G}^0_A)
\partial_{\theta_\alpha}\!\epsilon
({\cal G}^0_R-{\cal G}^0_A)
\partial_{\theta_\beta}\!\epsilon \Lambda u_\alpha \Lambda u_\beta\right),\quad\quad
\end{eqnarray}
where the `free particle Green function'
\begin{equation}\label{eq:276}
    {\cal G}^0_{R,A}\equiv\frac{1}{\epsilon\pm i}.
\end{equation}
Introducing the decomposition:
$u_\alpha=u^\parallel_\alpha+u^\perp_\alpha$, where $u^\parallel_\alpha$
($u^\perp_\alpha$) (anti)commutes with $\Lambda$, we
rewrite Eq.~(\ref{eq:S39}) as
\begin{eqnarray}
S_1=-\frac{1}{4}{\rm Str}\left(({\cal G}^0_R-{\cal G}^0_A)
\partial_{\theta_\alpha}\!\epsilon
({\cal G}^0_R-{\cal G}^0_A)
\partial_{\theta_\beta}\!\epsilon
u^\perp_\alpha u^\perp_\beta\right).\quad
\label{eq:S43}
\end{eqnarray}
Thanks to $u^\perp_\alpha u^\perp_\beta=-\frac{1}{4} T\nabla_\alpha Q
\nabla_\beta Q T^{-1}$, we further reduce it to \cite{note_coupling_constant}
\begin{equation}
S_1=-\frac{\sigma}{4}{\rm Str}(\nabla Q)^2,
\label{eq:S35}
\end{equation}
where
\begin{eqnarray}
\sigma&=&-\frac{1}{4}{\rm Tr}\left(({\cal G}^0_R-{\cal G}^0_A)
\partial_{\theta_1}\!\epsilon
({\cal G}^0_R-{\cal G}^0_A)
\partial_{\theta_1}\!\epsilon\right)\nonumber\\
&=&-\frac{1}{4}{\rm Tr}\left(({\cal G}^0_R-{\cal G}^0_A)
\partial_{\theta_2}\!\epsilon
({\cal G}^0_R-{\cal G}^0_A)
\partial_{\theta_2}\!\epsilon\right)
\label{eq:S34}
\end{eqnarray}
and $\nabla\equiv (\nabla_1,\nabla_2)$.
In deriving the second line of Eq.~(\ref{eq:S34}) we have
used the relations (\ref{eq:274}).
Substituting Eq.~(\ref{eq:276}) into Eq.~(\ref{eq:S34}) we obtain
\begin{equation}
\sigma = 2\int\!\!\!\!\int \frac{d\theta_1}{2\pi}\frac{d\theta_2}{2\pi}
    \frac{\partial_{\theta_1} \epsilon_i\partial_{\theta_1} \epsilon_i}{(\epsilon^2+1)^2}.
    \label{eq:3}
\end{equation}
This term is the same as that describing localization physics \cite{Efetov97} and the (inverse) coupling constant $\sigma$ mimics the unrenormalized (Drude) longitudinal conductivity in normal metals.

From Eq.~(\ref{eq:S35}) we see that $S_1$ is isotropic in $N$ space, in contrast to the anisotropicity of the microscopic system (\ref{eq:4}). This difference arises because this action is responsible for long-time but not short-time behavior. Due to strong
chaoticity the system loses memory on $\Theta$ after each kicking, and this leads to the isotropicity of $S_1$. It becomes clearer how this isotropic low-energy action emerges from an anisotropic microscopic Hamiltonian, when Eqs.~(\ref{eq:S35}) and (\ref{eq:S34}) are derived in an alternative way (see Appendix~\ref{sec:massive}). The alternative derivation shows that the isotropicity is washed out by short-time memory effects, and an action,
$-\frac{1}{4}
    (\sigma_{1} {\rm Str} (\nabla_1 Q)^2+
    \sigma_{2}{\rm Str} (\nabla_2 Q)^2),\, \sigma_1\neq \sigma_2$ results when these effects are taken into account. However, we emphasize that this anisotropicity is
weak and by appropriately rescaling
one can always recover an isotropic effective field theory. Therefore, we shall not discuss this issue further.

\subsubsection{Topological action}
\label{sec:topological_action}

We turn to the imaginary part. Similar to Eq.~(\ref{eq:S112}), $S_2$ is given by
\begin{eqnarray}
S_2&=&\frac{1}{2}{\rm Str}
\ln\Big( {\cal G}_+^{-1}
+i\partial_{\theta_\alpha} \epsilon\, u_\alpha \nonumber\\
&&-
\frac{1}{2}\partial_{\theta_\alpha \theta_\beta}^2 \epsilon\,
(u_\alpha u_\beta-\nabla_\alpha u_\beta)\Big)
-
c.c..
\label{eq:S113}
\end{eqnarray}
We then perform the hydrodynamic expansion up to the second order.
To this end we expand the logarithms in $u_\alpha$ up to the second order, obtaining
\begin{eqnarray}
S_2[Q] = S_2^{(1)}[Q] + S_2^{(2)}[Q],
\label{eq:S126}
\end{eqnarray}
with
\begin{eqnarray}
  S_2^{(1)}
  =\frac{i}{2}{\rm Str}\left(({\cal G}_+-{\cal G}_-)
\partial_{\theta_\alpha} \epsilon\, u_\alpha\right)
\label{eq:S46}
\end{eqnarray}
and
\begin{eqnarray}
S_2^{(2)}=\frac{1}{4}{\rm Str} \big(-({\cal G}_+-{\cal G}_-)
\partial_{{\theta_\alpha}{\theta_\beta}}^2\epsilon
(u_\alpha u_\beta-\nabla_\beta u_\alpha)\nonumber\\
+\left({\cal G}_+ \partial_{\theta_\alpha}\epsilon\,
u_\alpha{\cal G}_+ \partial_{\theta_\beta}\epsilon\, u_\beta -
{\cal G}_- \partial_{\theta_\alpha}\epsilon\, u_\alpha{\cal G}_-\partial_{\theta_\beta}\!\epsilon\,u_\beta\right)\big).
\label{eq:S47}
\end{eqnarray}
$S_2^{(1,2)}$ are both purely imaginary.

We first consider $S_2^{(1)}$. At first glance, it seems to be a first order hydrodynamic expansion and one might thereby expect that it vanishes. Yet, the boundary inevitably introduces
inhomogeneity effects, which must be investigated carefully.
In general, it deforms $\hat U$ in the way that $V_i(\Theta)$ acquires a parameter (denoted as $\mu$) dependence and
$\mu$ varies smoothly in $N$.
With this taken into account, Eq.~(\ref{eq:S46}) leads to a nonvanishing second order hydrodynamic expansion, which reads
\begin{eqnarray}
S_2^{(1)}&=&
-\frac{1}{2}{\rm Str}\big(
({\cal G}_R^0
\partial_\mu \epsilon\,
{\cal G}_R^0 \partial_{\theta_\alpha}\!\epsilon\,
{\cal G}_R^0 \partial_{\theta_\beta}\!\epsilon\nonumber\\
&&\quad \quad \quad - {\cal G}_A^0 \partial_\mu \epsilon\,
{\cal G}_A^0 \partial_{\theta_\alpha}\!\epsilon\,
{\cal G}_A^0 \partial_{\theta_\beta}\epsilon)
\Lambda u_\alpha \nabla_\beta\mu\big).\,\,
\label{eq:S60}
\end{eqnarray}
By using Stokes' theorem we rewrite it as
\begin{eqnarray}
S_2^{(1)} = \frac{\sigma_{\rm H}^I}{4} {\rm Str}\left(Q \nabla_1 Q \nabla_2 Q\right)
\label{eq:S68}
\end{eqnarray}
with
\begin{eqnarray}
\sigma_{\rm H}^I &=& - \int d\mu
{\rm Tr}
\big({\cal G}_R^0
\partial_\mu \epsilon\,
{\cal G}_R^0 \partial_{\theta_1} \epsilon\,
{\cal G}_R^0 \partial_{\theta_2} \epsilon \nonumber\\
&&\quad \quad \quad \quad - {\cal G}_A^0 \partial_\mu \epsilon\,
{\cal G}_A^0 \partial_{\theta_1}\epsilon\,
{\cal G}_A^0 \partial_{\theta_2}\epsilon\big).
\label{eq:280}
\end{eqnarray}
The expression (\ref{eq:280}) for $\sigma_{\rm H}^I$ resembles an expression for the quantum contribution to the (bare) Hall conductivity in conventional IQHE
\cite{Pruisken84b,Pruisken84c,Streda82}.

In Appendix \ref{sec:boundary_condition}, the deformation will be discussed in details. There, we further trade Eq.~(\ref{eq:280}) for an integral which is independent of the deformation, implying that $\sigma_{\rm H}^I$ is an intrinsic quantity and its value is unique.
We stress that the deformation is made only at the stage of deriving the effective field theory. It does not apply to the original system namely Eqs.~(\ref{eq:6}) and (\ref{eq:4}), and therefore does not affect numerical simulations below. In addition,
such deformation has no consequence on $S_1$ up to ${\cal O}(\nabla^2)$.

Next, we consider $S_2^{(2)}$. Because of $u_\alpha={\cal O}(\nabla)$
in performing the hydrodynamic expansion for Eq.~(\ref{eq:S47}) we need not consider the above deformation of $\hat U$, since its effects are of higher order. As a result,
\begin{eqnarray}
\label{eq:S50}
S_2^{(2)}=\frac{1}{4}{\rm Str}\left(({\cal G}^0_R+{\cal G}^0_A)
\partial_{\theta_\alpha}\epsilon
({\cal G}^0_R-{\cal G}^0_A)
\partial_{\theta_\beta}\epsilon u_\alpha \Lambda u_\beta\right).\quad
\end{eqnarray}
With the help of the identity:
\begin{eqnarray}
-2{\rm Str}(\Lambda (u_\alpha u_\beta-u_\beta u_\alpha))
={\rm Str}(Q\nabla_\alpha Q\nabla_\beta Q),
\label{eq:278}
\end{eqnarray}
Eq.~(\ref{eq:S50}) can be rewritten as
\begin{eqnarray}
S_2^{(2)}=\frac{\sigma_{\rm H}^{II}}{4}{\rm Str}(Q \nabla_1 Q\nabla_2 Q)
\label{eq:S53}
\end{eqnarray}
and the coefficient
\begin{eqnarray}
\sigma_{\rm H}^{II}=\frac{1}{2}{\rm Tr}\left(({\cal G}_R^0+{\cal G}^0_A)
\partial_{\theta_1}\epsilon\,({\cal G}^0_R-{\cal G}^0_A)
\partial_{\theta_2}\epsilon\right).
\label{eq:279}
\end{eqnarray}
Equation (\ref{eq:279}) resembles an expression for the contribution to the (bare) Hall conductivity arising from the Lorentz force in conventional IQHE \cite{Pruisken84c,Pruisken84b}.

Adding $S_2^{(1)}$ and $S_2^{(2)}$ together, we cast Eq.~(\ref{eq:S113}) to a topological action, namely, the theta term \cite{Pruisken84d,Efetov97,Pruisken84c},
\begin{eqnarray}
S_2[Q]=\frac{\sigma_{\rm H}}{4}{\rm Str}(Q \nabla_1 Q\nabla_2 Q),
\label{eq:S114}
\end{eqnarray}
with the coefficient,
\begin{eqnarray}
\sigma_{\rm H}=\sigma_{\rm H}^I+\sigma_{\rm H}^{II},
\label{eq:S27}
\end{eqnarray}
giving the unrenormalized topological theta angle, $2\pi\sigma_{\rm H}$. This term is topological in nature.
With the substitution of Eq.~(\ref{eq:276}), Eqs.~(\ref{eq:280}) and (\ref{eq:279}) are rewritten as
\begin{eqnarray}
\sigma_{\rm H}^{I} = 4\varepsilon^{ijk} \int\!\!\!\!\int \frac{d\theta_1}{2\pi}\frac{d\theta_2}{2\pi} \int d\mu \frac{\partial_\mu\epsilon_i\partial_{\theta_1} \epsilon_j
\partial_{\theta_2} \epsilon_k}{(\epsilon^2+1)^2}
\label{eq:S4}
\end{eqnarray}
and
\begin{eqnarray}
\sigma_{\rm H}^{II} = 4\varepsilon^{ijk} \int\!\!\!\!\int \frac{d\theta_1}{2\pi}\frac{d\theta_2}{2\pi}
\frac{\epsilon_i\partial_{\theta_1}\epsilon_j \partial_{\theta_2}\epsilon_k}{(\epsilon^2+1)^2},
 \label{eq:S3}
\end{eqnarray}
respectively.
In Eq.~(\ref{eq:S4}) the upper limit of the
$\mu$-integral takes the bulk value corresponding to an undeformed $\hat U$, while the lower limit corresponding to the integrable deformation of $\hat U$ depends on the
details of $V_i$. In Appendix \ref{sec:boundary_condition}, we show that Eq.~(\ref{eq:S4}) can be expressed in a form which is independent of the deformation.
Equations (\ref{eq:S4}) and (\ref{eq:S3}) justify Eq.~(\ref{eq:318}).

\subsubsection{Topological meanings of theta term}
\label{sec:physical_meanings_topological_term}

Let us gain some insights for the topological theta term (\ref{eq:S114}).
To this end it is sufficient to keep only the commuting components of $Z(N)$, i.e., $Z_{f,f}(N)$ and $Z_{b,b}(N)$.
One component, $Z_{f,f}(N)$, takes the value of unconstrained complex number. Therefore, it can be written as
\begin{eqnarray}\label{eq:54}
    Z_{f,f}(N)=\tan\left(\frac{1}{2}\theta_f(N)\right)e^{-i\varphi_f(N)},\quad\quad\\
    \theta_f \in [0,\pi],\,\,\varphi_f\in [0,2\pi).\quad\quad\quad\quad\nonumber
\end{eqnarray}
Consider the stereographic projection of $S^2$ from the south pole, with its Euclidean coordinate $(x,y,z)=(0,0,-1)$, onto the equator plane $z=0$. We see that the real and imaginary parts
of the parametrization (\ref{eq:54}) constitute the coordinate of this projection, i.e., $(\tan\frac{\theta_f}{2}\cos\varphi_f,\tan\frac{\theta_f}{2}\sin\varphi_f,0)$. This implies $Z_{f,f}\simeq S^2$. The other component, $Z_{b,b}(N)$, is constrained by $|Z_{b,b}(N)|<1$. Therefore, we can write it as
\begin{eqnarray}\label{eq:55}
    Z_{b,b}(N)=\tanh\left(\frac{1}{2}\theta_b(N)\right)e^{-i\varphi_b(N)},\quad\quad\\
    \theta_b\in \mathbb{R}^+,\,\,\varphi_b\in [0,2\pi).\quad\quad\quad\quad\nonumber
\end{eqnarray}
On the other hand, consider the upper sheet of the two-sheet hyperboloid $x^2+y^2-z^2=-1$, namely $H^2$. It can be parametrized as
$(\sinh\theta_b\cos\varphi_b,\sinh\theta_b\sin\varphi_b,\cosh\theta_b)$. Then, the real and imaginary parts of the parametrization (\ref{eq:55})
constitute the coordinate of the stereographic projection of $H^2$ from $(0,0,-1)$ onto the disk: $\{(x,y,z)|x^2+y^2<1,z=0\}$, which is $(\tanh\frac{\theta_b}{2}\cos\varphi_b,\tanh\frac{\theta_b}{2}\sin\varphi_b,0)$. This implies $Z_{b,b}\simeq H^2$.
Therefore, the field, $Z_{b,b}\times Z_{f,f}$, induces a mapping from $N$ space onto $H^2\times S^2$ discussed in Sec.~\ref{sec:origin_IQHE}.

Substituting Eqs.~(\ref{eq:54}) and (\ref{eq:55}) into
Eqs.~(\ref{eq:301}) and (\ref{eq:S114}),
we find
\begin{eqnarray}\label{eq:57}
    \frac{1}{8\pi}{\rm Str}\left(Q\nabla_1Q\nabla_2Q\right)
    =\frac{1}{4\pi}\int dN \boldsymbol{n}\cdot (\nabla_1\boldsymbol{n}\times \nabla_2\boldsymbol{n}),\,\,\,
\end{eqnarray}
where $\boldsymbol{n}\equiv(\sin\theta_f\cos\varphi_b,\sin\theta_f\sin\varphi_b,\cos\theta_f)$ is a three-dimensional unit vector.
The right-hand side of Eq.~(\ref{eq:57}) is the Brouwer degree of the mapping from the (compactified) $N$ space ($\simeq S^2$) to the target space ($\simeq S^2$ also) which is an integer.
For this mapping the degree is a complete homotopy invariant, implying $\pi_2(S^2)=\mathbb{Z}$ namely the last equality of Eq.~(\ref{eq:2}). Note that the noncompact component has no contributions.

\subsubsection{The frequency action}
\label{sec:frequency_term}

The finiteness of $\omega$
generates a third contribution, denoted as $S_\omega[Q]$, to the action.
As before, the fast modes and the $\boldsymbol{Z}$-components are ignored. Moreover, at large scale,
${\cal O}(\nabla)<h_e$, and low frequencies, $\omega\ll 1$, the terms $\sim\mathcal{O}(\omega \nabla^2)$ are irrelevant. That is, the inhomogeneity of $Z,\tilde Z$ can be ignored when we derive the frequency action.
As a result,
\begin{eqnarray}
\label{eq:S103}
  S_\omega[Q] &\approx& -2\left(\mathrm{Str}\ln(1-Z \tilde Z)
  -\mathrm{Str}\ln(1-e^{i\omega} Z \tilde Z)\right) \nonumber\\
  &\approx& -2i\omega\mathrm{Str} \frac{Z\tilde Z}{1-Z\tilde Z}\nonumber\\
  &=& -{i\frac{\omega}{2}} \mathrm{Str}\left(Q\Lambda\right),
\end{eqnarray}
where the factor of $2$ in the first line arises from the trace over the spin index.
(Recall that in Eq.~(\ref{eq:S103}) $Z,\tilde Z$ are $2\times 2$ supermatrices.) It is important that this action breaks the global rotation symmetry (\ref{eq:13}), exhibited by the zero-frequency action. The ensuing symmetry group is $U(1|1)\times U(1|1)\subset G=U(1,1|2)$. The rotation transformation representing this symmetry group leaves $\Lambda$ invariant.
As we will see in Sec.~\ref{sec:transport_parameter}, such symmetry breaking has significant consequences.

Adding Eqs.~(\ref{eq:S35}), (\ref{eq:S114}), and (\ref{eq:S103}) together, we find the total low-energy effective action,
\begin{eqnarray}\label{eq:111}
    S[Q]={1\over 4} \mathrm{Str} \left(-\sigma
    (\nabla Q)^2+\sigma_{\rm H}
Q \nabla_1 Q \nabla_2 Q
  -2i\omega Q\Lambda\right),\nonumber\\
\end{eqnarray}
which describes $2$D dynamics at large scales.
This is the supersymmetric version of the Pruisken-type field theory, previously obtained in studies of
conventional IQHE \cite{Efetov97,Pruisken84a,Pruisken84d,Pruisken84c,Pruisken84}. We stress that because the physical setup here does not exhibit any similarities to the conventional quantum Hall system, namely, a $2$D electron gas subjected to a magnetic field and strongly disordered potential, the derivation of the action (\ref{eq:111}) is totally different from that for the latter system. Comparing this action with the one for conventional IQHE, we find that, interestingly, the control parameters $\sigma$ and $\sigma_{\rm H}$ mimic the unrenormalized longitudinal and Hall conductivities, respectively. We should emphasize that this similarity, however, does not necessarily lead
to the IQHE-like transition. Whether and how it occurs still depends on the behavior
of $\sigma$ and $\sigma_{\rm H}$, and this is the main subject of the next two sections. Finally, we remark that in the absence of the topological theta term, this action is reduced to the one describing Anderson localization in a spinless quasiperiodic QKR \cite{Altland11}.

\subsection{The energy profile $E(t)$}
\label{sec:total_action}

Since in the effective field theory the supermatrix fields are all proportional to $\sigma^0$, the two-particle Green function is simplified to
\begin{eqnarray}
    K_\omega(Ns_+s_-,N's'_+s'_-)
    =\delta_{s_+s_-}\delta_{s'_+s'_-}\!\int\! D(Z,\tilde Z)e^{-S[Q]}\nonumber\\
    \times((1-Z \tilde Z)^{-1}Z)_{Nb,Nb} ((1-\tilde Z Z)^{-1}\tilde Z)_{N'b,N'b},\quad
    \label{eq:S111}
\end{eqnarray}
where $Z,\tilde Z$ are understood according to Eq.~(\ref{eq:300}). Exploiting the definition of $Q(N)$, namely Eqs.~(\ref{eq:301}) and (\ref{eq:36}), we further
express Eq.~(\ref{eq:S111}) as a functional integral over $Q$, which reads
\begin{eqnarray}\label{eq:S13}
&&K_\omega(Ns_+s_-,N's'_+s'_-)=-\frac{1}{4} \delta_{s_+s_-}\delta_{s'_+s'_-} \nonumber\\
&&\times\int D(Q)e^{-S[Q]}
Q(N)_{+b,-b}Q(N')_{-b,+b}.
\end{eqnarray}
The $Q$ integral above depends only on the difference of $(N-N')$, because the action (\ref{eq:111}) is translationally invariant.
Therefore, Eq.~(\ref{eq:S13}) can be rewritten as
\begin{eqnarray}
K_\omega(Ns_+s_-,N's'_+s'_-)
=\frac{1}{4} \delta_{s_+s_-}\delta_{s'_+s'_-} K_\omega(N-N'),\quad
\label{eq:302}
\end{eqnarray}
with the function $K_\omega(N)$ given by
\begin{eqnarray}
K_\omega(N)
\equiv -\int D(Q)e^{-S[Q]} Q(N)_{+b,-b}Q(0)_{-b,+b}.
\label{eq:S16}
\end{eqnarray}
Inserting Eqs.~(\ref{eq:302}) and (\ref{eq:S16}) into Eq.~(\ref{eq:7}), we find
\begin{eqnarray}
E(t)=\frac{1}{4}\sum_{N} n_1^2 \int\frac{d\omega}{2\pi}e^{-i\omega t}K_\omega(N).\qquad\qquad
\label{eq:S15}
\end{eqnarray}
Equations (\ref{eq:3}), (\ref{eq:S27}), (\ref{eq:S4}), (\ref{eq:S3}), (\ref{eq:111}), (\ref{eq:302}), (\ref{eq:S16}) and (\ref{eq:S15}) constitute the first-principles analytic formalism for calculating the energy profile.

\subsection{Universal scaling behavior of $\sigma$ and $\sigma_{\rm H}$ 
for small $h_e$}
\label{sec:asymptotic_behavior}

The effective field theory (\ref{eq:111}) is controlled by two
parameters, the unrenormalized (inverse) coupling constant $\sigma$ and topological angle $2\pi\sigma_{\rm H}$.
For a given potential $V$ they depend only on $h_e$.
Below we show that these two
parameters exhibit universal scaling behavior 
for small $h_e$, independent of the details of $V$.

First of all, by substituting Eq.~(\ref{eq:238}) into Eq.~(\ref{eq:3}) we obtain
\begin{eqnarray}\label{eq:24}
    \sigma &=& \frac{1}{2}\int\!\!\!\!\int \frac{d\theta_1}{2\pi}\frac{d\theta_2}{2\pi}\bigg(\left(\frac{\partial_{\theta_1}|V|}{h_e}\right)^2\nonumber\\
    &&-\sin^2\frac{|V|}{h_e}
    \frac{(\partial_{\theta_1}|V|)^2-\partial_{\theta_1}V_i\partial_{\theta_1}V_i}{|V|^2}\bigg).
\end{eqnarray}
This gives
\begin{eqnarray}\label{eq:25}
    \sigma \stackrel{h_e\ll 1}{\longrightarrow} \frac{1}{2h_e^2}\int\!\!\!\!\int
    \frac{d\theta_1}{2\pi}\frac{d\theta_2}{2\pi}\left(\partial_{\theta_1}|V|\right)^2 \propto h_e^{-2}
\end{eqnarray}
namely Eq.~(\ref{eq:291}). In fact, this rescaling exists also in the conventional QKR \cite{Altland11,Altland10}. It implies that $\sigma$ is proportional to the square of the mean free path, which is a manifestation of strongly chaotic motion at microscopic scales. As we will see later, $\sigma$ is the energy growth rate at short times.

Next, we analyze the scaling behavior of $\sigma_{\rm H}$.
To this end we rewrite Eq.~(\ref{eq:S4}) as
\begin{eqnarray}
\sigma_{\rm H}^I =
4\varepsilon^{ijk} \int\!\!\!\!\int \frac{d\theta_1}{2\pi}\frac{d\theta_2}{2\pi}
\int d\mu \frac{D_{\mu,i}D_{\theta_{1},j}D_{\theta_{2},k}}{\sin^{2}(|V|/2h_e)},
\label{eq:15}
\end{eqnarray}
where
\begin{equation}\label{eq:34}
    D_{\alpha,i}\equiv \sin\frac{|V|}{h_e}\frac{\partial_\alpha V_i}{2|V|}-\left(\frac{|V|}{h_e}+\sin\frac{|V|}{h_e}\right)\frac{V_i\partial_\alpha |V|^2}{4|V|^3},
\end{equation}
with the subscript $\alpha=\mu,\theta_{1,2}$. For $h_e\ll 1$ Eq.~(\ref{eq:15}) is simplified to
\begin{eqnarray}
\sigma_{\rm H}^I &\approx& -\frac{1}{h_e} \int\!\!\!\!\int \frac{d\theta_1}{2\pi}\frac{d\theta_2}{2\pi}
\int d\mu \left(1+\cos\frac{|V|}{h_e}\right)\nonumber\\
&&\qquad \qquad \times\frac{\varepsilon^{ijk} \partial_\mu V_i \partial_{\theta_1}V_j\partial_{\theta_2}V_k}{|V|^2}.
\label{eq:303}
\end{eqnarray}
For $h_e\ll 1$, the cosine function in the bracket oscillates rapidly in $|V|$ (and thereby $\Theta$) and the corresponding term is negligible. As a result, Eq.~(\ref{eq:303}) is simplified to
\begin{eqnarray}
&&\sigma_{\rm H}^I \stackrel{h_e\ll 1}{\longrightarrow} -\frac{1}{h_e} \int\!\!\!\!\int \frac{d\theta_1}{2\pi}\frac{d\theta_2}{2\pi}
\int d\mu \nonumber\\
&&\qquad \qquad \qquad \times\frac{\varepsilon^{ijk} \partial_\mu V_i \partial_{\theta_1}V_j\partial_{\theta_2}V_k}{|V|^2}.
 \label{eq:S11}
\end{eqnarray}
On the other hand, Eq.~(\ref{eq:S3}) can be written as
\begin{eqnarray}
\sigma_{\rm H}^{II}&=&\varepsilon^{ijk}\int\!\!\!\!\int \frac{d\theta_1}{2\pi}\frac{d\theta_2}{2\pi}
\frac{V_i\partial_{\theta_1}V_j\partial_{\theta_2}V_k}{|V|^{3}}\nonumber\\
&&\qquad\times\left(\sin\frac{|V|}{h_e}+\frac{1}{2}\sin\frac{2|V|}{h_e}\right).
 \label{eq:S10}
\end{eqnarray}
For the same reasons it is negligible for $h_e \rightarrow 0$. Taking this and Eq.~(\ref{eq:S11}) into account, we find
\begin{eqnarray}
\sigma_{\rm H} \stackrel{h_e\ll 1}{\longrightarrow} -\frac{1}{h_e} \int\!\!\!\!\int \frac{d\theta_1}{2\pi}\frac{d\theta_2}{2\pi}
\int d\mu \frac{\varepsilon^{ijk} \partial_\mu V_i \partial_{\theta_1}V_j\partial_{\theta_2}V_k}{|V|^2}.
 \label{eq:S12}
\end{eqnarray}
Since $V_i$ is $h_e$-independent, from Eq.~(\ref{eq:S12}) we obtain the universal scaling law (\ref{eq:9}), where the
proportionality coefficient generally does not vanish. In principle, corrections to Eq.~(\ref{eq:S12}) violate this scaling law. However, these corrections are small for $h_e\lesssim 1$ and therefore negligible. As we will see below,
this scaling law is crucial for establishing the universality of the Planck-IQHE pattern represented by Fig.~\ref{fig:5}. We recall that in conventional Hall systems the classical Hall conductivity increases linearly with the inverse magnetic field, when the magnetic field is strong \cite{Pruisken84a}. Comparing this law with Eq.~(\ref{eq:S12}) suggests an analogy between $h_e^{-1}$ and the filling fraction (or $h_e$ and the magnetic field). As we will show in Sec.~\ref{sec:microscopic_theory}, this is a key ingredient of the analogy between Planck- and conventional IQHE.

\section{Field theory of transport parameters}
\label{sec:transport_parameter}

Armed with the effective field theory (\ref{eq:111}), in this section we will calculate perturbative and nonperturbative contributions to the energy growth rate.
Moreover, the field theory allows us to introduce a virtual Hall conductivity. We will calculate its perturbative and nonperturbative parts as well. This virtual transport parameter enables us to uncover the hidden quantum number in the next section by using the RG method.
The calculation scheme of this section -- within the supersymmetry formalism -- is parallel in spirit to that developed by Pruisken and co-workers
for the replica field theory of conventional IQHE \cite{Pruisken87a,Pruisken87,Pruisken05}. However, the detailed treatments are very different.
In particular, it has not yet been reported in literatures whether and to what extent the relatively recent results \cite{Pruisken05} for the renormalization theory of conventional IQHE
could be extended to the supersymmetry formalism. On the other hand, there are principal reasons and
examples \cite{Tian05} showing that the agreement of perturbative results obtained from the replica and supersymmetry formalism does not gurantee
the agreement of nonperturbative results. In view of successes recently achieved in applications of the supersymmetry technique to spinless QKR \cite{Altland10,Altland11,Tian15}, it is natural to proceed to obtain explicit results from this technique namely the effective field theory (\ref{eq:111}). For these reasons (as well as for the self-contained purpose), we give
the details of the extension in the following, although some technical pieces are the same as earlier works \cite{Pruisken87a,Pruisken87,Pruisken05}, as they do not depend on specific formalism (replica or supersymmetry).

The results obtained in this section pave the way for RG analysis, which will be performed in the next section.
We stress that the treatments of this section are {\it not} exact. Rather, they are perturbative and nonperturbative single instanton analysis. We do not study the multi-instanton effect \cite{Zirnbauer88} which is far beyond the scope of the present work.

\subsection{Background field formalism}
\label{sec:electromagnetic_response}

Motivated by the similarity between the effective field theory (\ref{eq:111}) and that for conventional IQHE, we follow the field-theoretic treatment \cite{Pruisken84d,Pruisken84c,Pruisken84,Pruisken87a,Pruisken87,Pruisken05} of conventional IQHE to introduce a background field. This field, given by
\begin{eqnarray}
  \mathscr{U}&=&e^{i(n_1 j_1 \tau_1+n_2 j_2 \tau_2)}, \label{eq:116}\\
  \tau_i &=& \sigma_{ar}^{i} \otimes \mathbb{E}_{ff},\nonumber
\end{eqnarray}
varies smoothly in the $N$ space and is minimally coupled to the effective field theory (\ref{eq:111})
so that the gradients in the action (\ref{eq:111}) are replaced by the covariant derivatives,
\begin{eqnarray}\label{eq:47}
    \nabla_\alpha \rightarrow \nabla_\alpha + [\mathscr{U}\nabla_\alpha \mathscr{U}^{-1},\,].
\end{eqnarray}
Here $j_{1,2}$ are infinitesimal external parameters.
$\mathbb{E}_{\alpha\alpha'}$ is a projector in the {\it bf}-space which takes the value of unity for the
entry $(\alpha,\alpha')$ and is zero otherwise. Observing the structure of the exponent of $\mathscr{U}$, on general grounds, we expect that the response to this background field is characterized by two parameters, defined as
\begin{eqnarray}
    \tilde \sigma\equiv -\frac{1}{4\Omega}\partial^2_{j_1} {\cal Z}[\mathscr{U}]|_{j_{1,2}\rightarrow 0,\omega\rightarrow 0}
    \label{eq:48}
\end{eqnarray}
and
\begin{eqnarray}
    \tilde \sigma_{\rm H}\equiv\frac{1}{2i\Omega}\partial^2_{j_1j_2} {\cal Z}[\mathscr{U}]|_{j_{1,2}\rightarrow 0,\omega\rightarrow 0},
    \label{eq:50}
\end{eqnarray}
respectively, with $\Omega=\int dN$ being the system's volume
and the zero frequency limit of $\omega\rightarrow 0$ taken. Recall that $\omega$ is understood as $\omega+\frac{i{\eta}}{2}$, and the imaginary part, namely the infinitesimal positive ${\eta}$, is sent to zero only in the final results. Such term breaks the global $G$ symmetry of the zero-frequency action $S[Q]|_{\omega=0}$
and gives nonvanishing results for $\tilde \sigma$
and $\tilde \sigma_{\rm H}$. Although presently we are not aware of physical implications of this coupling to the original system (\ref{eq:6}),
formally the definitions bear a close analogy to
the genuine longitudinal and Hall conductivity, respectively \cite{Pruisken84a}. For this reason we dub $\tilde \sigma$ and $\tilde \sigma_{\rm H}$ `transport parameters'.

With the substitution of Eq.~(\ref{eq:111}) Eqs.~(\ref{eq:48}) and (\ref{eq:50}) are cast into the functional integral of $Q$,
\begin{eqnarray}
\tilde \sigma&=&-\frac{\sigma}{4\Omega}\langle
{\rm Str}\left(Q\tau_1Q\tau_1-\tau_1^2\right)
\rangle_{\eta}\nonumber\\
&&+\frac{\sigma^2}{4\Omega}\langle\left({\rm Str}\left(\tau_1Q\nabla_1Q\right)\right)^2\rangle_{\eta},
\label{eq:115}
\end{eqnarray}
and
\begin{eqnarray}
\tilde \sigma_{\rm H}&=& \sigma_{\rm H} -\frac{\sigma}{4\Omega}\langle
{\rm Str}\left(\tau_3Q\varepsilon_{\alpha\beta}n_\alpha\nabla_\beta Q\right)
\rangle_{\eta}\nonumber\\
&&+\frac{i\sigma^2}{2\Omega}\langle{\rm Str}\left(\tau_1Q\nabla_1Q\right){\rm Str}\left(\tau_2Q\nabla_2Q\right)\rangle_{\eta},
\label{eq:117}
\end{eqnarray}
where $\langle\cdot\rangle_{\eta} \equiv \int D(Q)(\cdot)e^{-S[Q]|_{\omega\rightarrow\frac{i{\eta}}{2}}}$ and $\varepsilon_{\alpha\beta}$ is
the totally antisymmetric tensor. In the final results
we first send $\Omega$ to infinity and then ${\eta}$ to zero.
We remark that these expressions are not invariant under the rotation: $Q\rightarrow T^{-1}QT$, where $[T,\Lambda]=0$. We do not know how to obtain from
Eqs.~(\ref{eq:115}) and (\ref{eq:117}) their equivalent and rotationally invariant expressions. Because of this, calculations below differ substantially from those performed in Refs.~\onlinecite{Pruisken87a,Pruisken87,Pruisken05}.

\subsection{Physical meanings of transport parameters}
\label{sec:perturbative_corrections}

While this section is devoted to explicit calculations of Eqs.~(\ref{eq:48}) and (\ref{eq:50}),
it may be useful to first obtain some insights into the physical meanings of these two transport parameters.

\subsubsection{Optical conductivity and physical meanings of $\sigma$}
\label{physical_meaning_sigma}

Because $\tilde \sigma$ and $\tilde \sigma_{\rm H}$ are functions of the parameters $\sigma$ and $\sigma_{\rm H}$ of the effective field theory (\ref{eq:111}), we first need to discuss the meaning of $\sigma$ and $\sigma_{\rm H}$.
To this end we derive a general result for $E(t)$. For the moment we restore the frequency term, i.e., finite $\omega$.
By the particle number
conservation law, the Fourier transformation of $K_\omega(N)$, denoted as $K_\omega({{\boldsymbol{\phi}}})$
with ${\boldsymbol{\phi}}\equiv (\phi_1,\phi_2)$, has to obey the limiting behavior,
$\lim_{{{\boldsymbol{\phi}}}\rightarrow 0} K_\omega({{\boldsymbol{\phi}}})= 2/(-i\omega)$.
Then, the most general low-${{\boldsymbol{\phi}}}$ asymptotic compatible with this requirement and the rotation symmetry must take the general form as follows,
\begin{equation}\label{eq:105}
    K_\omega({\boldsymbol{\phi}})=\frac{2}{-i\omega + \sigma(\omega) {\boldsymbol{\phi}}^2},
\end{equation}
where $\sigma(\omega)$ simulates the `optical conductivity' in condensed matter \cite{Ono85}.
Most importantly, it has a diffusive pole.
Substituting Eq.~(\ref{eq:105}) into Eq.~(\ref{eq:S15}) gives
\begin{equation}\label{eq:106}
    E(t)=-\int\frac{d\omega}{2\pi}\frac{e^{-i\omega t}}{\omega^2}\sigma(\omega).
\end{equation}
This is a general relation between the rotor's energy profile and $\sigma(\omega)$. It shows that the low-frequency behavior of $\sigma(\omega)$ governs the energy profile at long times. To be specific, if
$\sigma(\omega\rightarrow 0)$ is finite, then $E(t)\stackrel{t\rightarrow\infty}{\longrightarrow} t$ (`metal');
if
$\sigma(\omega\rightarrow 0)\sim -i\omega$, then $E(t)\stackrel{t\rightarrow\infty}{\longrightarrow} const.$ (`insulator')
with the saturation value characterizing the $2$D localization volume.

For sufficiently short times, which corresponds to $\omega$ much larger than a characteristic frequency $\sim e^{-4\pi\sigma^2}$ (which,
as we will discuss in the end of Sec.~\ref{sec:renormalization_group_flow}, is the inverse of the characteristic time for effecting quantum interference), the $Q$ field fluctuates weakly around $\Lambda$. Therefore, we can expand $Q$ in $Z,\tilde Z$. Substituting it into the expression (\ref{eq:S16}) of $K_\omega(N)$ and keeping the leading term gives
\begin{eqnarray}
K_\omega(N)=4\int D(Z,\tilde Z)e^{-S_0[Z,\tilde Z]} Z(N)_{b,b}\tilde Z(0)_{b,b},\quad
\label{eq:16}\\
S_0[Z,\tilde Z]=2{\rm Str}\left(Z(-\sigma\nabla^2-i\omega)\tilde Z\right).\quad\quad\quad
\nonumber
\end{eqnarray}
It is important to note that on the perturbation level the topological
term does not contribute to the action.
The Gaussian integral in Eq.~(\ref{eq:16}) can be readily calculated. The result is
$K_\omega({\boldsymbol{\phi}})=2/(-i\omega + \sigma {\boldsymbol{\phi}}^2)$.
Substituting it into Eq.~(\ref{eq:S15}) gives
\begin{equation}\label{eq:17}
    E(t)=-\frac{1}{4}\int\frac{d\omega}{2\pi}e^{-i\omega t}\partial_{\phi_1}^2|_{{\boldsymbol{\phi}}=0}
    K_\omega({\boldsymbol{\phi}})=\sigma t.
\end{equation}
We see that at early times ($\ll e^{4\pi\sigma^2}$) chaotic diffusion in the $N$ space dominates over localization effects arising from interference and $\sigma$ gives the short-time energy growth rate.
In fact, it is easy to show that the chaotic
diffusion occurs in both $n_1$- and $n_2$-directions. When
short-time correlations are negligible this $2$D chaotic diffusion is isotropic (see also discussions in Appendix~\ref{sec:massive}).

\subsubsection{Perturbative contributions to $\tilde \sigma$ and $\tilde \sigma_{\rm H}$}
\label{sec:perturbative_corrections_sigma_sigma_H}

We now discuss the physical meanings of $\tilde \sigma$ and $\tilde \sigma_{\rm H}$.
Let us make some observations of the perturbative parts of Eqs.~(\ref{eq:115}) and (\ref{eq:117}). Specifically, we
perform the $Z,\tilde Z$-expansion for these two expressions and keep the leading (quadratic) order expansion.
This gives $\tilde \sigma=\sigma$ and $\tilde \sigma_{\rm H}=\sigma_{\rm H}$ which, as discussed above, are valid only for
short times and totally exclude interference effects.
For longer times interference effects must dominate and strongly renormalize
$\sigma$ and $\sigma_{\rm H}$. To see this we calculate Eq.~(\ref{eq:115}) up to the two-loop order, which gives
\begin{eqnarray}
\delta\sigma_p=
\sigma\left(\frac{1}{2}-\frac{1}{d}\right)\langle 0|(-\sigma\nabla^2)^{-1}|0\rangle^2,
\label{eq:119}
\end{eqnarray}
with $d$ being the dimension and $\langle N|(-\sigma\nabla^2)^{-1}|N'\rangle$
the diffusive propagator. It exhibits infrared divergence which is a signature of strong interference effects at large scales. This is the well-known weak localization correction to $\sigma$ for systems with broken time-reversal symmetry.
Note that the one-loop correction vanishes as a result of the time-reversal symmetry breaking.
In contrast, $\sigma_{\rm H}$ does not receive any perturbative corrections,
\begin{eqnarray}
\delta\sigma_{{\rm H},p}=0,
\label{eq:120}
\end{eqnarray}
which reflects the nonperturbative nature of the topological term.

To cure the infrared divergence we resort to the RG method which is to be discussed in
Sec.~\ref{sec:microscopic_theory}. We note that the diffusive propagator suffers ultraviolet divergence. For this reason we cannot directly set $d=2$ in Eq.~(\ref{eq:119}). The ultraviolet divergence in Eq.~(\ref{eq:119}) can be readily cured by the dimensional regularization.

From these observations based on perturbative calculations, we may interpret $\tilde \sigma$ as the long-time energy growth rate or the quantum longitudinal conductivity. This will become clearer in the next part. Likewise, $\tilde \sigma_{\rm H}$ may be interpreted as the (virtual) quantum Hall conductivity.

\subsubsection{Zero-frequency limit of optical conductivity}
\label{sec:relation}

The absence of perturbative corrections $\delta \sigma_{{\rm H},p}$ signals that quantum interference gives rise to important
nonperturbative effects. Before turning to its quantitative analysis, which is the main subject of the remainder of this section, we
derive a general result for $\tilde \sigma$ showing that Eqs.~(\ref{eq:48}) and (\ref{eq:50}) indeed capture strong renormalization effects.
To this end we rewrite ${\cal Z}[\mathscr{U}]$ as
\begin{eqnarray}
&&{\cal Z}[\mathscr{U}] \label{eq:49}\\
&=&\int D[Q]e^{-{1\over 4} \mathrm{Str} \left(-\sigma
    (\nabla Q)^2+\sigma_{\rm H}
Q \nabla_1 Q \nabla_2 Q
  -2i\omega Q \mathscr{U}^{-1}\Lambda\mathscr{U}\right)}.
\nonumber
\end{eqnarray}
Upon substituting it into Eq.~(\ref{eq:48}) we obtain
\begin{eqnarray}
    \tilde \sigma =\lim_{\omega\rightarrow 0}\frac{1}{4}\omega^2\partial^2_{\phi_1}|_{{\boldsymbol{\phi}}=0} K_{\omega}({\boldsymbol{\phi}}).
\label{eq:51}
\end{eqnarray}
Inserting the general expression (\ref{eq:105}) of $K_\omega({\boldsymbol{\phi}})$ into Eq.~(\ref{eq:51})
we find
\begin{equation}\label{eq:258}
    \tilde \sigma=\sigma(\omega\rightarrow 0).
\end{equation}
So, $\tilde \sigma$ defined by Eq.~(\ref{eq:48}) is the energy growth rate in the long-time limit.

Similarly, Eq.~(\ref{eq:50}) gives the renormalization of the bare topological angle $2\pi\sigma_{\rm H}$.
Presently, we are not aware of how to directly probe $\tilde \sigma_{\rm H}$.
Nevertheless, this quantity has far-reaching physical consequences, which will become clear when we explicitly work out the two-parameter scaling theory in
Sec.~\ref{sec:microscopic_theory}.

\subsection{Nonperturbative instanton contributions}
\label{sec:nonperturbative_corrections}

We are ready to go beyond the perturbative results (\ref{eq:119}) and (\ref{eq:120}) where quantum corrections are organized as an expansion in $\frac{1}{\sigma}$.
Specifically, we will calculate nonperturbative instanton contributions to $\tilde \sigma$ and $\tilde \sigma_{\rm H}$.

\subsubsection{The single instanton approximation}
\label{sec:single_instanton_approximation}

The sufficient and necessary conditions leading to a stationary zero-frequency action are \cite{Zirnbauer88}
\begin{eqnarray}
\partial_{z^*}Z=0,\quad \partial_{z}\tilde Z=0
\label{eq:135}
\end{eqnarray}
corresponding to the instanton and
\begin{eqnarray}
\partial_{z}Z=0,\quad \partial_{z^*}\tilde Z=0
\label{eq:136}
\end{eqnarray}
to the anti-instanton, where we have identified the $N$ space as the complex plane with the coordinate $z\equiv n_1+in_2$ and $z^*$ being its complex conjugate.
As shown in Appendix \ref{sec:derivation_self_dual_equation}, they are equivalent to the self-duality equation,
\begin{equation}\label{eq:137}
    \nabla_\alpha Q_\pm\pm\varepsilon_{\alpha\beta} Q_\pm\nabla_\beta Q_\pm=0,
\end{equation}
with the $+(-)$ sign referring to the (anti-)instanton.

In general, the solutions of Eqs.~(\ref{eq:135}) and (\ref{eq:136}) give multi-instanton configurations.
The particular case of single instanton solution to Eq.~(\ref{eq:135}) (the so-called `dilute instanton gas') is given by
\begin{equation}\label{eq:142}
    Q_s(N)=T^{-1}\Lambda_s(N)T
\end{equation}
with $s=\pm$. Here $T\in G=U(1,1|2)$ is homogeneous in $N$ space generating a global rotation and
\begin{eqnarray}
\Lambda_s = R_s^{-1}\Lambda R_s,
  \label{eq:143}
\end{eqnarray}
where
\begin{eqnarray}
R_+ = R_-^*=\left(
              \begin{array}{cc}
                e_1^* & e_0 \\
                -e_0 & e_1 \\
              \end{array}
            \right)_{ar}\otimes \mathbb{E}_{ff}+\mathbbm{1}_{ar} \otimes \mathbb{E}_{bb},
\label{eq:233}
\end{eqnarray}
\begin{eqnarray}
e_0 = \frac{\lambda}{\sqrt{|z-z_0|^2+\lambda^2}},\, e_1 = \frac{z-z_0}{\sqrt{|z-z_0|^2+\lambda^2}}
\label{eq:320}
\end{eqnarray}
with $z_0\equiv n_{10}+in_{20}$ being the position of the instanton
and $\lambda$ the instanton size. According to Eq.~(\ref{eq:233}), the instanton configuration is nontrivial only in the fermionic-fermionic block, consistent with discussions above (cf. Eq.~(\ref{eq:54})). From now on we adopt the standard single instanton
approximation \cite{Efetov97,Pruisken84a,Pruisken10,Pruisken87a,Pruisken87,Pruisken05}.

Substituting Eqs.~(\ref{eq:142}) and (\ref{eq:143}) into the (zero-frequency) action, we obtain the stationary action (see
Appendix \ref{sec:derivation_instanton_action} for the derivation)
\begin{equation}\label{eq:148}
    S[Q_s]|_{\omega=0}=4\pi\sigma - s2\pi i\sigma_{\rm H}
\end{equation}
corresponding to the instanton configuration. We find from Eqs.~(\ref{eq:143}) and (\ref{eq:233})
that $\Lambda_s \rightarrow \Lambda$ at infinity of $N$ space, i.e., $|z-z_0|\rightarrow\infty$. In contrast, $R_s$ is not a constant at the boundary, i.e., \begin{eqnarray}
\label{eq:319}
  R_s\rightarrow e^{-i\vartheta \sigma_{ar}^3}\otimes \mathbb{E}_{ff}+\sigma_{ar}^0 \otimes \mathbb{E}_{bb}\equiv R_s(\vartheta),
\end{eqnarray}
which depends on the angle
\begin{eqnarray}
\vartheta \equiv {\rm Arg}(z-z_0) \in [0,2\pi).
\label{eq:314}
\end{eqnarray}
This has an important consequence. There exists a local $U(1)$ symmetry, i.e.,
$R_s^{-1}(\vartheta)\Lambda R_s(\vartheta)=\Lambda$, at the boundary. As shown in Appendix \ref{sec:derivation_instanton_action},
the instanton action (\ref{eq:148}) can be directly attributed to this local gauge symmetry.

The single-instanton solutions to the self-dual equation (\ref{eq:137}) have a structure
as $Q_s=U^{-1}\Lambda U,U=R_sT$ and constitute a manifold.
To explore the structure of this manifold we consider a subgroup $H\subset G$,
\begin{equation}\label{eq:155}
    H\equiv\{h|h\in G,\,h^{-1}\Lambda_s (N)h=\Lambda_s (N),\, \forall N\}.
\end{equation}
That is, $\Lambda_s (N)$ is invariant under the rotation transformation generated by $h\in H$.
By this definition the element $h$ has the general form, $h=\left(\begin{array}{cc}
                          e^{i\alpha_1} & 0 \\
                          0 & e^{i\alpha_2} \\
                        \end{array}
                      \right)_{ar}\otimes \mathbb{E}_{bb}+e^{i\gamma}\mathbbm{1}_{ar}\otimes \mathbb{E}_{ff}$
with $\alpha_{1,2},\gamma\in [0,2\pi)$, which implies $H\simeq U(1)\times U(1)\times U(1)$. The coset space,
\begin{eqnarray}\label{eq:154}
    G/H
    =\frac{U(1,1|2)}{U(1|1)\times U(1|1)}
    \times \left(\frac{U(1|1)}{U(1)\times U(1)}\right)^2
    \times U(1),\nonumber\\
\end{eqnarray}
then carries the degrees of freedom of the instanton. The first factor refers to the coset space associated with the supersymmetry $\sigma$ model of
unitary symmetry, and the subgroup $U(1|1)\times U(1|1)$ in the denominator generates rotations leaving $\Lambda$ invariant;
In the second factor, the subgroup $U(1)\times U(1)$ generates rotations leaving $\sigma_{bf}^3$
invariant, and the square accounts for the advanced-advanced and retarded-retarded blocks; The last factor generates the rotation in the complex plane, i.e.,
\begin{eqnarray}\label{eq:316}
    z-z_0\rightarrow e^{-i\phi} (z-z_0),\nonumber\\
    z^*-z^*_0\rightarrow e^{i\phi} (z^*-z^*_0),
\end{eqnarray}
where $\phi\in [0,2\pi)$. The first factor has $8$ generators, the second $2\times 2=4$, and the third $1$. On the other hand, $R_s$ has $3$ degrees of freedoms, i.e.,
$\{n_{10},n_{20},\lambda\}$. As a result, the total number of the instanton's degrees of freedom is
\begin{equation}\label{eq:272}
    8+4+1+3=16.
\end{equation}
We remark that the degrees of freedom
carried by $G/H$ are intrinsic to the zero-frequency limit, i.e., exist only if the frequency term $\sim {\rm Str} (Q\Lambda)$ in
the action (\ref{eq:111}) is absent. As we will see below, when this term is present, even for infinitesimal (imaginary) frequency $\omega\rightarrow\frac{i{\eta}}{2}$, the first factor
in Eq.~(\ref{eq:154}) is fully suppressed in the limit of $\Omega\rightarrow \infty$.

\subsubsection{Fluctuations and zero modes}
\label{sec:gaussian_fluctuation_and_zero_mode}

To calculate the nonperturbative instanton contributions to Eqs.~(\ref{eq:115}) and (\ref{eq:117})
we perform the semiclassical analysis. More precisely, for weak coupling,
$\sigma\gg 1$, the $Q$ functional integral is dominated by the Gaussian fluctuations around the instanton configurations.
To study these fluctuations we parametrize the $Q$ field as
\begin{eqnarray}
  Q = U^{-1}qU,\quad U=R_sT
\label{eq:144}
\end{eqnarray}
with
\begin{eqnarray}
  q = w+\Lambda\sqrt{1-w^2},\quad
  w = \left(
                                                 \begin{array}{cc}
                                                   0 & v \\
                                                   -\tilde v & 0 \\
                                                 \end{array}
                                               \right)_{ar},
\label{eq:313}
\end{eqnarray}
where the $N$-dependence is carried by $v$ and $R_s$ with $v,\tilde v$ being $2\times 2$ matrices in the {\it bf}-space.

Without loss of generality we focus on $s=+$.
Substituting Eqs.~(\ref{eq:144}) and (\ref{eq:313}) into Eq.~(\ref{eq:111}), we find
\begin{equation}\label{eq:234}
    S[Q]|_{\omega=0}\approx S[Q_+]|_{\omega=0}+\delta S_{np},
\end{equation}
where the fluctuation action
\begin{eqnarray}
\label{eq:146}
  \delta S_{np}[v,\tilde v] &=& \frac{\sigma}{2} \int dN \mu^2(N)(v_{bb}\hat O^{(0)}\tilde v_{bb}+v_{bf}\hat O^{(1)}\tilde v_{fb}\nonumber\\
  && \quad \quad \quad - v_{fb}\hat O^{(1)}\tilde v_{bf}-v_{ff}\hat O^{(2)}\tilde v_{ff}).\label{eq:146}
\end{eqnarray}
Here the operator $\hat O^{(a)}$ is defined as
\begin{eqnarray}
 \hat O^{(a)} \equiv -\frac{1}{\mu^{2}(N)}\left(\nabla_\alpha+\frac{ia\varepsilon_{\alpha\beta}(n_\beta-n_{\beta 0})}{|z-z_0|^2+\lambda^2}\right)^2-\frac{a}{2}
\label{eq:147}
\end{eqnarray}
with $\mu(N) = 2\lambda/(|z-z_0|^2+\lambda^2)$.
The fluctuation action (\ref{eq:146}) indicates
that the instanton configuration effectively introduces a curved space background and the corresponding measure
is $dN\mu^2(N)$, where
the Jacobian $\mu^2(N)$ arises from the nontrivial Riemannian metric.
The `$-$' sign of the last two terms of Eq.~(\ref{eq:146}) results from the supertrace definition.

To proceed further, we pass to the stereographic projection.The corresponding coordinates are denoted as $(\eta,\vartheta)$, with
\begin{eqnarray}
\label{eq:156}
  \eta \equiv \frac{|z-z_0|^2-\lambda^2}{|z-z_0|^2+\lambda^2}\in [-1,1]
\end{eqnarray}
and $\vartheta$ defined by Eq.~(\ref{eq:314}). In this coordinate
system, the space is flat because of $dN\mu^2(N)=d\eta d\vartheta$, and $\hat O^{(a)}$ has the following representation,
\begin{widetext}
\begin{eqnarray}
\label{eq:157}
  \hat O^{(a)} = - \left((1-\eta^2)\partial_\eta^2 - 2\eta\partial_\eta + \frac{1}{1-\eta^2}\partial_\vartheta^2
  - \frac{ia}{1-\eta}\partial_\vartheta -\frac{a^2}{4} \frac{1+\eta}{1-\eta} + \frac{a}{2}\right).
\end{eqnarray}
The eigenfunctions and eigenvalues of $\hat O^{(a)}$, satisfying
\begin{eqnarray}
    \hat O^{(a)} \Phi^{(a)}_{J,M}(\eta,\vartheta)=E_J^{(a)}\Phi^{(a)}_{J,M}(\eta,\vartheta),
    \label{eq:158}
\end{eqnarray}
are given by (see Appendix \ref{sec:eigenvalue_eigenfunction} for details)
\begin{eqnarray}
    E_J^{(a)}=J(J+a+1),\quad J=0,1,2,\cdots,
    \label{eq:159}
\end{eqnarray}
and
\begin{eqnarray}
    &&\Phi^{(a)}_{J,M}=
    \frac{1}{2^{M+1}}\sqrt{\frac{2J+a+1}{2^a\pi}\frac{\Gamma(J-M+1)\Gamma(J+M+a+1)}{\Gamma(J+1)\Gamma(J+a+1)}}
    e^{-iM\vartheta}(1-\eta^2)^{\frac{M}{2}}(1-\eta)^{\frac{a}{2}}P_{J-M}^{M+a,M}(\eta),\nonumber\\
    &&\qquad \qquad \qquad \qquad M=-J-a,-J-a+1,\cdots, J.
    \label{eq:160}
\end{eqnarray}
Here $P_{n}^{\alpha,\beta}(\eta)$ is a polynomial of degree $n$, defined as
\begin{eqnarray}
  P_{n}^{\alpha,\beta}(\eta) &\equiv& \frac{1}{n!} \sum_{\nu=0}^n C_\nu(n,\alpha,\beta) \left(\frac{\eta-1}{2}\right)^\nu,\label{eq:161}\\
  C_\nu(n,\alpha,\beta) &=& \Bigg\{\begin{array}{c}
                                     \noindent\left(\begin{array}{c}
                                             n \\
                                             \nu
                                           \end{array}
                                     \right) (n+\alpha+\beta+1)\cdots (n+\alpha+\beta+\nu)(\alpha+\nu+1)\cdots (\alpha+n),\quad \nu\neq 0,n,\\
                                     \noindent(\alpha+1)\cdots (\alpha+n),\quad \nu=0, \\
                                     \noindent(n+\alpha+\beta+1)\cdots (2n+\alpha+\beta),\quad \nu=n,
                                   \end{array}\nonumber
\end{eqnarray}
\end{widetext}
which extends the Jacobi polynomial defined for $\alpha,\beta>-1$ to arbitrary complex $\alpha,\beta$ (cf. Ref.~\onlinecite{Szego39}). The normalization
constant in Eq.~(\ref{eq:160}) is determined in the way such that
\begin{equation}\label{eq:162}
    \int_{-1}^1d\eta\int_0^{2\pi}d\vartheta \Phi^{(a)*}_{J',M'} (\eta,\vartheta)\Phi^{(a)}_{J,M} (\eta,\vartheta)=\delta_{JJ'}\delta_{MM'}.
\end{equation}
Equation (\ref{eq:160}) gives the degeneracy
\begin{equation}\label{eq:163}
    d^{(a)}(J)=2J+a+1
\end{equation}
for a given eigenvalue $E_J^{(a)}$.

For $J=0,-a\leq M\leq 0$, the eigenvalue vanishes and the corresponding eigenfunctions are
\begin{eqnarray}
  \Phi_{0,0}^{(0)} &=& \frac{1}{2\sqrt{\pi}},\label{eq:164}\\
  \Phi_{0,0}^{(1)} &=& \frac{1}{\sqrt{2\pi}}e_0,\,
  \Phi_{0,-1}^{(1)} = \frac{1}{\sqrt{2\pi}}e^*_1,\label{eq:165}\\
  \Phi_{0,0}^{(2)} &=& \sqrt{\frac{3}{4\pi}}e^2_0,\,
  \Phi_{0,-1}^{(2)} = \sqrt{\frac{3}{2\pi}}e_0e^*_1,\,
  \Phi_{0,-2}^{(2)} = \sqrt{\frac{3}{4\pi}}e^{*2}_1.\nonumber\\
  \label{eq:166}
\end{eqnarray}
These are the zero-mode bases. Fluctuations along these directions (in the Hilbert space) do not cost any action. According to Eq.~(\ref{eq:146}), $\Phi_{0,0}^{(0)}$ is associated with the bosonic-bosonic block, $\Phi_{0,M}^{(1)}$ with bosonic-fermionic or fermionic-bosonic block, and $\Phi_{0,M}^{(2)}$ with fermionic-fermionic block. As a result, the total number of zero modes is
\begin{equation}\label{eq:295}
    2\times(1+2\times 2+3)=16,
\end{equation}
which is the same as the instanton's degrees of freedom (\ref{eq:272}), as expected.

To understand better the meaning of zero modes let us consider a `motion' in the instanton manifold.
According to Eqs.~(\ref{eq:142}) and (\ref{eq:143}), the coordinates of the instanton manifold are composed of
${\boldsymbol c}\equiv \{n_{10},n_{20},\lambda\}$, which discriminate different $R_+$ matrices,
and the natural coordinates ${\boldsymbol \xi}$ of $T$. Then, the motion is defined as a displacement,
$Q_+\rightarrow Q_++ dQ_+$, generated by an infinitesimal coordinate change:
${\boldsymbol c}\rightarrow {\boldsymbol c}+d{\boldsymbol c}$,
${\boldsymbol \xi}\rightarrow {\boldsymbol \xi}+d{\boldsymbol \xi}$. Here,
\begin{eqnarray}
dQ_+=U^{-1}[U d U^{-1},\Lambda] U, \quad
U=R_+
T.
\label{eq:177}
\end{eqnarray}
Comparing this with the first order $w$-expansion of
Eq.~(\ref{eq:144}), we find that the coordinate change generates a $w$ field, $w=[U d U^{-1},\Lambda]$. More precisely,
\begin{equation}\label{eq:178}
    v=-2(UdU^{-1})_{+-},\quad \tilde v=-2(UdU^{-1})_{-+}.
\end{equation}
As shown in Appendix \ref{sec:discussions_zero_modes}, these
two special fields are spanned by the zero-mode bases (\ref{eq:164})-(\ref{eq:166}), with the expansion coefficients
being the position and size of the instanton and the generators of the coset space $G/H$.
Therefore, field configurations corresponding to the instanton manifold $Q_+=U^{-1}\Lambda U,U=R_+T$ have the same
action as (\ref{eq:148}) ($s=+$), as expected.

\subsubsection{Fluctuation determinant and zero-mode integration}
\label{sec:fluctuation_determinant}

Observing Eqs.~(\ref{eq:115}) and (\ref{eq:117}), we find that their nonperturbative parts have the general structure as follows,
\begin{eqnarray}
&&\sum_{s=\pm}\int DR_s\int_{G/H} DT\int D(v',\tilde v')
e^{-\delta S_{np}[v',\tilde v']}\nonumber\\
&\times&e^{-4\pi\sigma + s2\pi i\sigma_{\rm H}-{\eta}{\rm Str}(\Lambda Q_s)}(\cdot)|_{Q=Q_s}\equiv O_{np},
\label{eq:189}
\end{eqnarray}
where the fluctuations $v',\tilde v'$ exclude the zero-mode components, and $(\cdot)$ is a shorthand notation for the quantity inside
the bracket $\langle\cdot\rangle_{\eta}$. Equation (\ref{eq:189}) factorizes the functional integral into an integration over
the instanton manifold ($\int DR_sDT$) namely the zero-mode integration and a functional integral over non-zero modes ($\int D(v',\tilde v')$).

The measure of $v',\tilde v'$ is flat and the ensuing functional integral can be readily carried out, giving
\begin{eqnarray}
e^D \equiv \int D(v',\tilde v')
e^{-\delta S_{np}[v',\tilde v']}=\frac{({\rm det}'\hat O^{(1)})^2}{{\rm det}'\hat O^{(0)}{\rm det}'\hat O^{(2)}},\quad
\label{eq:191}
\end{eqnarray}
where the denominator arises from the integration over bosonic (complex number) fields $v_{bb},v_{ff}$
and the numerator over fermionic (Grassmannian) fields $v_{bf},v_{fb}$, and
the functional determinants ${\rm det}'\hat O^{(a)}$ exclude zero modes. Note that the overall factor $\frac{\sigma}{2}$ in Eq.~(\ref{eq:146}) is canceled out.
To calculate Eq.~(\ref{eq:191}) explicitly we use the spectral decomposition of $\hat O^{(a)}$. Using Eqs.~(\ref{eq:158})-(\ref{eq:163}), we obtain
\begin{eqnarray}
D=2D^{(1)}-D^{(2)},
\label{eq:192}
\end{eqnarray}
where
\begin{eqnarray}
D^{(1)}=\sum_{J=1}^\infty \left((2J+2)\ln E_J^{(1)}-(2J+1)\ln E_J^{(0)}\right),
\label{eq:296}
\end{eqnarray}
\begin{eqnarray}
D^{(2)}=\sum_{J=1}^\infty \left((2J+3)\ln E_J^{(2)}-(2J+1)\ln E_J^{(0)}\right).
\label{eq:297}
\end{eqnarray}
This result can also be obtained by using the replica method with the replica limit taken \cite{Pruisken05}.
Note that if we include the overall factor $\frac{\sigma}{2}$ in $\hat O^{(a)}$,
then $E_J^{(a)}$ is replaced by $\frac{\sigma}{2}E_J^{(a)}$. This factor is, however, canceled out and the value of $D$ is not affected.

Equations (\ref{eq:296}) and (\ref{eq:297}) show that $D^{(1,2)}$ suffer ultraviolet divergence. This can be cured by the regularization method of 't Hooft. The procedure and results are exactly
the same as that described in Ref.~\onlinecite{Pruisken87a} and here we shall give the results only, read
\begin{eqnarray}
  D^{(1)} \rightarrow D^{(1)}_{\rm reg}=-\ln \mathscr{M}+\frac{3}{2}-2\ln 2
\label{eq:193}
\end{eqnarray}
and
\begin{eqnarray}
  D^{(2)} \rightarrow D^{(2)}_{\rm reg}=-2\ln \mathscr{M}+4-3\ln 3-\ln 2,
\label{eq:193}
\end{eqnarray}
where $\mathscr{M}$ is a mass associated with the Pauli-Villars regulator fields. It is important that this mass cancels out in regularized $D$,
\begin{eqnarray}
\label{eq:195}
  D &\rightarrow& D_{\rm reg}=2D^{(1)}_{\rm reg}-D^{(2)}_{\rm reg}\nonumber\\
   &=& -1+3\ln\frac{3}{2}.
\end{eqnarray}
This cancelation has an important physical meaning as follows. Substituting Eq.~(\ref{eq:191}) into
Eq.~(\ref{eq:189}), we find that, if such divergence $\sim \ln \mathscr{M}$ existed, then it
would renormalize the instanton action namely $\sigma$; this would contradict the result of
Eq.~(\ref{eq:119}) showing that the one-loop perturbative correction to $\sigma$ vanishes.
Therefore, the cancelation of $\ln \mathscr{M}$ in Eq.~(\ref{eq:195}) reflects the well-known result of the vanishing of one-loop weak localization correction to $\sigma$ for
systems with broken time-reversal symmetry \cite{Efetov97}. If we keep the $v',\tilde v'$-expansion up to the fourth order,
then integrating out $v',\tilde v'$ gives rise to a nonvanishing
quantum correction $\sim{\cal O}(\frac{1}{\sigma})$ to the instanton action, i.e.,
\begin{equation}\label{eq:254}
    \delta S[Q_s]|_{\omega=0}=4\pi \frac{A_2}{\sigma},
\end{equation}
where the numerical constant $A_2$ is universal, to be given below.
With the substitution of Eqs.~(\ref{eq:191}), (\ref{eq:195}) and (\ref{eq:254}), Eq.~(\ref{eq:189}) is reduced to
\begin{eqnarray}
O_{np}&=&\frac{1}{e}\left(\frac{3}{2}\right)^3\sum_{s=\pm}e^{-4\pi\sigma\left(1+\frac{A_2}{\sigma^2}\right)+2\pi s i\sigma_{\rm H}}
\nonumber\\
&\times&\int DR_s\int_{G/H} DT e^{-{\eta}{\rm Str}(Q_s\Lambda)}(\cdot)|_{Q=Q_s}.\,\,\,\quad
\label{eq:194}
\end{eqnarray}
In what follows we will perform the integration over the instanton manifold.


Consider the length element
\begin{equation}\label{eq:212}
    ds^2\equiv\frac{1}{4}\int_{-1}^1 d\eta\int_0^{2\pi}d\vartheta {\rm str}(\tilde v v),
\end{equation}
where the zero-mode fields $v,\tilde v$ are generated by the change in the natural coordinates according to Eq.~(\ref{eq:178}).
In Appendix \ref{sec:discussions_zero_modes}, we express $v,\tilde v$ in terms of the coordinate change,
$d\lambda, dz_0,dz_0^*$, associated with $R_s$ and the generators,
\begin{eqnarray}
t_{\lambda\lambda'}^{\alpha\alpha'}\equiv(T
d
T^{-1})_{\lambda\alpha,\lambda'\alpha'},
\label{eq:240}
\end{eqnarray}
of $T$ (see Eqs.~(\ref{eq:181})-(\ref{eq:188})).
Substituting the expressions obtained for $v,\tilde v$ into Eq.~(\ref{eq:212}) gives
\begin{eqnarray}
ds^2&=& \overline{e_0^2|e_1|^2}\left(\frac{d\lambda}{\lambda}\right)^2 + \overline{e_0^2|e_1|^2} d\phi^2 \nonumber\\
&& + D{\bar \psi}_b g_b D\psi_b +
D{\bar \psi}_f g_f D\psi_f.
\label{eq:200}
\end{eqnarray}
Here,
\begin{eqnarray}
D\psi_b\equiv (dz_0,t_{+-}^{ff},t_{+-}^{bb})^{\rm T},
\label{eq:201}
\end{eqnarray}
\begin{eqnarray}
D{\bar \psi}_b\equiv (d{\bar z}_0,t_{-+}^{ff},t_{-+}^{bb}),
\label{eq:202}
\end{eqnarray}
\begin{eqnarray}
D\psi_f\equiv (t_{+-}^{bf},t_{+-}^{fb},t_{++}^{bf},t_{--}^{fb})^{\rm T},
\label{eq:203}
\end{eqnarray}
\begin{eqnarray}
D{\bar \psi}_f\equiv (t_{-+}^{fb},t_{-+}^{bf},t_{++}^{fb},t_{--}^{bf}),
\label{eq:204}
\end{eqnarray}
with the superscript `T' denoting the transpose, and
\begin{equation}\label{eq:273}
    id\phi\equiv t_{++}^{ff}-t_{--}^{ff}.
\end{equation}
The matrices $g_{b,f}$ are given by
\begin{eqnarray}
g_b\equiv \left(
              \begin{array}{ccc}
                \frac{1}{\lambda^2}\overline{ e_0^4} & \frac{1}{\lambda^2}\overline{ e_0^4} & 0 \\
                -\frac{1}{\lambda^2}\overline{ e_0^4} & -\left(\overline{e_0^4}+\overline{|e_1|^4}\right) & 0 \\
                0 & 0 & \overline{1} \\
              \end{array}
            \right)
\label{eq:205}
\end{eqnarray}
and
\begin{eqnarray}
g_f\equiv \left(
              \begin{array}{cccc}
                -\overline{|e_1|^2} & 0 & 0 & 0 \\
                0 & \overline{|e_1|^2} & 0 & 0 \\
                0 & 0 & -\overline{e_0^2} & 0 \\
                0 & 0 & 0 & \overline{e_0^2} \\
              \end{array}
            \right),
\label{eq:206}
\end{eqnarray}
respectively. The overline stands for $\int_{-1}^1d\eta\int_0^{2\pi}d\vartheta (\cdot)$.

By using Eq.~(\ref{eq:200}) we can factorize the measure $DR_sDT$ into four parts (see Appendix \ref{sec:splitting_measure} for the
proof): (i) $d\lambda dn_{10}dn_{20}$ associated with the
instanton's size and position, (ii) the measure of the $U(1)$ group represented by Eq.~(\ref{eq:316}), (iii) the measure of
the coset space $\frac{U(1,1|2)}{U(1|1)\times U(1|1)}$, and (iv) the measure of
the coset space $(\frac{U(1|1)}{U(1)\times U(1)})^2$. We parametrize $(\frac{U(1|1)}{U(1)\times U(1)})^2$ as
\begin{eqnarray}
  u
  \equiv
  \left(
                          \begin{array}{cc}
                            u_1 & 0 \\
                            0 & u_2 \\
                          \end{array}
                        \right)_{ar},
    \label{eq:196}
    \end{eqnarray}
where
    \begin{eqnarray}
  u_1 = \left(
                          \begin{array}{cc}
                            1+\frac{1}{2}{\zeta}_1{\zeta}_1^* & {\zeta}_1 \\
                            {\zeta}_1^* & 1-\frac{1}{2}{\zeta}_1{\zeta}_1^* \\
                          \end{array}
                        \right)_{bf},
                        \label{eq:217}
\end{eqnarray}
\begin{eqnarray}
u_2 = \left(
                          \begin{array}{cc}
                            1+\frac{1}{2}{\zeta}_2^*{\zeta}_2 & i{\zeta}_2^* \\
                            -i{\zeta}_2 & 1-\frac{1}{2}{\zeta}_2^*{\zeta}_2 \\
                          \end{array}
                        \right)_{bf},\label{eq:218}
\end{eqnarray}
with $\zeta_{1,2},\zeta_{1,2}^*$ being
Grassmannians \cite{Zirnbauer88}. The result of the factorization is
\begin{eqnarray}\label{eq:232}
    DR_sDT &=& \frac{4}{\pi^2}\frac{\overline{(e_0|e_1|)^2}\,\, \overline{e_0^4}\,\, \overline{|e_1|^4}\,\, \overline{1}}{\left(\overline{e_0^2}\,\,
    \overline{|e_1|^2}\right)^2} \nonumber\\
    &\times& \frac{d\lambda}{\lambda^3} d\phi dn_{10}dn_{20}d\tilde Td\zeta_1d\zeta_1^*d\zeta_2^*d\zeta_2,
\end{eqnarray}
with $\tilde T\in \frac{U(1,1|2)}{U(1|1)\times U(1|1)}$.
Corresponding to this factorization,
\begin{equation}\label{eq:235}
    Q_s=(u_0u\tilde T)^{-1}\Lambda_s(u_0u\tilde T),
\end{equation}
with
\begin{eqnarray}
u_0(\phi)\equiv e^{i\frac{\phi}{2}\sigma_{ar}^3}\otimes \mathbbm{E}_{ff}+\mathbbm{1}_{ar}\otimes \mathbbm{E}_{bb}.
\label{eq:236}
\end{eqnarray}
With the substitution
of Eqs.~(\ref{eq:232}) and (\ref{eq:235}), Eq.~(\ref{eq:194}) is reduced to
\begin{eqnarray}
O_{np}&=&\frac{4}{\pi^2 e}\sum_{s=\pm}e^{-4\pi\sigma+2\pi s i\sigma_{\rm H}}
\nonumber\\
&\times&\int\frac{d\lambda}{\lambda^3}dn_{10}dn_{20}d\phi\int d\tilde T
\int d\zeta_1d\zeta_1^*d\zeta_2^*d\zeta_2
\nonumber\\
&\times&e^{-{\eta}{\rm Str}(Q\Lambda)}(\cdot)\big|_{Q=(u_0u\tilde T)^{-1}\Lambda_s(u_0u\tilde T)}.
\label{eq:197}
\end{eqnarray}

To proceed further, we note that the frequency term has an important consequence. To see this we substitute
\begin{eqnarray}
&&{\eta}{\rm Str}(Q\Lambda)|_{Q=(u_0u\tilde T)^{-1}\Lambda_s(u_0u\tilde T)}\nonumber\\
&=&
{\eta} \Omega {\rm str}(\Lambda \tilde T^{-1}\Lambda\tilde T)\nonumber\\
&&-2\pi \lambda^2 {\eta}\ln\frac{\Omega
}{\lambda^2}{\rm str}(\Lambda (u\tilde T)^{-1}(\sigma_{ar}^3\otimes \mathbbm{E}_{ff})(u\tilde T))\quad
\label{eq:249}
\end{eqnarray}
into the exponent of $e^{-{\eta}{\rm Str}(Q\Lambda)}$. Then, the first term in Eq.~(\ref{eq:249}) implies that in the limit of $\Omega\rightarrow\infty$ the
$\tilde T$ integration in Eq.~(\ref{eq:197}) is restricted to $\tilde T = \mathbbm{1}$ and the integration over fluctuations around $\tilde T = \mathbbm{1}$ gives a factor of unity
due to supersymmetry. The second term in Eq.~(\ref{eq:249}) is reduced correspondingly to $4\pi \lambda^2 {\eta}\ln\frac{\Omega
}{\lambda^2}$, which, formally, diverges in the infrared (i.e., $\Omega\rightarrow\infty$). However,
as discussed in Appendix \ref{sec:constrained_instanton}, such divergence is unphysical because the
instanton solution described by Eqs.~(\ref{eq:142})-(\ref{eq:233}) is valid only for instanton size $\lambda \lesssim \sqrt{\sigma/{\eta}}$. In other words, $\sqrt{\sigma/{\eta}}$ serves as an infrared cutoff of the integral over $\lambda$ and $N$. Taking this into account we find that the second term
is finite, which is $4\pi \lambda^2 {\eta}\ln\frac{\sigma
}{{\eta}\lambda^2}$, and vanishes in the limit of ${\eta}\rightarrow 0$. So, a physical meaning of the scale $\sqrt{\sigma/{\eta}}$ can be given as follows. When the instanton size reaches
this scale the instanton action $4\pi\sigma$ becomes comparable to the frequency term. At this scale the derivation
of the instanton solution given by Eqs.~(\ref{eq:142})-(\ref{eq:233}) is invalid, since it ignores the frequency term. This suggests that the instanton behavior at large $N$ would be very different from that described by Eqs.~(\ref{eq:143}) and (\ref{eq:233}).
Instead, the so-called constrained instanton \cite{Pruisken05} is involved and we refer the readers to Appendix \ref{sec:constrained_instanton} for further discussions.

Returning to Eq.~(\ref{eq:197}) and taking the discussions above into account,
we eventually reduce Eq.~(\ref{eq:197}) to
\begin{eqnarray}
O_{np}&=&\frac{4}{\pi^2 e}\sum_{s=\pm}\int\frac{d\lambda}{\lambda^3}dn_{10}dn_{20}d\phi
\int d\zeta_1d\zeta_1^*d\zeta_2^*d\zeta_2\nonumber\\
&\times&
e^{-4\pi\sigma\left(1+\frac{A_2}{\sigma^2}\right)+2\pi s i\sigma_{\rm H}}
(\cdot)|_{Q=(u_0u)^{-1}\Lambda_s(u_0u)}
\label{eq:199}
\end{eqnarray}
upon sending ${\eta}$ to zero.

\subsubsection{Instanton contribution to $\tilde \sigma$ and $\tilde \sigma_{\rm H}$}
\label{sec:single_instanton_contribution_longitudinal_conductivity}

Applying Eq.~(\ref{eq:199}) to Eq.~(\ref{eq:115}), we find
\begin{eqnarray}\label{eq:124}
    \delta\sigma_{np}&=&-\frac{
    32\pi}{e} \int \frac{d\lambda}{\lambda}(\sigma^2+{\cal O}(\sigma))\nonumber\\
    &\times& e^{-4\pi\sigma\left(1+\frac{A_2}{\sigma^2}\right)}\cos 2\pi \sigma_{\rm H}.
\end{eqnarray}
The term $\sim {\cal O}(\sigma)$ in the bracket arises from the first term of Eq.~(\ref{eq:115}) and
higher order contributions of fluctuations (around instanton configurations) to the second term of Eq.~(\ref{eq:115}).
Applying Eq.~(\ref{eq:199}) to Eq.~(\ref{eq:117}), we find that the first term vanishes and
the second term gives
\begin{eqnarray}\label{eq:125}
    \delta\sigma_{{\rm H},np}&=&-\frac{
    64\pi}{e} \int \frac{d\lambda}{\lambda}(\sigma^2+{\cal O}(\sigma))\nonumber\\
    &\times& e^{-4\pi\sigma\left(1+\frac{A_2}{\sigma^2}\right)}\sin 2\pi \sigma_{\rm H}.
\end{eqnarray}
Here the term $\sim {\cal O}(\sigma)$ arises from
higher order contributions of fluctuations (around instanton configurations).

Equations (\ref{eq:119}), (\ref{eq:120}), (\ref{eq:124}) and (\ref{eq:125}) are the main results of this section. Note that the integral over $\lambda$ suffers an infrared divergence. The treatments of this divergence is the main subject of the next section.

\section{Two-parameter scaling theory of Planck-IQHE}
\label{sec:microscopic_theory}

We have shown that at short times the energy always grows linearly with time, and the growth rate grows quadratically with $h_e^{-1}$ for 
small $h_e$. What happens to the energy growth at long times? Having obtained the leading perturbative and instanton contributions to the transport parameters, we are now ready to answer this question. In this section we will show that the Planck's quantum drives a dynamical analog of IQHE.

\subsection{RG equations}
\label{sec:renormalization_group_flow}

As mentioned before, the perturbative and nonperturbative instanton contributions to the transport parameters suffer infrared divergence. The idea to circumvent this, is to consider the transport parameters
at a finite scale size $\tilde \lambda$, denoted as $\tilde\sigma(\tilde \lambda)$ and $\tilde\sigma_{\rm H}(\tilde \lambda)$ accordingly, and find the RG equations satisfied by them. Inheriting from the structure of $\tilde \sigma=\sigma+\delta\sigma_{p}+
\delta\sigma_{np}$ and $\tilde\sigma_{\rm H}=\sigma_{\rm H}+\delta\sigma_{{\rm H},np}$, these equations have the general form,
\begin{eqnarray}
    \frac{d\tilde \sigma}{d\ln \tilde\lambda}=
    \beta_{{\rm L},p}(\tilde \sigma) +\beta_{{\rm L},np}(\tilde \sigma,\tilde \sigma_{\rm H})\equiv \beta_{\rm L}(\tilde \sigma,\tilde \sigma_{\rm H}),
    \label{eq:133}
\end{eqnarray}
where the RG function $\beta_{\rm L}$ is composed of perturbative
($\beta_{{\rm L},p}$) and nonperturbative ($\beta_{{\rm L},np}$) parts, with the former depending only on $\tilde \sigma$ and
is an expansion in $\frac{1}{\tilde\sigma}$, and the latter on both $\tilde \sigma$ and $\tilde \sigma_{\rm H}$, and
\begin{eqnarray}
    \frac{d\tilde \sigma_{\rm H}}{d\ln \tilde\lambda}=\beta_{{\rm H}}(\tilde \sigma,\tilde \sigma_{\rm H}),
    \label{eq:134}
\end{eqnarray}
where $\beta_{\rm H}$ consists of nonperturbative ($\beta_{{\rm H},np}$) part only.
Then, we solve these two equations and find the fixed points of RG flow (i.e., $\tilde \lambda \rightarrow \infty$).

To find $\beta_{\rm L,H}$ explicitly we recall a perturbative one-loop calculation within the replica formalism \cite{Pruisken05}. According to this calculation, the one-loop correction to the instanton action, i.e., $\sigma$ in Eq.~(\ref{eq:148}), is identically the same as the perturbative one-loop
expansion of $\tilde \sigma$. In particular, (within the replica formalism)
these two one-loop results have the same ultraviolet divergence structure. This calculation leads to the ansatz that
perturbative loop expansions for $\tilde \sigma$ and the instanton action are identically the same.
Because such loop expansions are of perturbative nature, we expect this ansatz to be valid also for the supersymmetry technique. (It is well known that on the perturbation level the replica and supersymmetry techniques give the same results \cite{Tian05,Efetov97}.) Indeed, we have already shown explicitly that the perturbative one-loop contribution to $\tilde \sigma$ and the instanton action both vanish (comparing Eqs.~(\ref{eq:119}) and (\ref{eq:254})),
in agreement with the replica limit of the results obtained in Ref.~\onlinecite{Pruisken05}. Taking this ansatz into account, we find
\begin{eqnarray}\label{eq:252}
    \tilde\sigma
    &=&\left(\sigma+\beta_{{\rm L},p}(\sigma)\ln (\mathscr{M}e^\gamma)\right)\nonumber\\
    &&-\frac{
    32\pi}{e} \int^{\tilde \lambda}
    \frac{d\lambda}{\lambda}\sigma^2 e^{-4\pi(\sigma+\beta_{{\rm L},p}(\sigma)\ln (\mathscr{M}e^\gamma))}\cos 2\pi \sigma_{\rm H}\nonumber\\
\end{eqnarray}
and
\begin{eqnarray}\label{eq:253}
    \tilde\sigma_{\rm H}
    &=&\sigma_{\rm H}\nonumber\\
    &&-\frac{
    64\pi}{e} \int^{\tilde\lambda} \frac{d\lambda}{\lambda}\sigma^2e^{-4\pi(\sigma+\beta_{{\rm L},p}(\sigma)\ln (\mathscr{M}e^\gamma))}\sin 2\pi \sigma_{\rm H}\nonumber\\
\end{eqnarray}
from Eqs.~(\ref{eq:119}), (\ref{eq:120}), (\ref{eq:124}) and (\ref{eq:125}), where the Euler constant$\gamma\approx 0.577$.

Note that in Sec.~\ref{sec:fluctuation_determinant} the Pauli-Villars regularization was performed for the field theory in $(\eta,\vartheta)$ space. This corresponds to the
introduction of a spatially varying quantity, $\mu^2(N)\mathscr{M}$, in
the flat $N$ space. Following Ref.~\onlinecite{Pruisken05}, upon passing
to the $N$ space, we make the following replacement,
\begin{equation}\label{eq:255}
    \mathscr{M}\rightarrow\frac{1}{4}e^2\lambda \mu_0
\end{equation}
for the Pauli-Villars mass, with $\mu_0^{-1}$
being a flat microscopic angular momentum scale.

In Appendix \ref{sec:beta_function} we derive the perturbative RG function,
\begin{equation}\label{eq:130}
    \beta_{{\rm L}, p}(\tilde \sigma)=-\frac{1}{
    8\pi^2\tilde \sigma},
\end{equation}
which gives the coefficient $A_2=-\frac{1}{8\pi^2}$ in Eq.~(\ref{eq:254}).
We substitute Eqs.~(\ref{eq:255}) and (\ref{eq:130}) into Eqs.~(\ref{eq:252}) and (\ref{eq:253}). Noticing that
$\sigma$ has the same form of the quantum corrections in all scales of $\tilde \lambda$ and so does $\sigma_{\rm H}$, we obtain
the following self-consistent equations,
\begin{eqnarray}\label{eq:256}
    \tilde\sigma(\tilde \lambda)
    &=&\tilde\sigma(\lambda_0)-\int^{\tilde \lambda}_{\lambda_0}\frac{d\lambda}{\lambda}\bigg(\frac{1}{
    8\pi^2\tilde \sigma(\lambda)}+\nonumber\\
    &&\frac{
    32\pi}{e}
    (\tilde\sigma(\lambda))^2 e^{-4\pi\tilde\sigma(\lambda)}\cos (2\pi \tilde\sigma_{\rm H}(\lambda))\bigg)
\end{eqnarray}
and
\begin{eqnarray}\label{eq:257}
    \tilde\sigma_{\rm H}(\tilde\lambda)
    &=&\tilde\sigma_{\rm H}(\lambda_0)-\nonumber\\
    &&\frac{
    64\pi}{e} \int^{\tilde\lambda}_{\lambda_0}
    \frac{d\lambda}{\lambda}\sigma^2e^{-4\pi\tilde\sigma(\lambda)}\sin (2\pi \tilde\sigma_{\rm H}(\lambda)),\quad
\end{eqnarray}
with $\lambda_0$ being the renormalization reference point. These two equations suggest that $\tilde \lambda$ can be interpreted as the size of a (large) background instanton. From Eq.~(\ref{eq:256}) we obtain Eq.~(\ref{eq:292}), with
\begin{equation}\label{eq:131}
    \beta_{\rm L}=-\frac{1}{
    8\pi^2\tilde \sigma}-\frac{
    32\pi}{e} \tilde \sigma^2 e^{-4\pi\tilde \sigma}\cos 2\pi \tilde \sigma_{\rm H},
\end{equation}
where the second term is the instanton contribution $\beta_{{\rm L},np}$.
From Eq.~(\ref{eq:257}) we obtain Eq.~(\ref{eq:293}), with
\begin{equation}\label{eq:132}
    \beta_{\rm H}=-\frac{
    64\pi}{e} \tilde \sigma^2 e^{-4\pi\tilde\sigma}\sin 2\pi \tilde \sigma_{\rm H}.
\end{equation}
Although these results are derived for the weak coupling regime (i.e., large $\tilde \sigma$),
as we will see below, they turn out to capture well the system's behavior in the strong coupling regime (i.e., for small $\tilde \sigma$), even quantitatively. The RG equations (\ref{eq:292}) and (\ref{eq:293}) constitute a two-parameter scaling theory.

Here we make a remark. With the substitution of Eqs.~(\ref{eq:255}) and (\ref{eq:130}) into the first line of Eq.~(\ref{eq:252}), we find that at a length scale of $\mu_0^{-1} e^{8\pi^2\sigma^2}$ the perturbative contribution to $\tilde \sigma$ is comparable to $\sigma$, signalling that localization physics begins to dominate. This length scale can be translated into a characteristic time scale $\sim e^{4\pi^2\sigma^2}$. At this time localization effects dominate over chaotic diffusion.

\subsection{RG flow and quantum phase structures}
\label{sec:IQHE}

\begin{figure}[h]
\includegraphics[width=8.6cm]{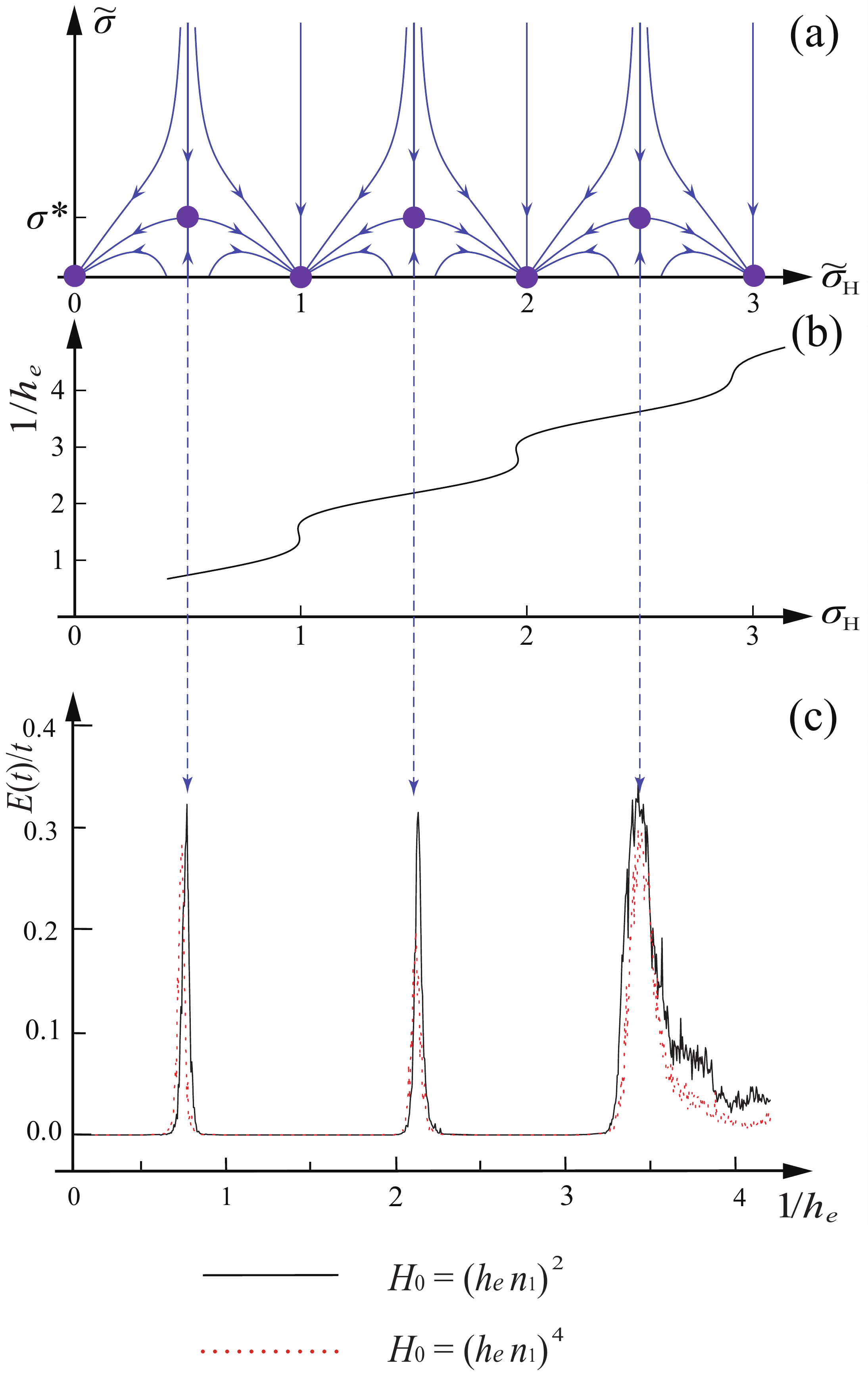}
\caption{(a) The RG flow line structure of
two-parameter RG equations. The fixed points (solid circles) at $\tilde \sigma_{\rm H}=n$ ($n\in \mathbb{Z}$) correspond to insulating phases and
at $\tilde \sigma_{\rm H}=n+\frac{1}{2}$ to metal-insulator transitions.
(b) The analytic prediction for the critical $h_e$-values is made by letting the unrenormalized transport parameter $\sigma_{\rm H}$
(solid line) be half integers. Namely, the solid and dashed lines intersect at the critical $h_e$-values.
(c) Long-time simulations confirm the transitions with the critical values in excellent
agreement with analytic predictions, and the
critical points are evenly spaced along the $h_e^{-1}$-axis; They also confirm that the critical metallic phase has an energy
growth rate order of unity.
The simulation results of $E(t)/t$ are for $t=1.2\times 10^6$ and for the specific forms of $H_0=(-ih_e\partial_{\theta_1})^\alpha$, with the exponent $\alpha=2$ (black solid line) and $\alpha=4$ (red dashed line), respectively, and $V$ given by Eqs.~(\ref{eq:52}),
(\ref{eq:58}) and (\ref{eq:59}).
Simulations confirm that the critical value is robust against changing $\alpha$.}
\label{fig:3}
\end{figure}

The RG flow lines given by Eqs.~(\ref{eq:292}) and (\ref{eq:293}) are shown in Fig.~\ref{fig:3} (a). In spite of the absence of the Landau bands in the present system, this RG flow line structure is identical to that responsible for conventional magnetic field-driven IQHE in strongly disordered environments
\cite{Pruisken10,Pruisken87a,Pruisken87,Pruisken05,Pruisken84a,Khmelnitskii83}. Below we summarize the main features of the RG flow lines.

First,
thanks to the periodicity of the cosine (sine) function in Eq.~(\ref{eq:131})
(Eq.~(\ref{eq:132})), the RG flow lines are periodic in $\tilde\sigma_{\rm H}$, i.e., invariant under the shift: $\tilde\sigma_{\rm H}\rightarrow \tilde\sigma_{\rm H}+1$.

Second, the RG flow has two types of fixed points which are the zeros of $\beta_{\rm L,H}$ (solid circles in Fig.~\ref{fig:3} (a)). One type of fixed points, located at
$(\sigma_{\rm H}^*=n,0)$, are stable. They correspond to quantum phases with a vanishing zero-frequency conductivity,
$\sigma(\omega\rightarrow 0)=\tilde \sigma=0$. (Recall Eq.~(\ref{eq:258}).)
This is a characteristic of insulator. These insulating phases correspond to the plateau regimes in Fig.~\ref{fig:3}(a). They are distinguished by
the plateau value $\sigma_{\rm H}^*$;
this number or more precisely $2\pi\sigma_{\rm H}^*$ is the renormalization of the bare topological angle $2\pi\sigma_{\rm H}$ and, therefore, of topological nature.
The other type of fixed points, located at $(n+\frac{1}{2},\sigma^*)$, are stable in the $\tilde\sigma$-direction but unstable in the $\tilde \sigma_{\rm H}$-direction: these are critical fixed points. They correspond to quantum phases
with a zero-frequency conductivity $\sigma(\omega\rightarrow 0)=\tilde \sigma=\sigma^*$,
which is a main characteristic of metals. Substituting this zero-frequency conductivity into Eq.~(\ref{eq:106}) gives $E\stackrel{t\rightarrow \infty}{\longrightarrow}\sigma^*t$. The critical lines are located at $\tilde \sigma_{\rm H}=n+\frac{1}{2}$. Passing through each of these lines, the system exhibits a plateau transition: $\sigma_{\rm H}^*$ changes by unity. Simultaneously, a metal-insulator transition occurs.

Inheriting from the universality and $\tilde \sigma_{\rm H}$-periodicity of $\beta_{\rm L}$, the value of $\sigma^*$ namely the zero of $\beta_{\rm L}|_{\tilde \sigma_{\rm H}=n+\frac{1}{2}}$ is universal. Specifically, this value is insensitive to system's details,
e.g., $H_0$, $V_i$ and the critical $h_e$-values. It is important to note that this universal value is much smaller than the
short-time energy growth rate $\sigma\sim h_e^{-2}$ (for 
small $h_e$). This substantial difference reflects both quantum and topological nature of the critical metal, as discussed in Sec.~\ref{sec:quantum_stochastic_web}.


\subsection{Planck's quantum-driven phase transitions}
\label{sec:microscopic_theory_IQHE}

We have analyzed the structure of the RG flow lines. A variety of possible quantum phases
are predicted, which are the fixed points of this RG flow. However, no information has been provided regarding the phase diagram
as the Planck's quantum varies, which is the subject of this subsection.

\subsubsection{Universal Planck-IQHE pattern 
for small $h_e$}
\label{sec:transition}

According to Fig.~\ref{fig:3}(a),
the bare Hall conductivity $\sigma_{\rm H}$ determines the quantum phase, namely,
the fixed point where the RG flow line ends.
For small $h_e
$, a universal scaling law,
$\sigma_{\rm H}(h_e)\sim h_e^{-1}$ (namely Eq.~(\ref{eq:9})), follows.
Consequently, as $h_e^{-1}$ increases the system successively passes through the critical lines in Fig.~\ref{fig:3}(a),
corresponding to half-integer $\sigma_{\rm H}(h_e)$ (cf. Eq.~(\ref{eq:60})), and sequential quantum phase transitions are triggered.
At the critical value of $h_e$, determined by Eq.~(\ref{eq:60}), the system is metallic and the quantum number $\sigma_{\rm H}^*$ jumps by unity;
between two nearest critical $h_e$-values the system is insulating and
$\sigma_{\rm H}^*$ does not change. In addition, because of the linear scaling (\ref{eq:9}) the critical points are evenly spaced along the $h_e^{-1}$-axis. This feature is insensitive to the details of the potential, although the value of the spacing does depend on the potential via Eq.~(\ref{eq:S12}). These results lead to the pattern in Fig.~\ref{fig:5}, which bears a close resemblance to conventional IQHE.

We stress that the Planck-IQHE pattern is robust. It is independent of the modification of the free rotation Hamiltonian
$H_0$ and the kicking potential $V_i$, as long as strong chaoticity and symmetry are not destroyed. Moreover, we note that the microscopic expressions (\ref{eq:3}), (\ref{eq:S4}) and (\ref{eq:S3}) for the bare transport parameters $\sigma$ and $\sigma_{\rm H}$
are independent of $H_0$. This leads to a remarkable result. That is, the critical point determined by Eq.~(\ref{eq:60}) is not shifted when $H_0$ is modified.
Correspondingly, the pattern in Fig.~\ref{fig:5} is totally unaffected by this modification.

We remark that there are some very special cases (see Appendix \ref{sec:integral_limit_case_study} for example) for which the proportionality coefficient of the scaling law (\ref{eq:9}) vanishes. In this case no transitions occur at small $h_e$. Whether and how transitions occur at large $h_e$ then depends on the details of $V_i$, and we discuss this issue below.

\subsubsection{Nonuniversal pattern for large $h_e$}
\label{sec:quantum_anomalous_Hall_effect}

We have seen that the universality of the scaling law (\ref{eq:9}) is responsible for that of Planck-IQHE pattern for small $h_e$. 
For large $h_e
$, this scaling generally breaks down. In contrast to Eq.~(\ref{eq:9}),
the behavior of $\sigma_{\rm H}$ is non-monotonic as $h_e^{-1}$ varies. As a result, given an integer $n$
Eq.~(\ref{eq:60}) generally have several solutions, depending on the details of $V$. Corresponding to this, unlike the blue line in Fig.~\ref{fig:5}, as $h_e^{-1}$ increases $\sigma_{\rm H}^*$ exhibits reentrant behavior, i.e., the system reenters into the same insulating phase.
So, the $\sigma_{\rm H}^*$ pattern is composed of several pieces,
for each of which $\sigma_{\rm H}^*$ monotonically increases or decreases.
Note that at every critical point the change of $\sigma_{\rm H}^*$ is unity, i.e., $\Delta \sigma_{\rm H}^*=\pm 1$. Since the linear scaling of $\sigma_{\rm H}$ breaks down, the critical points are no longer
evenly spaced along the $h_e^{-1}$-axis. Therefore, the $\sigma_{\rm H}^*$ pattern
for large $h_e$ is nonuniversal, depending on specific $V$.

Here we make two remarks. First, as before
the critical values of $h_e$ are not shifted by the modification of $H_0$. Second, it is possible that for certain potentials the pattern in Fig.~\ref{fig:5} extrapolates well to the large-$h_e$ regime, as long as the linear scaling (\ref{eq:9}) holds there.

\subsubsection{Estimate of $\sigma^*$ and critical exponent}
\label{sec:critical_exponent}

Although the RG functions (\ref{eq:131}) and (\ref{eq:132}) are derived for weak coupling, we may extrapolate them to the strong coupling regime to quantitatively assess various properties of quantum criticality. Specifically, by setting $\tilde \sigma_{\rm H}=n+\frac{1}{2}$ in Eq.~(\ref{eq:131}) and letting $\beta_{\rm L}$ vanish, the zero gives an estimate of $\sigma^*$. The result is
\begin{equation}\label{eq:260}
    \sigma^*\approx 0.44.
\end{equation}
This value is consistent with the value of $0.25$ as one expects based on the well-known result of the critical universal conductivity in the conventional IQHE (see, e.g., Refs.~\onlinecite{Wei86,Pruisken06,Weng98,Ruzin94,Bhatt93,Bhatt95,Zhang92} and references therein), although we do not know how to generalize the critical theory of conventional IQHE \cite{Zhang92} to the present QKR systems. Note that the longitudinal conductivity defined by Eq.~(\ref{eq:48}) differs from that in conventional IQHE by a factor of $2$ (cf. Ref.~\onlinecite{note_coupling_constant}).

With the help of Eq.~(\ref{eq:260}), we find from Eq.~(\ref{eq:132}) the critical exponent $\nu$ for the localization length $\xi$, i.e.,
\begin{eqnarray}
&&\xi\sim |\sigma_{\rm H}(h_e)-\sigma_{\rm H}(h_e^*)|^{-\nu},\nonumber\\
&&\nu=\left(\frac{d\beta_{\rm H}}{d\tilde\sigma_{\rm H}}\right)^{-1}\bigg|_{(\tilde\sigma_{\rm H}=n+\frac{1}{2},
\tilde\sigma=\sigma^*)}\approx 2.75,
\label{eq:259}
\end{eqnarray}
where $h_e^*$ is the critical value of Planck's quantum.
Equations (\ref{eq:260}) and (\ref{eq:259}) are valid, no matter whether the critical value is small or large.

\subsection{Discussions on potential's topology}
\label{sec:discussions_effects_topology_potential}

It is important that, unlike many topological phenomena in condensed matter systems, the Planck-IQHE here is {\it not} triggered by changes in the potential's topology, i.e., the Pontryakin index,
$-\frac{1}{4\pi}\int\!\!\!\!\int d\theta_1d\theta_2 \hat V \cdot (\partial_{\theta_1} \hat V \times \partial_{\theta_2} \hat V)$ with $\hat V\equiv (V_1,V_2,V_3)/|V|$. Indeed, this index is independent of $h_e$. If it were applied to characterize topological phases in present system, a conclusion can be immediately reached. That is, the Planck's quantum drives no transitions, which contradicts the results established above. Therefore, our results are, both conceptually and phenomenologically, different from an earlier finding in QKR \cite{Beenakker11} (see Sec.~\ref{sec:modification_kicking_potential} for further discussions). In particular, counterintuitively, our results remain valid even when the Pontryakin index defined above vanishes.

\begin{figure}[h]
\includegraphics[width=8.6cm]{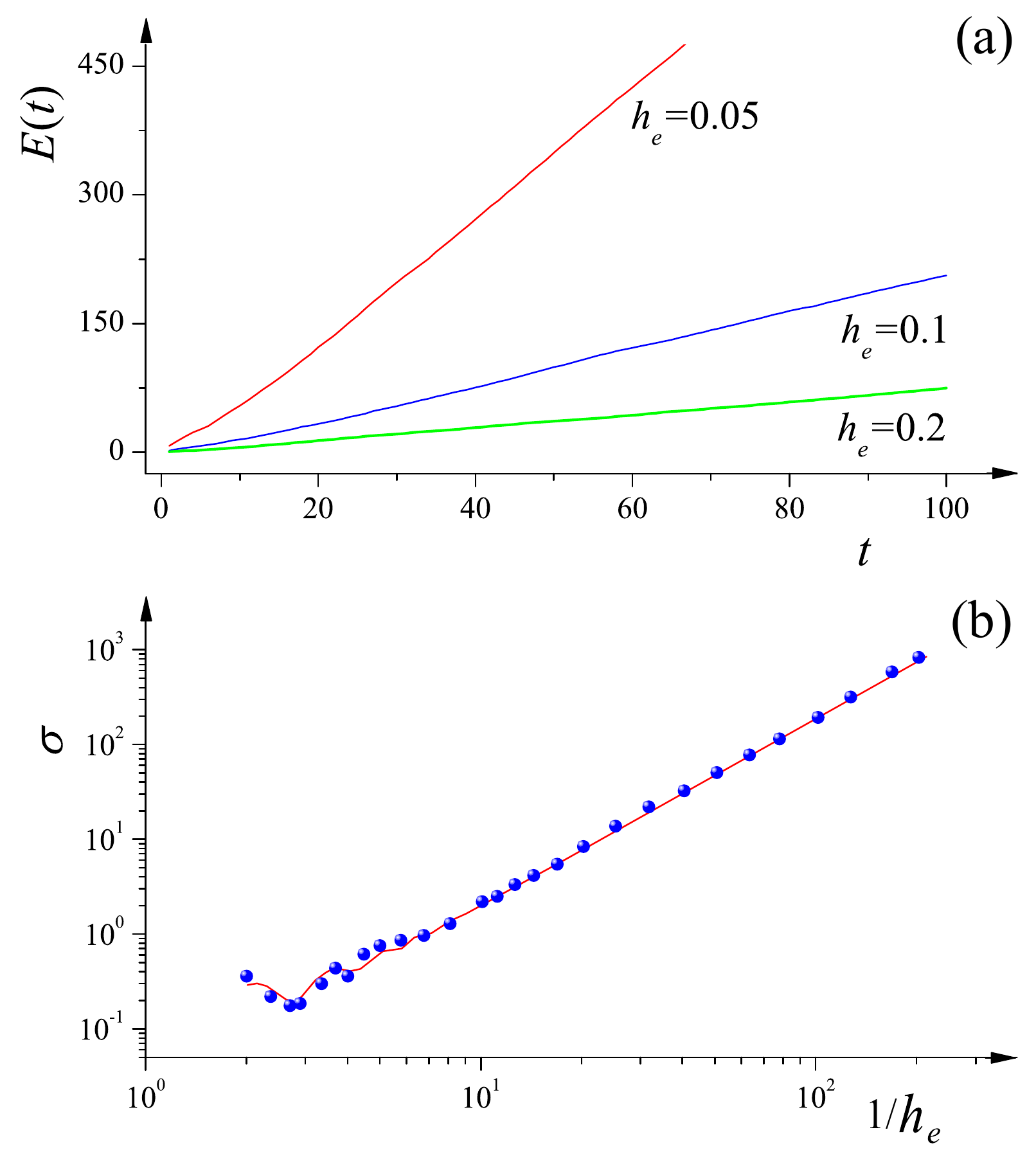}
\caption{(a) Simulations of the quantum evolution (\ref{eq:6}) show that at short times
the rotor's energy grows linearly irrespective of the value of $h_e$. (b) The numerical (solid circles) and analytical (solid line)
results for the rate of this short-time energy growth are in excellent agreement.}
\label{fig:2}
\end{figure}

\section{Numerical confirmation of Planck-IQHE}
\label{sec:numerical_test_incommensurate}

In this section we put analytic results shown in Sec.~\ref{sec:microscopic_theory} into numerical tests.

\subsection{The model and short-time dynamics}
\label{sec:chaoticity}

We first consider a specific form of Eq.~(\ref{eq:6}) with
\begin{eqnarray}
H_0=(-ih_e\partial_{\theta_1})^2
\label{eq:281}
\end{eqnarray}
and
\begin{eqnarray}
V(\Theta)=
\frac{2\arctan Kd}{d} \boldsymbol{d}\cdot\boldsymbol\sigma,
\label{eq:52}
\end{eqnarray}
where
\begin{eqnarray}
\boldsymbol{d}=(\sin\theta_1, \sin\theta_2, \beta
(\mu
-\cos\theta_1-\cos\theta_2))
\label{eq:58}
\end{eqnarray}
and $d=|\boldsymbol{d}|$. This model was introduced in Ref.~\onlinecite{Beenakker11} for a numerical study of $\mu$-driven topological phenomenon, where $h_e$ is fixed to unity.
In contrast to that work, we are interested in the physics sensitive to $h_e$. Thus, we fix all the parameters in $V(\Theta)$, and $h_e$ is the only tuning parameter. To be specific, in most parts of this section (except Sec.~\ref{sec:modification_kicking_potential}) we set
\begin{equation}\label{eq:59}
    K=2, \quad \beta=0.8,\quad \mu=1.
\end{equation}
In addition, throughout this section we set $\tilde \omega=2\pi/\sqrt{5}$ for Eq.~(\ref{eq:6}) without loss of generality \cite{footnote_1}.

We apply the fast Fourier transformation technique to simulate the quantum evolution (\ref{eq:6}) at integer times, i.e., $\tilde \psi_t=(\prod_{s=1}^t {\hat U}'_s) \tilde \psi_0$, where
$\hat U'_s \equiv e^{-h_e{\hat
n}_1^2 }e^{-\frac{i}{h_e}V(\theta_1,\theta_2+{\tilde \omega}s)}$.
The angular momentum ($n_1$) space is
of
$
16384$ sites and the periodic boundary condition is imposed. Note that as far as $\hat U'_s$ is concerned, $\theta_2$ is understood as an external parameter.

\subsubsection{Linear energy growth}
\label{sec:numerical_diffusion}

We first simulate the short-time evolution for a broad range of $h_e$ varying from $5\times 10^{-3}$ to $5\times 10^{-1}$. Figure \ref{fig:2}(a) and (b) are representative simulation results for the energy profile and the energy growth rate, respectively. These results are obtained by averaging over $10^2$ values of $\theta_2$. As shown in Fig.~\ref{fig:2}(a), the rotor's energy grows linearly at short times, irrespective of the value of $h_e$. This is in agreement with the analytic prediction (\ref{eq:17}). The simulation results for the energy growth rate, $E(t)/t$, are shown in Fig.~\ref{fig:2}(b) (solid circles). On the other hand, substituting Eqs.~(\ref{eq:52}) and (\ref{eq:58}) into Eq.~(\ref{eq:238}) we find
\begin{equation}\label{eq:23}
    \epsilon_i=\frac{\cot\varphi}{d}d_i,\quad \varphi= \frac{\arctan
Kd}{h_e}.
\end{equation}
With the substitution of Eq.~(\ref{eq:23}) into Eq.~(\ref{eq:3}) we find the analytic expression of $\sigma$ as a function of $h_e$ for $V$ given above. The results are shown in Fig.~\ref{fig:2}(b) (red line). We see that the analytic and simulation results are in excellent agreement. This confirms the physical meaning of $\sigma$ as the short-time energy growth rate, namely Eq.~(\ref{eq:17}). Figure \ref{fig:2}(b) also confirms the scaling, $\sigma \stackrel{h_e\rightarrow 0}{\sim} h_e^{-2}$, predicted by Eq.~(\ref{eq:291}). Similar to conventional QKR \cite{Izrailev90}, the early linear growth in rotor's energy is a
manifestation of chaotic motion in the angular momentum space.

\begin{figure}[h]
\includegraphics[width=8.6cm]{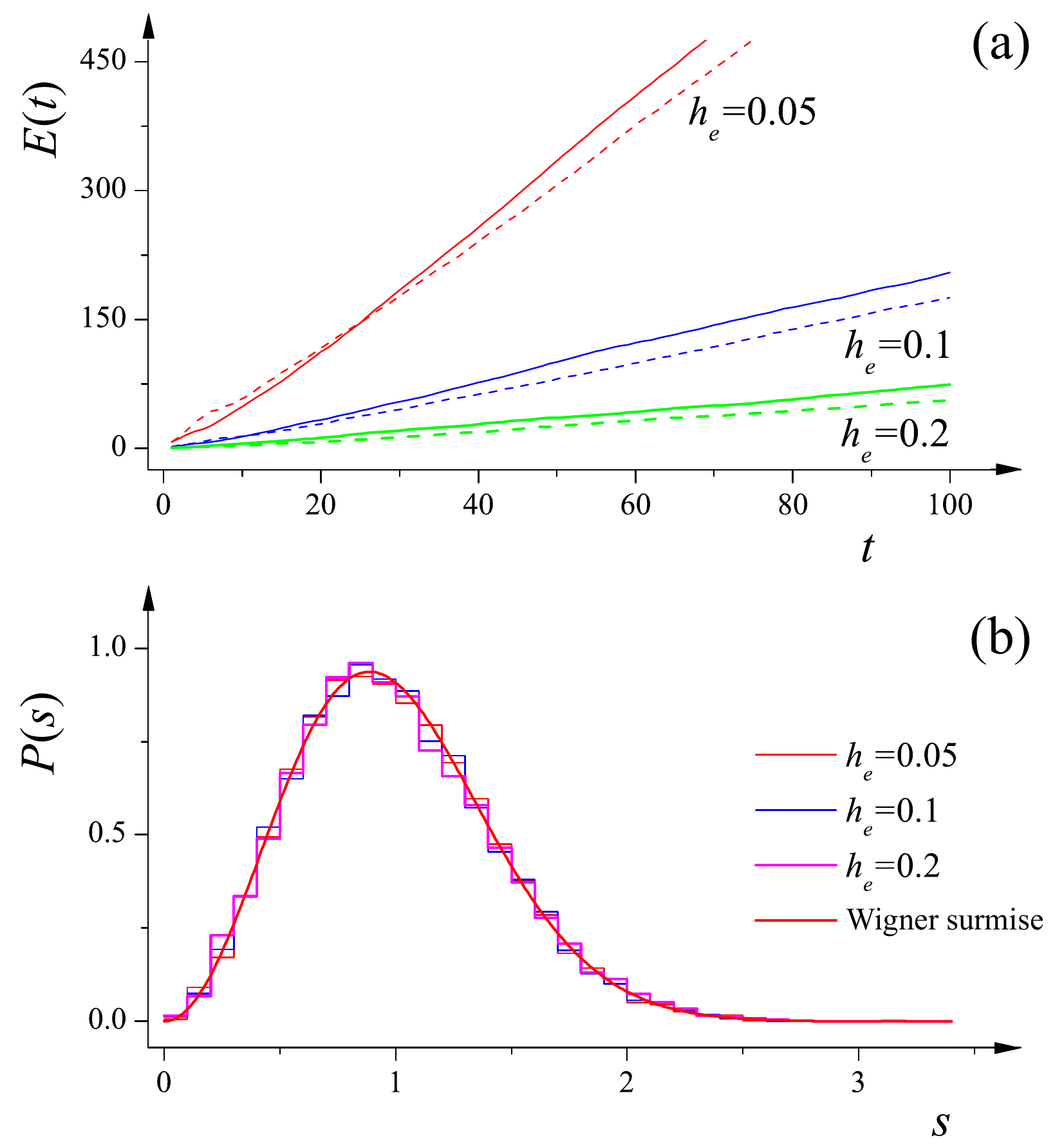}
\caption{Further numerical evidences of strong chaoticity. (a) Simulations of the equivalent $2$D evolution show that the system exhibits short-time diffusion in both $n_1$- and $n_2$-directions, i.e., both $E_1(t)$ (solid lines) and $E_2(t)$ (dashed lines) grow linearly at short times.
(b) The Floquet operator $\hat U$ governing the $2$D evolution has a chaotic quasi-energy spectrum when the angular momentum lattice is small.
The level spacing distribution (histograms) obeying the Wigner surmise of unitary type (solid line). These results are irrespective of the value of $h_e$.
}
\label{fig:7}
\end{figure}

\subsubsection{Further tests of strong chaoticity}
\label{sec:chaoticity}

Note that the analytical derivation of Eq.~(\ref{eq:3}) starts from the equivalent
$2$D motion (\ref{eq:5}) and is based on the fact that this motion is
chaotic in both $n_1$- and $n_2$-directions.
Therefore, the agreement between simulation and analytic results for the
early energy growth rate indicates that such strong chaoticity indeed exists for $V$ given above. To confirm the latter, we directly simulate the equivalent $2$D evolution and compute $E_{1,2}(t)\equiv\frac{1}{2} \langle \psi_t|\hat n_{1,2}^2 |\psi_t\rangle
$. As shown in Fig.~\ref{fig:7}(a), they both grow linearly at early times, reflecting chaotic motion in both $n_1$- and $n_2$-directions. In addition,
we study the quasi-energy spectrum of the Floquet operator, $\hat U$, governing the $2$D evolution
by numerical diagonalization.
The Hilbert space is composed of wave functions
on the angular momentum lattice of $64\times 64$ sites
and subject to periodic boundary conditions. This lattice size is much smaller than the length scale at which interference effects begin to dominate.
The quasi-energy spectrum is found to be chaotic. As shown in Fig.~\ref{fig:7}(b),
the level spacing distribution, $P(s)$, satisfies the Wigner surmise for
the circular unitary ensemble \cite{Haake}.
These results fully confirm that the equivalent $2$D motion is chaotic in all directions of $N$ space. Note that the small deviations between the slopes of $E_1(t)$ and $E_2(t)$ for given $h_e$ result from short-time correlation effects (cf. Appendix \ref{sec:massive}) and reflect the anisotropic nature of the first exponent of $\hat U$ (see the definition given in Eq.~(\ref{eq:5})).

\subsection{Long-time dynamics: Planck-IQHE}
\label{sec:confirmation_IQHE}

\subsubsection{Analytic predictions for critical points}
\label{sec:critical_points}

We turn to the long-time behavior of quantum evolution (\ref{eq:6}). First of all, we apply the developed analytic theory to the specific model with $H_0$ and $V$ given by Eqs.~(\ref{eq:281})-(\ref{eq:58}) and the corresponding parameters by Eq.~(\ref{eq:59}). This enables us to make analytic predictions for the critical points. For this $V$ the parameter $\mu$ in Eq.~(\ref{eq:S4}) can be identified as $\mu$ in Eq.~(\ref{eq:58}), and the lower limit of the corresponding integral set to $+\infty$ (see Appendix~\ref{sec:integral_limit_case_study} for discussions). Taking this into account and substituting Eq.~(\ref{eq:23}) into Eqs.~(\ref{eq:S4}) and (\ref{eq:S3}), we obtain
\begin{eqnarray}
    \sigma_{\rm H}^{I}=
    \frac{4\beta}{Kh_e}
    \int\!\!\!\!\int \frac{d\theta_1}{2\pi}\frac{d\theta_2}{2\pi}
    \!\!\int_1^{+\infty}\!\! d\mu \frac{\cos^2\varphi_{\mu} \cos\theta_1\cos\theta_2}{d^2_{\mu}(d^2_{\mu}+
    K^{-2}
    )}\,\,
\label{eq:S5}
\end{eqnarray}
and
\begin{eqnarray}
    \sigma_{\rm H}^{II} &=&
    4\beta
    \int\!\!\!\!\int \frac{d\theta_1}{2\pi}\frac{d\theta_2}{2\pi}
    \frac{\sin\varphi\cos^3\varphi}{d^3}
    \nonumber\\
    && \times
    (\cos\theta_1\cos\theta_2-\cos\theta_1-\cos\theta_2),
    \label{eq:S1}
\end{eqnarray}
where
\begin{eqnarray}
\label{eq:311}
  d_{\mu}&=&(\sin^2\theta_1+\sin^2\theta_2+\beta^2(\mu-\cos\theta_1-\cos\theta_2)^2)^{1/2},\nonumber\\
  \varphi_{\mu}&=&\frac{\arctan Kd_{\mu}}{h_e},
\end{eqnarray}
with $\varphi_{\mu=1}\equiv \varphi$ and $d_{\mu=1}\equiv d$.
In deriving Eq.~(\ref{eq:S5}) we have used the identity,
\begin{eqnarray}
\label{eq:53}
  &&\partial_\mu d^2 {\boldsymbol d}\cdot \partial_{\theta_1} {\boldsymbol d} \times \partial_{\theta_2} {\boldsymbol d}+
  \partial_{\theta_1} d^2 {\boldsymbol d}\cdot \partial_{\theta_2} {\boldsymbol d} \times \partial_\mu {\boldsymbol d}\nonumber\\
  &&+\partial_{\theta_2} d^2 {\boldsymbol d}\cdot \partial_\mu {\boldsymbol d} \times \partial_{\theta_1} {\boldsymbol d}
  = 2d^2 \beta \cos\theta_1\cos\theta_2
\end{eqnarray}
for $\boldsymbol{d}$ defined in Eq.~(\ref{eq:58}).
Carrying out numerical evaluations of the integrals in these expressions
(see Appendix~\ref{sec:quantum_critical_points} for details), we obtain $\sigma_{\rm H}(h_e)$ as shown in Fig.~\ref{fig:3}(b).
The linear scaling is clearly seen, consistent with Eq.~(\ref{eq:9}). The slight deviation is due to that the value of $h_e$ in Fig.~\ref{fig:3}(b) is not small enough.

According to the result, when $h_e^{-1}$ takes the value of
$0.73$, $2.19$, and $3.60$, the bare Hall conductivity $\sigma_{\rm H}$
is $\frac{1}{2}$, $\frac{3}{2}$, and $\frac{5}{2}$,
respectively. Combining with the RG flow lines shown in Fig.~\ref{fig:3}(a), we predict that there are three topological quantum phase transitions for $h_e \geq 0.2$. To be specific, $h_e^{-1}=0.73$ corresponds to a plateau transition
(cf. blue line in Fig.~\ref{fig:5}) from $\sigma_{\rm H}^*=0$ to $1$,
$h_e^{-1}=2.19$ from $\sigma_{\rm H}^*=1$ to $2$, and
$h_e^{-1}=3.60$ from $\sigma_{\rm H}^*=2$ to $3$.
At these critical points $\tilde \sigma=\sigma^*$.

\subsubsection{Numerical confirmation}
\label{sec:confirmation}

To probe the predicted transitions we simulate the quantum evolution (\ref{eq:6}) up to $t=1.2\times 10^6$.
The value of $h_e$ varies from $0.2$ to $10.0$ (i.e., $0.1\leq h_e^{-1}\leq 5.0$).
For such long times the profile of $E(t)/t$ is found to converge well (cf. Appendix \ref{sec:finite_time_effects} for discussions).
Figure \ref{fig:3}(c) shows the simulation result (black solid line).
We find that $E(t)/t$ exhibits three sharp peaks
at $h_e^{-1}=0.77, 2.13$ and $3.42$, each of which corresponds to a topological transition predicted above. These critical values are in excellent agreement with
analytic predictions. The peak values are nearly the same. They are small ($\approx 0.3$) and closed to the analytic estimation (\ref{eq:260}).
More sophisticated numerical studies of the peak value will be presented in Sec.~\ref{sec:quantum_growth_rate}.
Here we only stress that this smallness of the (asymptotic) energy growth rate confirms the quantum nature of
the rotor metal.
Between these peaks,
$E(t)/t$ is fully suppressed which is a characteristic of rotor insulator. Moreover, the peaks are approximately evenly spaced with a spacing
$\approx 1.33$. This agrees with the analytic prediction and is an evidence of the $\tilde \sigma_{\rm H}$-periodicity of the RG flow lines. Thus, simulations confirm the Planck-IQHE.

\begin{figure}[h!]
\vskip-.21cm \hskip-.5cm
\includegraphics[width=8.6cm]{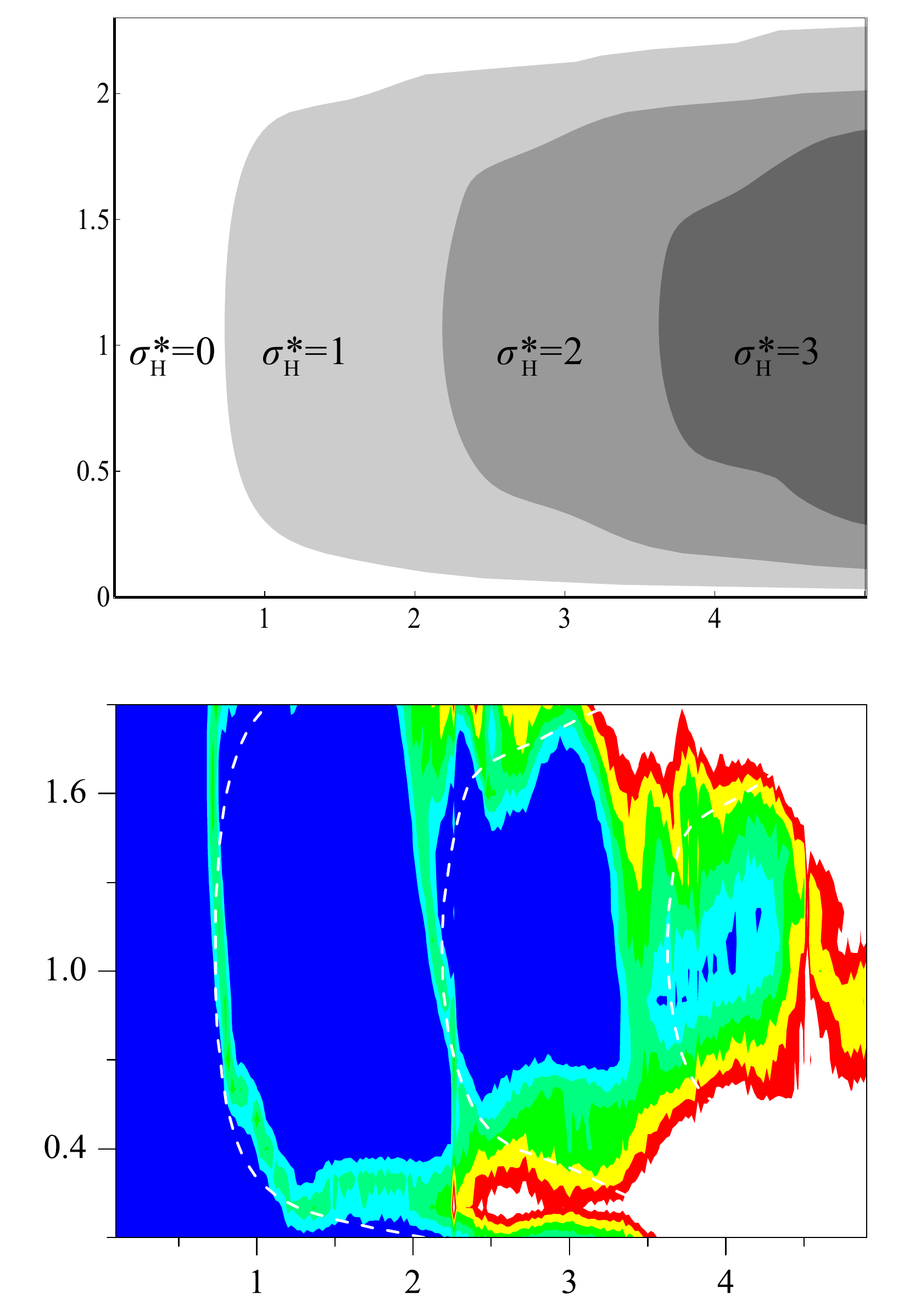}
\vskip-.1cm\hskip-.5cm
\includegraphics[width=8.1cm]{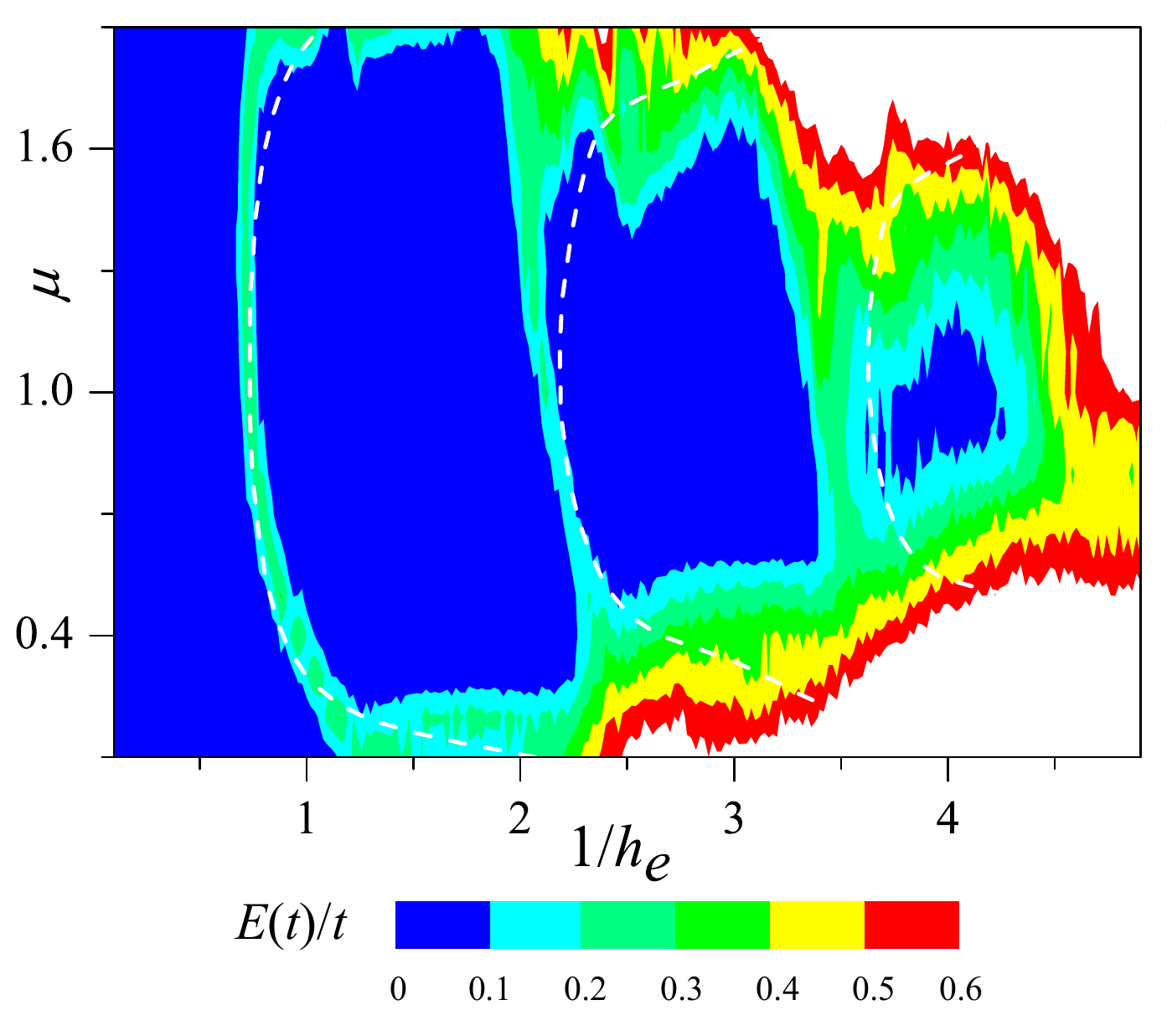}
\vskip-.5cm
\caption{Analytic prediction for the $\mu$-$h_e^{-1}$ phase diagram (top panel). Simulation results of the energy growth rate at $t=10^5$ for $H_0=(-ih_e\partial_{\theta_1})^\alpha$, with $\alpha=2$ (middle panel) and $4$ (lower panel). The dashed lines are the phase boundaries predicted analytically.}
\label{fig:8}
\end{figure}

\subsection{Robustness of Planck-IQHE}
\label{sec:robustness}

A straightforward result implied by the topological meaning of $\sigma_{\rm H}^*$ is the robustness of Planck-IQHE. In this subsection we will test this result numerically. We recall that, like conventional QKR \cite{Fishman84,Altland11,Altland10,Tian04}, for $\hat U$ defined in Eq.~(\ref{eq:5}) $\Theta$ mimics the electron velocity in normal metals and $V(\Theta)$ the free electron Hamiltonian, while the first exponent of $\hat U$, i.e., $e^{-\frac{i}{h_e}(H_0(h_e {\hat n}_1)+h_e{\tilde \omega} {\hat n}_2)}$, plays the role of impurities. Keeping this in mind, below we will modify $H_0(h_e\hat n_1)$ -- the `disordered potential' -- and $V(\Theta)$ -- the `free electron Hamiltonian', respectively and examine the robustness of Planck-IQHE.

\subsubsection{Modifying free rotation Hamiltonian $H_0$}
\label{sec:modification_free_rotation_Hamiltonian}

First of all, we choose another form of free rotation Hamiltonian,
read $H_0=(-ih_e\partial_{\theta_1})^4$,
and keep the kicking potential, $V$, unchanged. (Recall that for this Hamiltonian, exactly speaking, the meaning of $E(t)$ is the variance of angular momentum.)
The ensuing equivalent $2$D system remains strongly chaotic, i.e., the memory of $\Theta$ is quickly lost. Therefore, the results of Sec.~\ref{sec:microscopic_theory} apply. In particular, according to what was discussed in the end of Sec.~\ref{sec:transition}, the critical values of $h_e^{-1}$ are not shifted under such modification. We simulate this modified system, with the procedures being exactly the same as before. The results are shown by the red dashed line in Fig.~\ref{fig:3}(c). It is clear that the critical points are not shifted. Besides, the peak values are only slightly changed.

\subsubsection{Modifying kicking potential $V$}
\label{sec:modification_kicking_potential}

Next, we modify the kicking potential, $V$, while fixing the free rotation Hamiltonian to $H_0=(-ih_e\partial_{\theta_1})^\alpha$. To be specific, we tune the parameter $\mu$ in Eq.~(\ref{eq:58}) and study how Planck-IQHE evolves with this parameter.

Analytically, for each {\it fixed} $\mu$ we apply the developed effective field theory to such modified model. In this way we obtain a phase structure where the $h_e^{-1}$-axis is divided into an infinite number of intervals. Each interval represents an insulting phase characterized by a quantum number $\sigma_{\rm H}^*$. Upon tuning $\mu$ this phase structure is turned into a $\mu$-$h_e^{-1}$ phase diagram (Fig.~\ref{fig:8}, upper panel). The phase diagram is composed of an infinite number of shells, each of which corresponds to an insulating phase characterized by $\sigma_{\rm H}^*$. For the $n$th shell $\sigma_{\rm H}^*=n$, with $n=0,1,2,\cdots$ counted from the outermost shell. The boundary between the shells of $\sigma_{\rm H}^*=n$ and $\sigma_{\rm H}^*=n+1$ is determined by the solution to Eq.~(\ref{eq:60}), with its left-hand side given by
\begin{eqnarray}
    &&\sigma_{\rm H}(
    h_e) \label{eq:61}\\
    &=&4\beta
    \int\!\!\!\!\int \frac{d\theta_1}{2\pi}\frac{d\theta_2}{2\pi}
    \bigg(\int_\mu^{+\infty}\!\! d\mu \frac{\cos^2\varphi_{\mu} \cos\theta_1\cos\theta_2}{Kh_e d^2_{\mu}(d^2_{\mu}+
    K^{-2}
    )}\nonumber\\
    &+&\frac{\sin\varphi_{\mu}\cos^3\varphi_{\mu}}{d^3_\mu}
    (\mu\cos\theta_1\cos\theta_2-\cos\theta_1-\cos\theta_2)\bigg).\nonumber
\end{eqnarray}
As before the parameters $K=2$ and $\beta=0.8$. The shells (insulating phases) correspond to
a vanishing (asymptotic) energy growth rate, i.e., $E(t)/t\stackrel{t\rightarrow \infty}{\longrightarrow} 0$, while the boundaries between two shells (metallic phases) to an energy growth rate $E(t)/t\stackrel{t\rightarrow \infty}{\longrightarrow} \sigma^*$.

\begin{figure}[h]
\includegraphics[width=8.6cm]{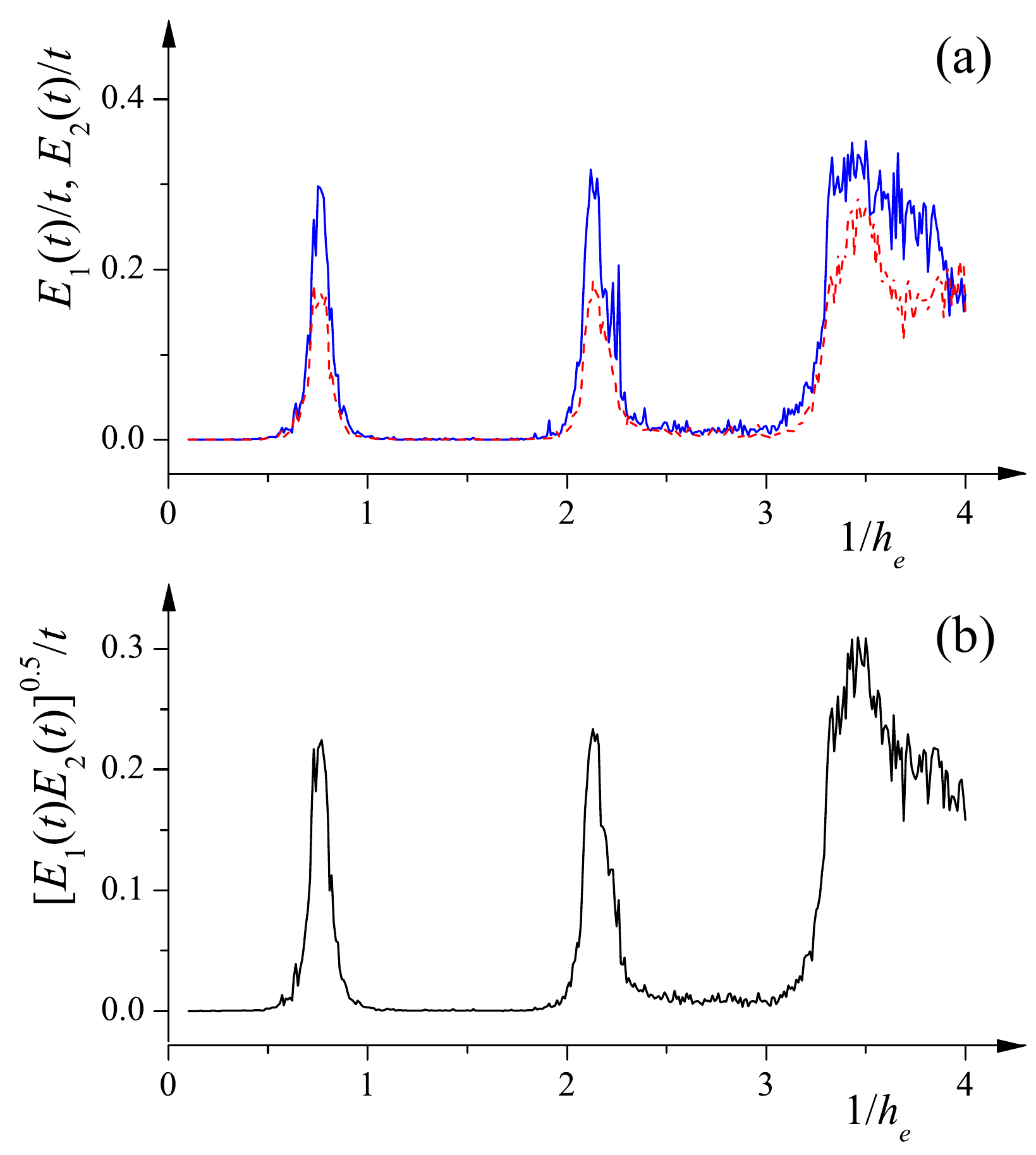}
\caption{(a) Simulations of $2$D dynamics show that the long-time diffusion rates of rotor metal in $n_1$- (blue solid line) and $n_2$-directions (red dashed line) are different, indicating the anisotropicity of the $2$D model. For this model $H_0=(-ih_e\partial_{\theta_1})^2$ and $V(\Theta)$ is given by Eqs.~(\ref{eq:52})-(\ref{eq:59}). (b) The geometric mean of the two diffusion rates. The results are for $t=10^4$.}
\label{fig:9}
\end{figure}

As done in Sec.~\ref{sec:confirmation_IQHE}, we simulate the long-time quantum evolution and obtain the energy profile for different values of $\mu$ and $h_e^{-1}$. The simulation results for $E(t)/t, t=10^5$ are shown in the middle (for $\alpha=2$) and (for $\alpha=4$) panels of Fig.~\ref{fig:8}. Four insulating phases (dark blue regimes) corresponding to $\sigma_{\rm H}^*=0,1,2,3$ (compared with upper panel) are clearly seen. Three narrow phase boundaries are also clearly seen. Their locations are in good agreement with analytic predictions (dashed lines). We see that these boundaries correspond to a small energy growth rate and thereby to quantum metallic phases. Note that the phase boundaries at large $h_e^{-1}$ are smeared because for such values of $h_e^{-1}$ finite-time effects are strong \cite{note_special_case}.

\begin{figure}[h]
\includegraphics[width=8.6cm]{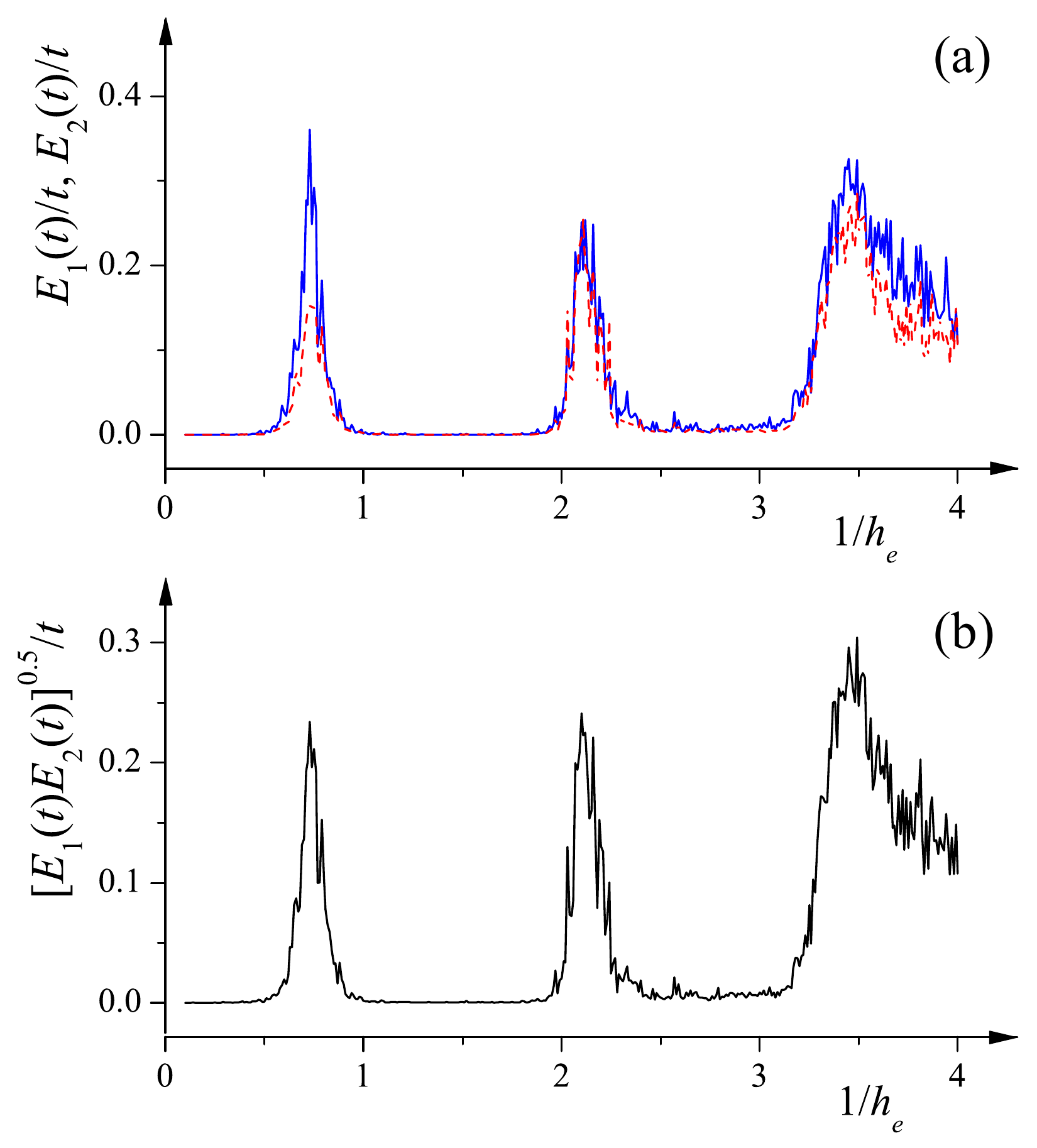}
\caption{The same as Fig.~\ref{fig:9} but $H_0=(-ih_e\partial_{\theta_1})^4$.}
\label{fig:10}
\end{figure}

\subsection{Simulation results for $\sigma^*$}
\label{sec:quantum_growth_rate}

From simulation results above for various models, we have seen that the peak values are consistently closed to each other ($\approx 0.3$). However, they are obtained by $1$D simulations, and do not take into account effects of residual anisotropicity of the equivalent $2$D model. The latter modifies the numerical value of $\sigma^*$. On the other hand, such anisotropicity generally exists for realistic models. Indeed, it has already been seen in simulation results of the short-time energy growth rate (Fig.~\ref{fig:7}(a)). So, to confirm fully the universality of the long-time energy growth rate at critical points we simulate the equivalent $2$D model as well.

Figures \ref{fig:9} and \ref{fig:10} show the simulation results for two $2$D models, whose $1$D equivalent lead to the result represented by the black solid and red dashed lines in Fig.~\ref{fig:3} (c), respectively.  From panel (a) of Figs.~\ref{fig:9} and \ref{fig:10}, we find that the quantum diffusion rates in $n_1$- and $n_2$-directions are different, i.e., $\frac{E_1(t)}{t}\neq \frac{E_2 (t)}{t}$. This indicates the anisotropicity of long-time $2$D dynamics. In these $2$D simulations, the quantum diffusion rates in different directions exhibit peaks at the same critical $h_e$-values. Moreover, these critical values are identical to those obtained from $1$D simulations. Note that because of computer limitation we can simulate the $2$D dynamics only up to $t=10^4$. This time is much shorter than that in $1$D simulations and causes relatively large finite-time effects, especially the broadening of the third peak in Figs.~\ref{fig:9} and \ref{fig:10}.

Because of the anisotropicity the effective quantum diffusion rate is the geometric mean of the long-time quantum diffusion rates in $n_1$- and $n_2$-directions (cf. Eq.~(\ref{eq:283})). The simulation results of this effective diffusion rate are shown in Figs.~\ref{fig:9} (b) and \ref{fig:10} (b) and the corresponding peak values, i.e., $\sigma^*$, given in Table \ref{Table3}. The values listed are all closed to $0.25$, as one expects based on the common belief of the critical universal conductivity in conventional IQHE. (Recall that the longitudinal conductivity in the present work, defined based on the effective field theory, differs from that in conventional IQHE by a factor of $2$.) They are in good agreement with the data for conventional IQHE systems \cite{Weng98,Bhatt93,Bhatt95,Wei86,Pruisken06}. Note that because finite-time effects are enhanced by increasing $h_e^{-1}$, the third peak in Figs.~\ref{fig:9}(b) and \ref{fig:10}(b) gives relatively large values of $\sigma^*$.

\begin{table}[htbp]
\caption{\label{Table3} Simulation results for $\sigma^*$ taking anisotropicity of $2$D dynamics into account.}
\begin{tabular}{c|c|c|c}
  \hline\hline
  $H_0=(-ih_e\partial_{\theta_1})^\alpha$
   & $1^{\rm st}$ peak
   & $2^{\rm nd}$ peak
   & $3^{\rm rd}$ peak \\
   \hline
  $\alpha=2$ & $0.22$ & $0.23$ & $0.30$ \\
  \hline
  $\alpha=4$ & $0.23$ & $0.24$ & $0.30$ \\
  \hline\hline
\end{tabular}
\end{table}

\section{Disappearance of Planck-IQHE at rational $\tilde \omega/(2\pi)$}
\label{sec:absence_IQHE}

The effective field theory (\ref{eq:111}) holds only for $\tilde \omega$ incommensurate with $2\pi$. Indeed, provided this condition is violated, i.e., $\tilde \omega=2\pi p/q$ with $p,q\in \mathbb{N}$ co-prime, the Floquet operator $\hat U$ governing the $2$D evolution is invariant under the translation, $n_2\rightarrow n_2+q$; this translation symmetry is not respected by the effective field theory (\ref{eq:111}). In this section we analytically show and numerically confirm that for rational $\tilde \omega/(2\pi)$ the Planck-IQHE does not occur and, instead, the rotor's energy saturates at long times irrespective of the value of $h_e$, i.e., the system is always an insulator as conventional QKR \cite{Fishman84,QKR79}.

\subsection{A case study: $q=1$}
\label{sec:case_study}

To understand better the situation we first consider the simplest case of $q=1$. (The qualitative behavior of the system is not affected by $p$.) In this case the Floquet
operator defined in Eq.~(\ref{eq:5}) is simplified to $
\hat U=e^{-\frac{i}{h_e}H_0(h_e {\hat n}_1)}e^{-\frac{i}{h_e}V(\theta_1,\theta_2)}\equiv \hat U_{\theta_2}$.
Since the first exponent does not depend on $\hat n_2$, i.e., the system is translationally invariant in $n_2$-direction with a (spatial) period of unity, the ensuing $2$D autonomous stroboscopic dynamics conserves the `velocity' component $\theta_2$. That is, $\theta_2$ is a good quantum number which implies that the dynamics exhibits partial regularity.
The subscript of $\hat U_{\theta_2}$ is the bookkeeping of this conservation law.
Because of this conservation law the
$2$D dynamics is reduced to a family of $1$D
dynamics each of which is controlled by this good quantum
number. Such reduced $1$D system is exactly same as the original $1$D system (\ref{eq:6}).

From the general expression (\ref{eq:7}) we find
\begin{equation}\label{eq:30}
    E(t)=\frac{1}{2}
    \int\frac{d\omega}{2\pi}e^{-i\omega t}\langle{\rm Tr} (\hat n_1^2 K_{\omega,\theta_2} {\psi}_0\otimes {\psi}_0^\dagger)\rangle_{\theta_2},
\end{equation}
where
\begin{eqnarray}
&&K_{\omega,\theta_2}(n_1s_+s_-,n'_1s'_+s'_-)\nonumber\\
&\equiv& \langle\!\langle n_1s_+|\frac{1}{1-e^{i\omega_+}{\hat U}_{\theta_2}}| n_1's'_+\rangle\nonumber\\
&\times& \langle n_1's'_-|\frac{1}{1-e^{-i\omega_-}{\hat U}^\dagger_{\theta_2}}|n_1s_-\rangle\!\rangle_{\omega_0}
\label{eq:35}
\end{eqnarray}
describes interference between the advanced and retarded quantum amplitudes corresponding to the reduced $1$D dynamics governed by ${\hat U}_{\theta_2}$.

\subsubsection{Effective field theory}
\label{sec:theory_commensurate}

We follow the procedures described in Sec.~\ref{sec:field_theory} to calculate Eq.~(\ref{eq:35}). Thanks to the $1$D nature of the reduced dynamics, the results turn out to be totally different. Specifically, we find
\begin{eqnarray}
    K_{\omega,\theta_2}(n_1s_+s_-,n_1's'_+s'_-)\qquad\qquad\qquad\qquad\label{eq:S20}\\
    =-\frac{1}{4}\delta_{s_+s_-}\delta_{s'_+s'_-}
    \int\!\! D(Q)e^{-S
    }
Q(n_1)_{+b,-b}Q(n_1')_{-b,+b},
\nonumber
\end{eqnarray}
where $Q$ depends only on $n_1$ and the effective action is
\begin{eqnarray}\label{eq:S21}
    S[Q]={1\over 4} \mathrm{Str} \left(-\sigma
    (\nabla_1 Q)^2-2i\omega Q\Lambda\right).
\end{eqnarray}
The (bare) conductivity
is
\begin{eqnarray}
\sigma
=-\frac{1}{4}{\rm Tr}\left(({\cal G}^0_R-{\cal G}^0_A)
\partial_{\theta_1}\!\epsilon
({\cal G}^0_R-{\cal G}^0_A)
\partial_{\theta_1}\!\epsilon\right).
\label{eq:S22}
\end{eqnarray}
Recall that the trace `Tr' includes both the spin and the angular ($\theta_1$) indices of the theory.
The Green functions ${\cal G}^0_{R,A}$ as well as $\epsilon$ are formally defined in the same way as Eq.~(\ref{eq:276}), but with the conserved velocity component $\theta_2$ understood as an external parameter. As a result, $\sigma$ (and thereby $S$) are $\theta_2$-dependent.
Most importantly, compared to the action (\ref{eq:111}) there is no topological term in Eq.~(\ref{eq:S22}), which can be attributed to the trivial homotopy group (\ref{eq:31}).

The effective field theory (\ref{eq:S21}) describes Anderson localization
of quasi $1$D disordered systems of unitary symmetry \cite{Efetov97}.
With the help of Eqs.~(\ref{eq:S20}) and (\ref{eq:S21}) it can be shown that
Eq.~(\ref{eq:106}) is still valid. Physically, this validity is a result of the diffusive pole which reflects the particle conservation law and is irrespective of the presence or absence of topological term in the effective field theory. Furthermore, from Eqs.~(\ref{eq:S20}) and (\ref{eq:S21}) we find
\begin{equation}\label{eq:37}
    \sigma(\omega)=-\zeta(3)i\omega \xi^2
\end{equation}
for $\omega\ll \sigma/\xi^2$,
where $\zeta(x)$ is the Riemann $\zeta$-function. Note that the localization length $\xi=4\sigma$ parametrically depends on $\theta_2$. With the substitution of Eq.~(\ref{eq:37}) into Eq.~(\ref{eq:106}) we obtain
\begin{eqnarray}
E(t)\stackrel{t\rightarrow \infty}{\sim} \langle\xi^2\rangle_{\theta_2}
\sim \langle\sigma^2\rangle_{\theta_2}.
\label{eq:S26}
\end{eqnarray}
This implies that irrespective of the value of $h_e$ the energy saturates at long times, i.e., the Planck-IQHE is washed out.

\subsubsection{Numerical confirmation}
\label{sec:numerical_test_commensurate}

We put these analytic predictions on numerical tests. The model for simulations is the same as that used in Sec.~\ref{sec:numerical_test_incommensurate} described by Eqs.~(\ref{eq:281})-(\ref{eq:59}) except that the value of $\tilde \omega$ is changed. Simulations show that irrespective of the value of $h_e$ the rotor's energy saturates after a transient growth,
in full agreement with the analytic prediction of Eq.~(\ref{eq:S26}). The solid lines in Fig.~\ref{fig:S1}(a) are representative results for $\tilde \omega=2\pi$. These energy profiles are qualitatively the same as those of insulating phases for irrational $\tilde \omega/(2\pi)$. The saturation value increases with $h_e^{-1}$, which will be discussed in details in Sec.~\ref{sec:general_cases}. These profiles are in sharp contrast to those of critical metallic growth at irrational $\tilde \omega/(2\pi)$, for which a linear energy growth persists in the entire course of time. The dashed lines in Fig.~\ref{fig:S1}(a) are representative examples which correspond to the critical values of $h_e^{-1}=0.77$, $ 2.13$, and $3.42$, respectively and $\tilde \omega=2\pi/\sqrt{5}$.

\subsubsection{Mechanism}
\label{sec:absence_IQHE_discussions}

\begin{figure}[h]
\includegraphics[width=8.6cm]{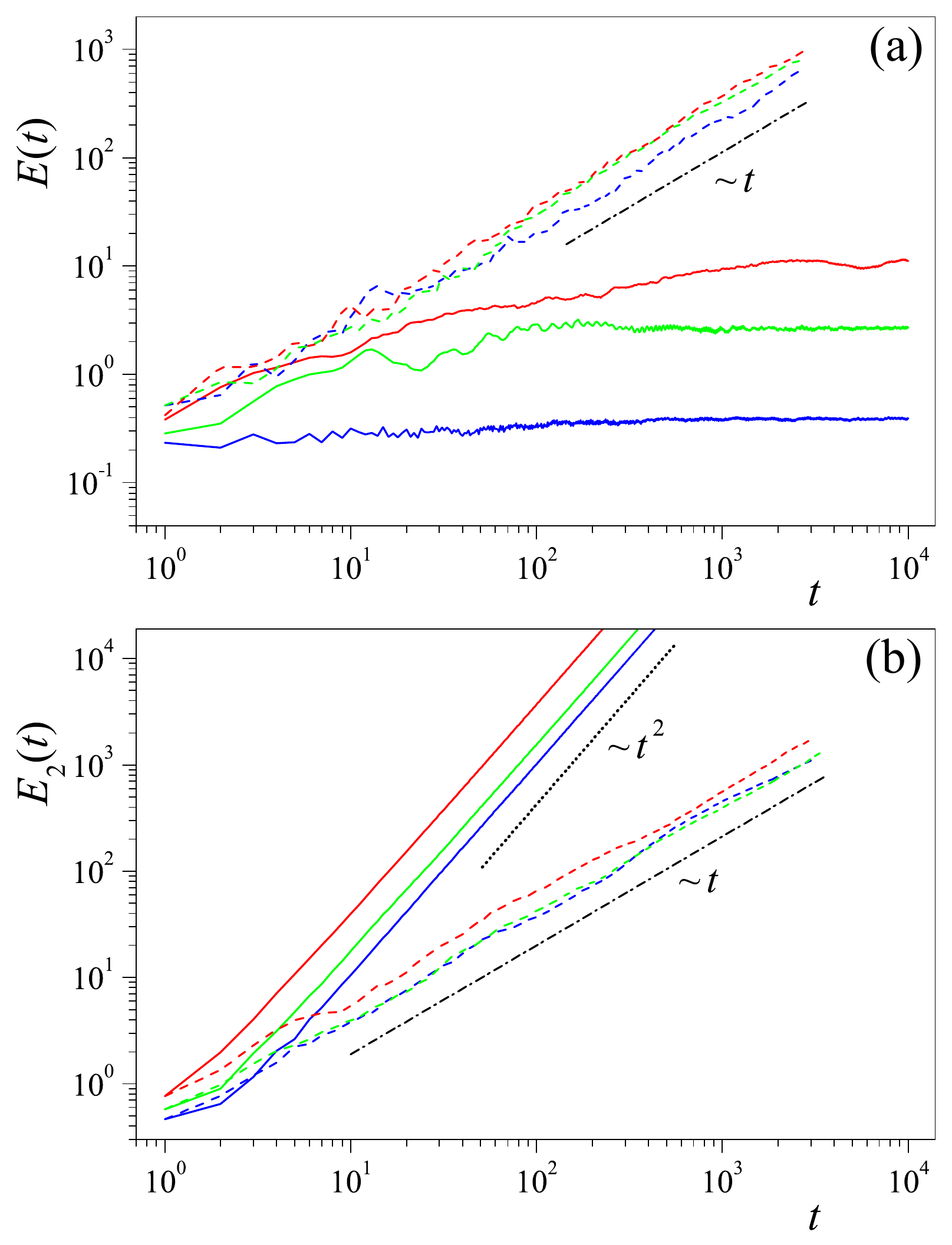}
\caption{(a) For $\tilde \omega=2\pi$ the rotor's energy saturates following a transient growth, irrespective of the value of $h_e$. The representative simulation results are shown by the solid lines with $h_e^{-1}=0.77$ (blue), $ 2.13$ (green), and $3.42$ (red), respectively. These energy profiles are in sharp contrast to those represented by the dashed lines for which the values of $h_e$ are the same but $\tilde \omega=\frac{2\pi}{\sqrt{5}}$. The latter exhibits linear growth in the entire course of time since the system is at the
critical metallic phase. (b) The motion in $n_2$-direction is ballistic for $\tilde \omega=2\pi$ (solid) while diffusive for $\tilde \omega=\frac{2\pi}{\sqrt{5}}$ (dashed). Two lines with the same color correspond to the same value of $h_e$.}
\label{fig:S1}
\end{figure}

To investigate the mechanism we simulate the equivalent $2$D evolution and compute the profile of $E_2(t)$. The simulation results show that
it grows quadratically after a transient process (Fig.~\ref{fig:S1}(b), solid lines). This implies a ballistic motion in $n_2$-direction. Therefore, the memory of the velocity component $\theta_2$ is never lost. In other words, the regular motion in $n_2$-direction is restored at $\tilde \omega=2\pi$. In sharp contrast, at $\tilde \omega=2\pi/\sqrt{5}$
chaoticity leads to quick loss of the memory of $\theta_2$, manifesting in that $E_2(t)$ grows linearly in short times irrespective of the value of $h_e$ (Fig.~\ref{fig:S1}(b), dashed lines). These results are in agreement with the aforementioned analysis that for Planck-IQHE to occur it is necessary that the motion is chaotic in both $n_1$- and $n_2$-directions.

\subsection{General $q$}
\label{sec:general_cases}

For general $q$ the $1$D system (\ref{eq:6}) is {\it time-periodic} with a period of $q$. For sufficiently short times this periodicity does not play any roles and the system behaves essentially the same as the one with $\tilde \omega/(2\pi)$ being irrational ($\approx p/q$). In particular, provided $q$ is sufficiently large we expect to observe certain signatures of Planck-IQHE in intermediate time scales. Then, similar to the $q=1$ case, at longer times the time periodicity completely changes the system's behavior. To study this change it is sufficient to consider the $1$D evolution (\ref{eq:6}) at times of multiple $q$. This evolution is described by
\begin{equation}\label{eq:39}
    \tilde \psi_{\tilde t}={\hat U}'^{\tilde t} \tilde \psi_0,\quad \tilde \psi_{\tilde t}\equiv \tilde \psi_{\tilde tq},
\end{equation}
with $\tilde t\in \mathbb{Z}$ and
\begin{equation}\label{eq:38}
    {\hat U}'=\prod_{s=1}^q \hat U'_{s}
\end{equation}
being the effective Floquet operator. Note that this operator parametrically depends on $\theta_2$.
By using the definition (\ref{eq:26}) we obtain
\begin{equation}\label{eq:41}
    E(\tilde tq)=\frac{1}{2}\int\frac{d\omega}{2\pi}e^{-i\omega \tilde t}\langle{\rm Tr} (\hat n_1^2 K_{\omega}' {\tilde \psi}_0\otimes {\tilde \psi}_0^\dagger)\rangle_{\theta_2}
\end{equation}
for times of multiple $q$, where
\begin{eqnarray}
&&K_{\omega}'(n_1s_+s_-,n'_1s'_+s'_-) \nonumber\\
&\equiv&\langle\!\langle n_1s_+|\frac{1}{1-e^{i\omega_+}{\hat U}'}| n_1's'_+\rangle\nonumber\\
&\times&\langle n_1's'_-|\frac{1}{1-e^{-i\omega_-}{\hat U}'^\dagger}|n_1s_-\rangle\!\rangle_{\omega_0}.
\label{eq:40}
\end{eqnarray}
As before, Eqs.~(\ref{eq:41}) and (\ref{eq:40}) lay down a foundation for field-theoretic treatments of the energy profile.

\subsubsection{Effective field theory}
\label{sec:theory_commensurate_general}

Repeating the procedures of deriving Eqs.~(\ref{eq:8})-(\ref{eq:12}), we can rewrite Eq.~(\ref{eq:40}) as
\begin{eqnarray}
&&K_{\omega}'(n_1s_+s_-,n'_1s'_+s'_-) = \int D(Z,\tilde Z)e^{-S}\label{eq:42}\\
&&((1-Z \tilde Z)^{-1}Z)_{n_1s_+b,n_1s_-b}((1-\tilde Z Z)^{-1}\tilde Z)_{n'_1s'_-b,n'_1s'_+b},
\nonumber
\end{eqnarray}
with the action
\begin{eqnarray}
  \label{eq:43}
  S = -\mathrm{Str}\ln(1- Z \tilde Z)+\mathrm{Str}\ln(1-e^{i\omega} {\hat U}' Z
  {\hat U}'^\dagger \tilde Z).\quad
\end{eqnarray}
Similar to discussions in Sec.~\ref{sec:origin_IQHE}, $Z$ describes the coherent propagation of
the advanced and retarded quantum amplitudes corresponding to the $1$D time-periodic evolution (\ref{eq:39}). The only but crucial difference is that this motion takes place in $n_1$ instead of $N$ space. Similar to situations discussed before, the off-diagonal components of $Z$ in this space, $Z_{n_1,n_1'}$ ($n_1\neq n_1'$), carry the information on the relaxation of the velocity of the coherent propagation. Because of the chaoticity of the evolution (\ref{eq:39}) the memory on the velocity
is lost quickly. As before, this eliminates the off-diagonality of $Z$ in $n_1$ space,
yielding $Z_{n_1,n'_1}\propto \delta_{n_1n'_1}Z_{n_1,n_1}$. Then, the homotopy group for the mappings from the $n_1$
into
target space is the same as (\ref{eq:31}). This implies the absence of a topological term in the effective field theory.

\begin{figure}
\centering
  \centerline{\includegraphics[width=8.6cm]{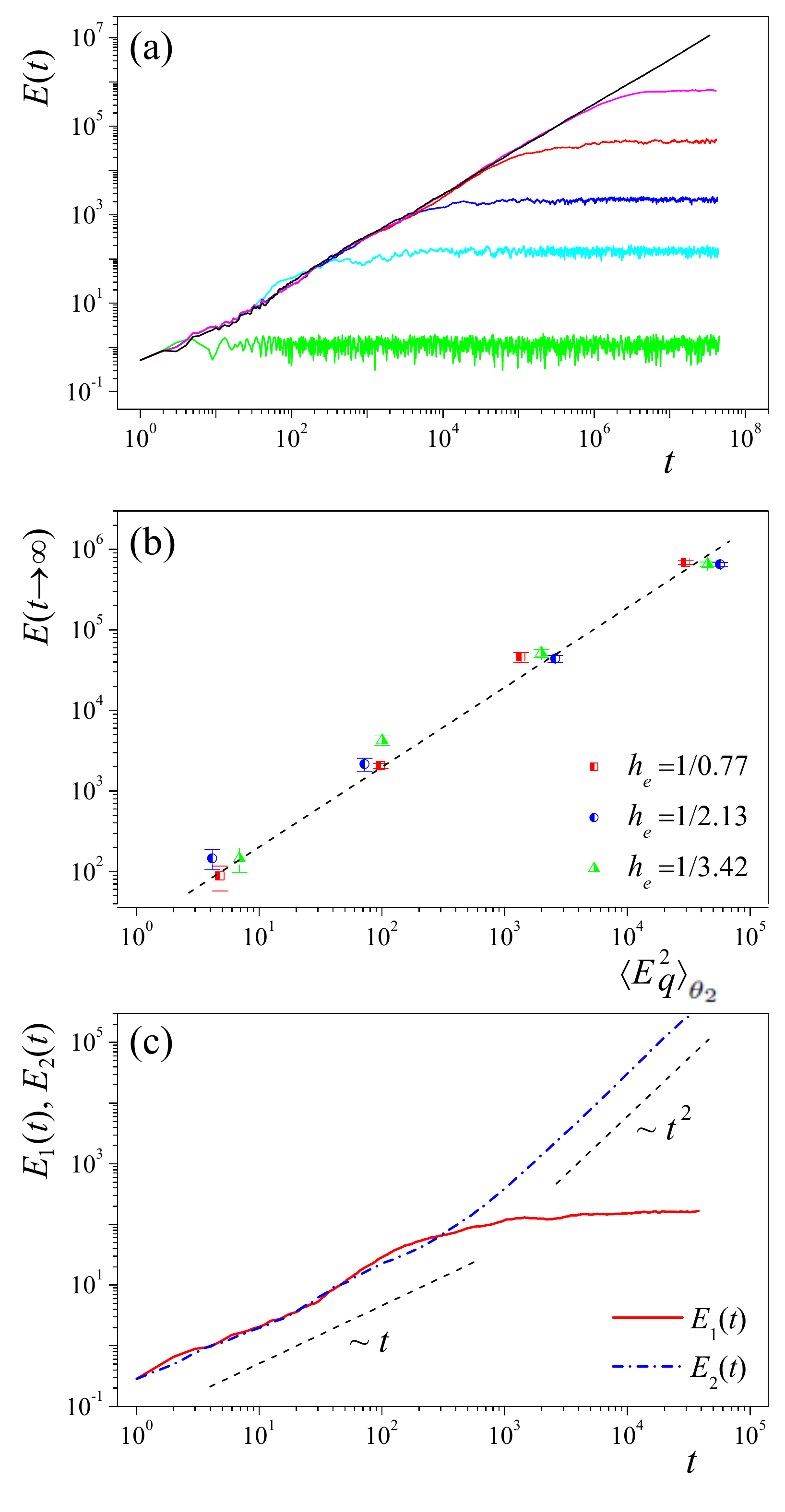}}
\caption{(a) The metallic growth at the topological transition
for $\frac{\tilde \omega}{2\pi}=\frac{1}{\sqrt{5}}$ (black)
is replaced by a crossover from a linear energy growth
to saturation, when $\frac{1}{\sqrt{5}}$ is approximated by a rational number $p/q$. This number is set to $1/2$ (green), $4/9$ (light blue), $17/38$ (dark blue), $72/161$ (red), and $ 305/682$ (pink) from bottom to top. Here $h_e^{-1}=2.13$.
(b) Simulation results confirm the scaling (\ref{eq:46}) for the saturation value. For each value of $h_e^{-1}$ we choose different value of $p/q$. From left to right, the four clusters of data points are for $p/q=4/9, 17/38, 72/161$, and $305/682$, respectively.
(c) Simulations of the equivalent $2$D systems show that the saturation of
$E_1(t)$ at large times (red solid), is associated with the restoration of regular dynamics in $n_2$-direction manifesting in an
asymptotic ballistic growth of $E_2(t)$ (blue dashed). Here $h_e^{-1}=2.13$ and $p/q=4/9$.}
\label{fig:6}
\end{figure}

To find the effective action explicitly we first note that, as before, in the action (\ref{eq:43}) the $\boldsymbol{Z}$-components are massive and thereby negligible. Moreover, at $\omega=0$ such action must bear the same symmetry as (\ref{eq:13}), with $Q(n_1)$ and $T(n_1)$ defined in the same way as Eq.~(\ref{eq:301}). Then, it is more convenient to follow the procedures described in Appendix~\ref{sec:massive} to derive the effective action, since the topological term is now absent. Specifically, as $Q(n_1)$ varies smoothly in $n_1$, we can rotate $Q(n_1)$ back to $\Lambda$ and perform the $Z$-expansion around it. The quadratic action thereby obtained has a rotationally invariant generalization which describes fluctuations around any $Q(n_1)$ (see Ref.~\onlinecite{Altland10} for details). The result is
\begin{eqnarray}\label{eq:44}
    S|_{\omega=0} = -\frac{E_q}{4} {\rm Str} (\nabla_1 Q)^2,
\end{eqnarray}
where $E_q$ characterizes the energy growth within a single period, i.e., from time $t=1$ to $q$, for fixed parameter $\theta_2$. Adding the frequency term to the zero-frequency action (\ref{eq:44}) gives the effective action as
\begin{eqnarray}\label{eq:45}
    S
    ={1\over 4} \mathrm{Str} \left(-E_q
    (\nabla_1 Q)^2-2i\omega Q\Lambda\right).
\end{eqnarray}
Combining Eqs.~(\ref{eq:42}) and (\ref{eq:45}) we find
\begin{eqnarray}
E(\tilde tq)\stackrel{\tilde t\rightarrow \infty}{\sim}\langle E_q^2\rangle_{\theta_2},
\label{eq:46}
\end{eqnarray}
implying the absence of Planck-IQHE. In the special case of $q=1$ Eq.~(\ref{eq:46}) is reduced to Eq.~(\ref{eq:S26}).

\subsubsection{Numerical confirmation}
\label{sec:numerical_test_commensurate_general}

We put these analytic predictions on numerical tests and consider different values of $q$. For small $q$ (e.g., $q=3$) we find that the system behaves essentially in the same way as that at $q=1$. That is, irrespective of the value of $h_e$ the rotor's energy saturates following a transient growth.
For large $q$ (with $p/q\approx 1/\sqrt{5}$) we observe that the peaks in Fig.~\ref{fig:5} show up in intermediate times, but are suppressed at longer times and eventually disappear. Therefore, the system is insulating irrespective of the value of $h_e$. In particular, the quantum phase transitions shown in Fig.~\ref{fig:3}(c) are replaced by a crossover from linear energy growth to saturation. Figure \ref{fig:6}(a) shows some representative energy profiles with $h_e^{-1}=2.13$, which is the critical value for the central peak in Fig.~\ref{fig:3}(c). Simulations also confirm the scaling (\ref{eq:46}) of the saturation value, as shown in Fig.~\ref{fig:6}(b). Therefore, the absence of Planck-IQHE for rational $\tilde \omega/(2\pi)$ is confirmed.

\subsubsection{Mechanism}
\label{sec:mechanism}

To investigate the mechanism we also carry out simulations of the equivalent $2$D system. Specifically, we
compute $E_{1,2}(t)$ and the results for $h_e^{-1}=2.13$ and $p/q=4/9$ are shown in Fig.~\ref{fig:6} (c). We see that
following a linear growth -- which is a remnant
of the metallic phase at the topological transition
for irrational $\tilde \omega/(2\pi)=1/\sqrt{5}$ -- $E_1(t)$ saturates while, simultaneously, $E_2(t)$
crosses over to a ballistic growth. Therefore, similar to
the special case of $q=1$, the partial restoration of
regular dynamics corrupts the metallic phase and thereby washes out the Planck-IQHE. Interestingly, the simultaneous occurrence of localization (namely energy saturation) in one direction and ballistic motion in the other has been observed in a spinless quasiperiodic QKR \cite{Altland11}.

\section{Conclusions}
\label{sec:conclusion}

Summarizing, we have shown analytically and confirmed numerically a dynamical phenomenon, namely, the Planck-IQHE, in a large class of spin-$\frac{1}{2}$ quasiperiodic QKR. The phenomenon is driven by the Planck's quantum, and topological in nature. Strikingly, it is found to emerge from strong chaos.
The phenomenon bears a firm analogy to conventional IQHE, occurring in $2$D electronic systems, such as MOSFET, which are fundamentally different from QKR. Specifically, the inverse Planck's quantum and the asymptotic energy growth rate of the rotor mimic the filling fraction and the longitudinal conductivity, respectively.
Moreover, the rotor insulating phase, for which the rotor's energy saturates at long times,
is characterized by a hidden quantum number $\sigma_{\rm H}^*$; this number mimics the quantized Hall conductivity in conventional IQHE.
For the first time, we find that a topological theta angle emerges from strongly chaotic motion at microscopic scales. The renormalization of this topological angle gives the hidden quantum number. On the other hand, when the dynamics is (partially) regular, the topological theta angle does not show up and therefore the Planck-IQHE does not occur. This is also confirmed by numerical simulations.

The system considered here differs from conventional quasiperiodic QKR in having a spin degree of freedom. This and the angular degrees of freedom are coupled to each other upon kicking. The roles of this coupling are two-fold. First, it is the origin of strong chaos, from which the topological theta angle emerges. Second, it gives rise to a universal linear scaling law ($\sim h_e^{-1}$) for the unrenormalized theta angle for large $h_e^{-1}$. As a result, the renormalized theta angle has a quantization spectrum $\mathbb{Z}$, and as $h_e^{-1}$ increases infinitely many insulating phases result, with $\sigma_{\rm H}^*$ increasing one by one. In other words, the renormalization of the linear scaling law leads to the universal stair-like pattern in Planck-IQHE.


Because the system here is non-integrable, the Planck-IQHE is beyond the canonical TKNN paradigm, where the topological invariant is expressed as
an integral over conserved quantities (e.g., Bloch momentum).
Therefore, it is different from
topological phenomena found in other driven systems
(see, e.g., Refs.~\onlinecite{Arovas90,Galitski11,Gong12,Gong15}).
Those phenomena fall into the TKNN paradigm;
and the quantum phase transition there is associated
with the change in the topological structure of (effective) Floquet bands that
are formed due to the presence of translation symmetry and(or) adiabatic parameter cycles.

In the present work we do not study details of quantum criticality, which require substantially more efforts, both analytically and numerically. The physical implication of the hidden quantum number $\sigma_{\rm H}^*$ remains largely unexplored. Another prominent issue is the exact role of spin. Does the Planck-IQHE exist in higher spin cases? These issues are currently under investigation.

Our findings suggest that rich quantum topological phenomena can emerge from chaos. Many interesting questions are thereby opened. First of all, in this work we have focused on spinful quasiperiodic QKR with a single modulation frequency. Our preliminary investigation has shown that when there are more modulation frequencies, chaos can trigger the even more interesting Planck's quantum-driven topological transitions. Second, the interplay between nonlinearity and Anderson-like localization is currently under intense investigations \cite{Delande14}. To the best of our knowledge, so far no attention has been paid to potential effects of topology. A natural question is: how does the nonlinearity affect the Planck-IQHE? Third, we have seen that chaos triggers the IQHE analog, even when a magnetic field and fermi statistics are both absent. This motivates us to explore analogs of the fractional quantum Hall effect in chaotic systems. Finally, it seems very promising to confirm the Planck-IQHE by cold-atom experiments on quasiperiodic QKR \cite{Deland08}, previously used to confirm Anderson transition.

\section*{Acknowledgements}

We would like to thank B. L. Altshuler, J. B. Gong, I. Guarneri
and Y. S. Wu for stimulating discussions, and to
G. Casati and
S. Fishman
for useful conservations.
This work is supported by the NSFC (Nos. 11174174, 11275159, 11335006 and 11535011).

\appendix

\section{Effects of short-time correlation}
\label{sec:massive}

In this appendix we will present an alternative derivation of fluctuation action given by Eqs.~(\ref{eq:S35}) and (\ref{eq:S34}), following Ref.~\onlinecite{Altland10}. Most importantly, this derivation helps us to understand better effects of short-time correlation encoded in the massive modes.

Thanks to the gauge invariance (\ref{eq:13}), it is sufficient to first analyze the fluctuations near the reference point $Z=\tilde Z=0$ and derive the corresponding zero-frequency action. Then, the rotationally invariant generalization of the latter gives the full action. To facilitate this procedure we suppose that in the parametrization (\ref{eq:36})
the angular momentum-dependent fields $Z,\tilde Z$ are small. As a result, the zero-frequency action is simplified to
\begin{eqnarray}
  \label{eq:80}
  S[Z,\tilde Z]|_{\omega=0}=\mathrm{Str}\,\left(\tilde Z (1-{\rm Ad}_{\hat U}) Z\right).
\end{eqnarray}
Recall that $Z,\tilde Z$ are $2\times 2$ supermatrices and the supertrace includes the spin index.

Next, we introduce the slow and fast mode decomposition for $Z,\tilde Z$,
\begin{eqnarray}
    Z=B+C,\quad B\equiv (1-\hat \pi) Z,\, C \equiv \hat \pi Z,
    \label{eq:81}
\end{eqnarray}
\begin{eqnarray}
    \tilde Z=\tilde B+\tilde C,\quad \tilde B\equiv (1-\hat \pi) \tilde Z,\, \tilde C \equiv \hat \pi \tilde Z,\label{eq:81}
\end{eqnarray}
where $\hat\pi$ is the fast mode projector and we refer to
Ref.~\onlinecite{Altland10} for its exact definition. Here $B,\tilde B$ are the slow modes and $C,\tilde C$ the fast modes.
Inserting this decomposition into Eq.~(\ref{eq:80}), we obtain
\begin{eqnarray}
\label{eq:71}
  S[B,C]|_{\omega=0}
  &=& {\rm Str}\left((\tilde B+\tilde C)(1-{\rm Ad}_{\hat U})(B+C)\right)\nonumber\\
  &=& {\rm Str}\big(\tilde B (1-{\rm Ad}_{\hat U})B +
  \tilde C (1-\hat \pi {\rm Ad}_{\hat U})C  \nonumber\\
  &&-\tilde C \hat \pi\, {\rm Ad}_{\hat U}B-\hat \pi\, {\rm Ad}_{{\hat U}^\dagger}\tilde B C\big).
\end{eqnarray}
The first (second) term in the second equality shows that the $B$ ($C$) field is massless (massive). In the presence of strong chaos at microscopic scales, the $C$ field describes short-time velocity ($\Theta$) correlation and its effects diminish rapidly. As we will show below, these effects feed back into the effective action via introducing short-time correlation corrections to the (bare) control parameters of the action.

Substituting Eq.~(\ref{eq:71}) into the functional integral and performing the Gaussian integral of the fast modes, we obtain
an effective fluctuation action of the slow mode $B$,
\begin{eqnarray}
  S[B,\tilde B]|_{\omega=0} = S^0[B,\tilde B]+S^m[B,\tilde B],
\label{eq:294}
\end{eqnarray}
where
\begin{eqnarray}
  S^0[B,\tilde B] =
  {\rm Str}\left(\tilde B (1-{\rm Ad}_{\hat U_0})B\right),
  \label{eq:73}
\end{eqnarray}
and
\begin{eqnarray}
  S^m[B,\tilde B] &=& -{\rm Str} \left(\tilde B {\rm Ad}_{\hat U}
  \frac{\hat\pi\, {\rm Ad}_{\hat U}}{1-\hat\pi\,{\rm Ad}_{\hat U}}B\right)\nonumber\\
  &=&-\sum_{k=1}^\infty{\rm Str} \left(\tilde B {\rm Ad}_{\hat U}
  (\hat\pi\, {\rm Ad}_{\hat U})^kB\right),
  \label{eq:74}
\end{eqnarray}
with $\hat U_0=e^{-\frac{i}{h_e}V(\Theta)}$. This result is formally the same as that for conventional QKR \cite{Altland10} but the details, namely the operators $\hat U$ and $\hat U_0$, not. Most importantly, (i) the leading contribution, $S^0[B,\tilde B]$,
results solely from the slow modes and describes
that the system loses memory on $\Theta$ after each kicking
which accounts for the reduction, $\hat U\rightarrow \hat U_0$, in Eq.~(\ref{eq:73}); (ii) the fast modes introduce corrections to the action $S^0[B,\tilde B]$
and as we will see below, they renormalize the bare longitudinal conductivity; and (iii) these renormalization corrections
account for short-time memory effects and
arise from the expansion in the second equality of Eq.~(\ref{eq:74}).

Consider the action (\ref{eq:73}). We introduce the Fourier transformations:
$B_{N}=\int\!\!\!\!\int\frac{d^2{\boldsymbol{\phi}}}{(2\pi)^2}e^{iN\boldsymbol{\phi}}B(\boldsymbol{\phi})$ and
$\tilde B_{N}=\int\!\!\!\!\int\frac{d^2\boldsymbol{\phi}}{(2\pi)^2}e^{iN\boldsymbol{\phi}}\tilde B(\boldsymbol{\phi})$ and
substitute it into the action. Since the slow modes are composed of $B(\boldsymbol{\phi})$ with very small
$\boldsymbol{\phi}$, we can expand the ensuing action in $\boldsymbol{\phi}$. Keeping the expansion up to the second
order, we obtain
\begin{widetext}
\begin{eqnarray}\label{eq:77}
    S[B,\tilde B]|_{\omega=0}&\approx& \frac{1}{2}\int\!\!\!\!\int\!\!\frac{d^2\boldsymbol{\phi}}{(2\pi)^2}\left({\rm Tr} \left(\partial_{\theta_1}{\hat {U}}_0^\dagger
   \partial_{\theta_1}{\hat {U}}_0\right)\phi^2_1+
    {\rm Tr}\left(\partial_{\theta_2}{\hat {U}}_0^\dagger
    \partial_{\theta_2}{\hat {U}}_0\right)\phi^2_2\right)
    {\rm str}[B(\boldsymbol{\phi})\tilde B(-\boldsymbol{\phi})]\nonumber\\
    &=& \frac{1}{2}\left({\rm Tr} \left(\partial_{\theta_1}{\hat {U}}_0^\dagger
    \partial_{\theta_1}{\hat {U}}_0\right)
    {\rm Str}(\nabla_1 \tilde B \nabla_1 B) +
    {\rm Tr}\left(\partial_{\theta_2}{\hat {U}}_0^\dagger
    \partial_{\theta_2}{\hat {U}}_0\right)
    {\rm Str}(\nabla_2 \tilde B \nabla_2 B)\right).
\end{eqnarray}
This action was obtained by the quadratic expansion around the reference configuration
$Q=\Lambda$. Its rotationally invariant generalization is
\begin{eqnarray}\label{eq:78}
    S[B,\tilde B]|_{\omega=0}\rightarrow S_1[Q]= -\frac{1}{16}
    \left({\rm Tr} \left(\partial_{\theta_1}{\hat {U}}_0^\dagger
    \partial_{\theta_1}{\hat {U}}_0\right) {\rm Str} (\nabla_1 Q)^2+
    {\rm Tr}\left(\partial_{\theta_2}{\hat {U}}_0^\dagger
    \partial_{\theta_2}{\hat {U}}_0\right) {\rm Str} (\nabla_2 Q)^2\right).
\end{eqnarray}
\end{widetext}
By using Eq.~(\ref{eq:S34}) it can be shown that $\sigma=\frac{1}{4}{\rm Tr} \left(\partial_{\theta_{1,2}}{\hat {U}}_0^\dagger
    \partial_{\theta_{1,2}}{\hat {U}}_0\right)$ after straightforward calculations.
Therefore, Eq.~(\ref{eq:78}) is identical to Eq.~(\ref{eq:S35}).

From the above derivation we see that the isotropicity of the longitudinal conductivity
arises from the isotropic nature of $V(\Theta)$ and the ignorance of the short-time memory effects.
In order to take short-time memory effects into account we need to
compute the corrections (\ref{eq:74}) to the action, which must be of the general form,
$\int\!\!\!\!\int \frac{d^2{\boldsymbol{\phi}}}{(2\pi)^2}{\rm str}
(\tilde B(\boldsymbol{\phi})(\delta \sigma_{11}\phi_1^2+\delta \sigma_{22}\phi_2^2)B(\boldsymbol{\phi}))$.
This gives rise to the renormalization of
longitudinal conductivity (or inverse coupling constant),
\begin{equation}\label{eq:83}
    \sigma\rightarrow \sigma_{1}=\sigma+\delta \sigma_{11},\quad
    \sigma\rightarrow \sigma_{2}=\sigma+\delta \sigma_{22}.
\end{equation}
Note that this renormalization arises from short-time memory effects and does not contain any localization physics, which is a long-time effect.
Because the operator $\hat {U}$ is anisotropic, the values of $\sigma_{1}$ and $\sigma_{2}$ are different, giving
\begin{eqnarray}\label{eq:84}
    S_1[Q]\rightarrow -\frac{1}{4}
    \left(\sigma_{1} {\rm Str} (\nabla_1 Q)^2+
    \sigma_{2}{\rm Str} (\nabla_2 Q)^2\right)
\end{eqnarray}
Therefore, the effective field theory is anisotropic when the
short-time memory effects are taken into account.

With the rescaling,
\begin{equation}\label{eq:282}
    (n_1,n_2)\rightarrow (\sqrt{\sigma_2/\sigma_1}n_1,n_2)\equiv (n_1',n_2'),
\end{equation}
the action (\ref{eq:84}) becomes isotropic again,
\begin{eqnarray}\label{eq:283}
    S_1[Q]\rightarrow -\frac{\sqrt{\sigma_{1}\sigma_{2}}}{4}
    \left({\rm Str} (\nabla'_1 Q)^2+
    {\rm Str} (\nabla'_2 Q)^2\right),
\end{eqnarray}
with $\sqrt{\sigma_{1}\sigma_{2}}$ being the effective (inverse) coupling constant. $\nabla_{1,2}'$ stand for the gradient with respect to the rescaled angular momenta $n_{1,2}'$ and the supertrace includes these rescaled angular momenta.

\section{Derivation of Eq.~(\ref{eq:S30})}
\label{sec:derivation_1}

Keeping the hydrodynamic expansion of
the commutator, $[\epsilon
, T^{-1}]$, up to the second order, we obtain
\begin{eqnarray}
[\epsilon, T^{-1}]\approx
\partial_{{\theta_\alpha} } \epsilon
[{\theta}_\alpha, T^{-1}] -\frac{1}{2} \partial^2_{\theta_\alpha \theta_\beta} \epsilon
[{\theta}_\alpha,[{\theta}_\beta, T^{-1}]].
\label{eq:27}
\end{eqnarray}
As a result,
\begin{eqnarray}
&&T\epsilon T^{-1}=\epsilon +T[\epsilon,T^{-1}]\nonumber\\
&\approx&\epsilon+T\partial_{\theta_\alpha}\epsilon[{\theta}_\alpha,T^{-1}]-\frac{1}{2}
T\partial^2_{\theta_\alpha \theta_\beta}\epsilon[{\theta}_\alpha,[{\theta}_\beta,T^{-1}]]\nonumber\\
&=&\epsilon+\partial_{\theta_\alpha}\epsilon T[{\theta}_\alpha,T^{-1}]
-[\partial_{\theta_\alpha}\epsilon,T][{\theta}_\alpha,
T^{-1}]\nonumber\\
&&-\frac{1}{2}T\partial^2_{\theta_\alpha\theta_\beta}
\epsilon[{\theta}_\alpha,[{\theta}_\beta,T^{-1}]].
\label{eq:29}
\end{eqnarray}
For the commutator $[\partial_{\theta_1}\epsilon,T]$, we find
\begin{equation}\label{eq:28}
    [\partial_{\theta_1}\epsilon,T]\approx
    \partial^2_{\theta_1}\epsilon [\theta_1,T]+
    \partial^2_{\theta_1 \theta_2}\epsilon [\theta_2,T]
\end{equation}
and likewise for the commutator $[\partial_{\theta_2}\epsilon,T]$.
Substituting them into Eq.~(\ref{eq:29}) and
using the identity:
\begin{eqnarray}\label{eq:277}
    &&T[ \theta_\alpha,[\theta_\beta,T^{-1}]]\nonumber\\
    &=&-[ \theta_\alpha,T][\theta_\beta,T^{-1}]+
[ \theta_\alpha,T[\theta_\beta,T^{-1}]],
\end{eqnarray}
we obtain Eq.~(\ref{eq:S30}).

\section{Boundary-induced deformation of $\hat U$}
\label{sec:boundary_condition}

When we derived the topological action, we discussed that as a boundary effect $\hat U$ is deformed in $N$ space, which gives partial topological action namely Eq.~(\ref{eq:S68}). In this appendix we make further discussions on this deformation.

\subsection{Integrable limit of deformation}
\label{sec:integrable_limit_general_discussions}

The kicking potential can be expressed in a general form,
\begin{eqnarray}
V(\Theta)=
\frac{f(d)}{d} {{\boldsymbol d}}\cdot\boldsymbol{\sigma}, \quad {\boldsymbol d}\equiv \{d_i(\Theta)\}.
\label{eq:263}
\end{eqnarray}
Here $f(d)$ and $d_i(\Theta)$ are generic continuous functions. In addition, $d_i(\Theta)$ satisfies the same symmetry as $V_i(\Theta)$ (Table \ref{Table2}) and is periodic in $\Theta$ also. An example is given by Eqs.~(\ref{eq:52}) and (\ref{eq:58}).
Thanks to the periodicity,
$\{d_i(\Theta)\}$ generates a mapping from the torus $T^2=S^1\times S^1$ onto a closed oriented surface in $\mathbb{R}^3$, denoted as ${\cal S}$.
This surface is composed of several pieces, $\{{\cal S}_\gamma\}$, with the same orientation as ${\cal S}$ and $\forall \gamma\neq \gamma':\, {\cal S}_\gamma\cap {\cal S}_{\gamma'}=\Gamma$, where $\Gamma$ is their common boundary line.

Then, the deformation of $\hat U$ in $N$ space consists of two steps. First, we let ${\boldsymbol d}$ acquire a smooth parameter ($\mu$) dependence,
\begin{equation}\label{eq:309}
    \boldsymbol{d}(\Theta)\rightarrow \boldsymbol{d}(\Theta;\mu),
\end{equation}
such that the deformed surfaces, ${\cal S}_\gamma \rightarrow {\cal S}_\gamma(\mu)$, do not intersect and $\boldsymbol{d}(\Theta;\mu)$ has the same symmetry and periodicity properties as $\boldsymbol{d}(\Theta)$. Second, we let $\mu$ smoothly vary in $N$. Since the particle cannot escape from the boundary, at the boundary $\hat U$ is deformed into an integrable operator diagonal in $N$ index. Thanks to
\begin{eqnarray}\label{eq:264}
    \hat U&=&e^{-\frac{i}{h_e}\left(H_0(h_e {\hat n}_1)+h_e{\tilde \omega} {\hat n}_2\right)}\nonumber\\
    &&\times\left(\cos\frac{f(d)}{h_e}-i\sin \frac{f(d)}{h_e}\frac{\boldsymbol{d}\cdot\boldsymbol{\sigma}}{d}\right),
\end{eqnarray}
the diagonality implies
\begin{eqnarray}\label{eq:304}
    f(d)=const.,\,\, \frac{d_i}{d}=const..
\end{eqnarray}
Because $d_1|_{\theta_1\rightarrow 0}=d_2|_{\theta_2\rightarrow 0}=0$ arising from the symmetry and continuity properties of $d_{1,2}$, the conditions (\ref{eq:304}) can be met only if
\begin{eqnarray}
d=const.,\,\, \frac{d_{1,2}}{d}=0,\,\, \frac{d_{3}}{d}=\pm 1.
\label{eq:265}
\end{eqnarray}
Equivalently (without loss of generality we take the positive sign for the last equation above.),
\begin{eqnarray}
\forall \Theta:\quad \frac{d_{1,2}}{d_3}=0,\quad d_{3}=+\infty.
\label{eq:266}
\end{eqnarray}
At this limit, an integrable deformation results,
\begin{eqnarray}\label{eq:270}
&&\hat U \rightarrow e^{-\frac{i}{h_e}\left(H_0(h_e {\hat n}_1)+h_e{\tilde \omega} {\hat n}_2\right)}\nonumber\\
&&\qquad \times\left(\cos\frac{const.}{h_e}-i\sin \frac{const.}{h_e}\sigma^3\right),
\end{eqnarray}
with the constant given by $f(d\rightarrow \infty)$. Note that this operator is diagonal in spin index also.

Due to the condition (\ref{eq:266}), ${\cal S}_{\Gamma}\equiv\cup_\mu \Gamma(\mu)$ with $\Gamma(\mu)$ being deformed $\Gamma$, has an infinite extension in positive $d_3$-direction. Moreover, each family of deformed surfaces $\{{\cal S}_\gamma(\mu)\}$ generate a manifold ${\mathscr{D}_\gamma}
$. For $\gamma\neq \gamma'$, $\partial{\mathscr{D}_\gamma}\cap \partial{\mathscr{D}_{\gamma'}}={\cal S}_\Gamma$.

\subsection{Invariant form of $\sigma_{\rm H}^I$}
\label{sec:uniqueness_Hall_conductivity}

For Eq.~(\ref{eq:S4}) namely Eq.~(\ref{eq:280}) the deformation (\ref{eq:309}) is implied. Because for given ${\mathscr{D}_\gamma}$ the mapping,
\begin{equation}\label{eq:310}
    \boldsymbol{d}(\Theta;\mu):\, (\theta_1,\theta_2,\mu)
    \rightarrow (d_1,d_2,d_3),
\end{equation}
is bijective, we have
\begin{eqnarray}\label{eq:306}
    &&\varepsilon^{ijk} d\theta_1d\theta_2d\mu \partial_\mu\epsilon_i\partial_{\theta_1} \epsilon_j
\partial_{\theta_2} \epsilon_k\nonumber\\
&=&(-1)^{s_\gamma}\varepsilon^{ijk} d(d_1)d(d_2)d(d_3)\partial_{d_1} \epsilon_i\partial_{d_2} \epsilon_j\partial_{d_3}\epsilon_k,
\end{eqnarray}
where $s_\gamma$ is the sign of the $d_3$-component of the normal direction of ${\cal S}_\gamma$. Substituting it into Eq.~(\ref{eq:S4}) gives
\begin{eqnarray}
\sigma_{\rm H}^{I}
=\frac{\varepsilon^{ijk}}{\pi^2} \sum_{\gamma}
\int\!\!\!\!\int\!\!\!\!\int_{\mathscr{D}_\gamma} d(d_1)d(d_2)d(d_3)\frac{\partial_{d_1} \epsilon_i\partial_{d_2} \epsilon_j\partial_{d_3}\epsilon_k}{(\epsilon^2+1)^2}.\nonumber\\
\label{eq:308}
\end{eqnarray}
Because ${\cal S}$ is closed, the numbers of positive and negative $s_\gamma$ must be the same. Taking this into account, we can rewrite Eq.~(\ref{eq:308}) as
\begin{eqnarray}
\sigma_{\rm H}^{I}
=\frac{\varepsilon^{ijk}}{\pi^2} \int\!\!\!\!\int\!\!\!\!\int_{\mathscr{D}} d(d_1)d(d_2)d(d_3)\frac{\partial_{d_1} \epsilon_i\partial_{d_2} \epsilon_j\partial_{d_3}\epsilon_k}{(\epsilon^2+1)^2},
\label{eq:269}
\end{eqnarray}
where $\mathscr{D}$ is the volume enclosed by the closed oriented surface ${\cal S}$. Equation (\ref{eq:269}) trades Eq.~(\ref{eq:S4}) namely Eq.~(\ref{eq:280}) for a $\boldsymbol{d}$-integral within the volume $\mathscr{D}$, which is invariant under the change of deformation. This shows that $\sigma_{\rm H}^I$ is an intrinsic quantity and its value is unique.

\subsection{Applications}
\label{sec:integral_limit_case_study}

When the surface ${\cal S}$ has a complicated geometry, in general, it is difficult to use Eq.~(\ref{eq:269}) to calculate $\sigma_{\rm H}^{I}$. In this case, the invariance of Eq.~(\ref{eq:269}) allows us to calculate $\sigma_{\rm H}^{I}$ by an alternative method. That is, we find a simple deformation and perform the ensuing $\mu$-integral in Eq.~(\ref{eq:S4}). This method is particularly useful when the deformation can be easily found. For example, consider the kicking potential given by
Eqs.~(\ref{eq:52})-(\ref{eq:59}). A `natural' deformation is to identify $\mu$ defined in Eq.~(\ref{eq:58}) as the deformation
parameter $\mu$, and let it vary from $1$ -- as specified in Eq.~(\ref{eq:59}) -- to the boundary value $+\infty$ at which $d_3=+\infty$. Accordingly, we deform $d_3$, while let $d_{1,2}$ be unchanged, i.e.,
\begin{eqnarray}\label{eq:307}
&&(\sin\theta_1,\sin\theta_2, 0.8(1-\cos\theta_1-\cos\theta_2))\nonumber\\
&\rightarrow&(\sin\theta_1,\sin\theta_2, 0.8(\mu-\cos\theta_1-\cos\theta_2)).
\end{eqnarray}
In this case, the constant in the integrable operator (\ref{eq:270}) is $\pi$. Substituting this deformation into Eq.~(\ref{eq:S4}) we obtain Eq.~(\ref{eq:S5}).

\begin{figure}[h]
\includegraphics[width=8.6cm]{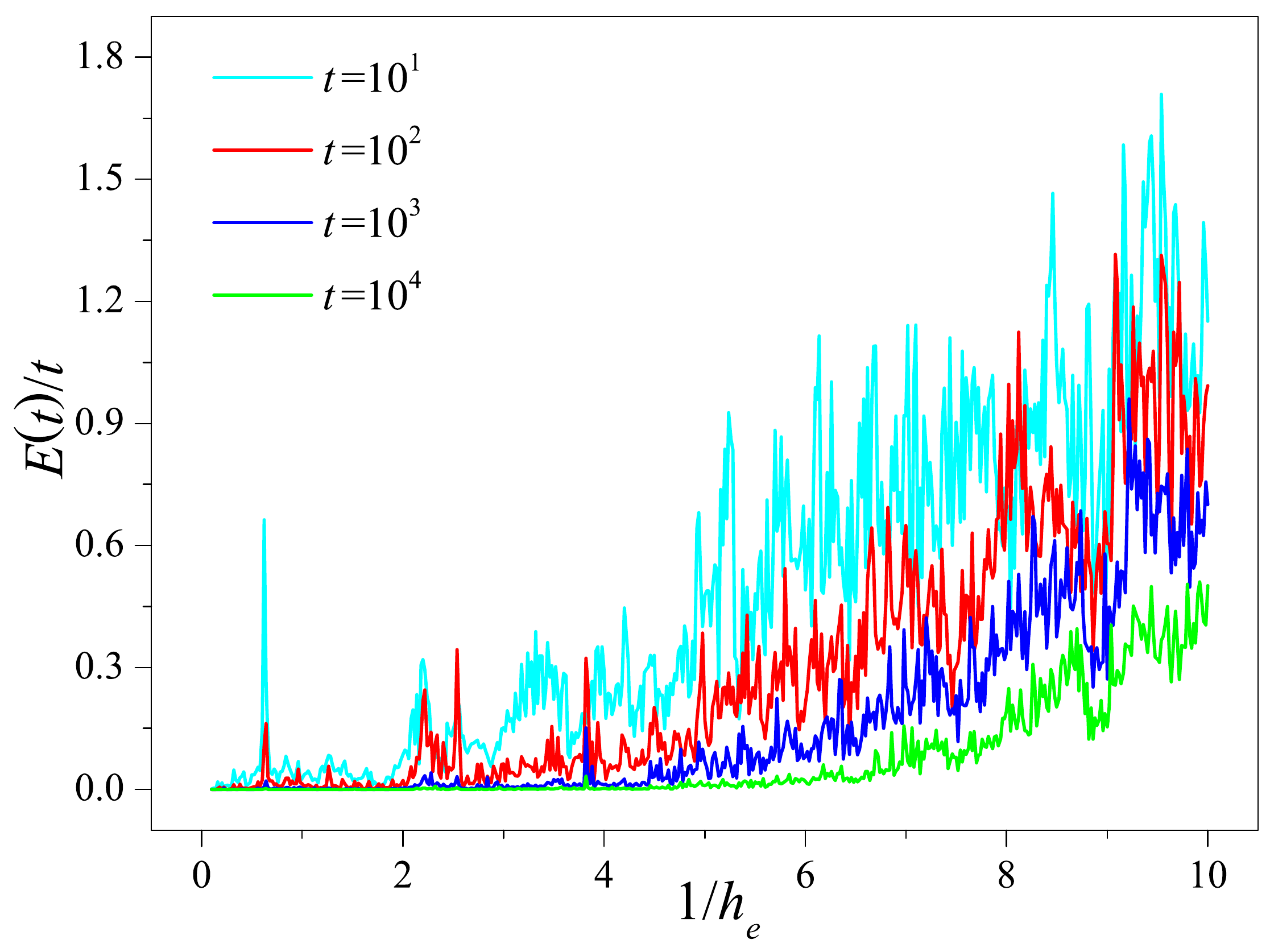}
\caption{Numerical simulations show that when $d_3=0.8(1-\cos\theta_1-\cos\theta_2)$ (cf. Eq.~(\ref{eq:58})) is replaced by $d_3=0.8$ and other conditions of the model used for obtaining Fig.~\ref{fig:3}(c) are not changed, the system is always insulting and no transition occurs, in agreement with analytic predictions.}
\label{fig:13}
\end{figure}

Interestingly, when the volume $\mathscr{D}$ enclosed by ${\cal S}$ is zero, Eq.~(\ref{eq:269}) vanishes. In this case, no transitions can occur at large $h_e^{-1}$. An example can be obtained by replacing $\beta(\mu-\cos\theta_1-\cos\theta_2)$ in Eq.~(\ref{eq:58}) by $\beta\mu$. For this modified model, we find no transitions for $0\leq h_e^{-1}\leq 10$ numerically (Fig.~\ref{fig:13}), consistent with theoretical predictions.

\section{Self-dual equations}
\label{sec:derivation_self_dual_equation}

Without loss of generality we focus on the case of $s=+$ and suppress the subscript $+$. By using the parametrization given in Eqs.~(\ref{eq:301}) and (\ref{eq:36}) we find
\begin{eqnarray}
\label{eq:138}
  && \partial_zQ-Q\partial_z Q \nonumber\\
  &=& 4 (1+iW)^{-1} \left(
                      \begin{array}{cc}
                        0 & 0 \\
                        \partial_z \tilde Z & 0 \\
                      \end{array}
                    \right)_{ar}
  (1-iW)^{-1}=0,
\end{eqnarray}
where we have used $\partial_z \tilde Z=0$. Similarly,
\begin{eqnarray}
\label{eq:139}
  && \partial_{z^*}Q+Q\partial_{z^*} Q \nonumber\\
  &=& -4 (1+iW)^{-1} \left(
                      \begin{array}{cc}
                        0 & \partial_{z^*} Z \\
                        0 & 0 \\
                      \end{array}
                    \right)_{ar}
  (1-iW)^{-1}=0,\quad \quad
\end{eqnarray}
where we have used $\partial_{z^*} Z=0$. From these two equations we obtain
\begin{eqnarray}
\label{eq:140}
  && \partial_1Q+iQ\partial_2 Q \nonumber\\
  &=& (\partial_{z}Q-Q\partial_{z^*} Q)+(\partial_{z^*}Q+Q\partial_{z^*} Q)=0.
\end{eqnarray}
Multiplying both sides by $-iQ$ gives
\begin{eqnarray}
\label{eq:141}
  \partial_2 Q-iQ \partial_1Q=0.
\end{eqnarray}
Equations (\ref{eq:140}) and (\ref{eq:141}) are the self-dual equation (\ref{eq:137}) corresponding to the `$+$' sign.

\section{Stationary instanton action}
\label{sec:derivation_instanton_action}

By using the self-dual equation (\ref{eq:137}), we obtain
\begin{eqnarray}
\label{eq:149}
  -\frac{\sigma}{4}{\rm Str}(\nabla Q_s)^2 = s i \frac{\sigma}{2} {\rm Str} (Q_s \nabla_1 Q_s \nabla_2 Q_s).
\end{eqnarray}
Therefore, in the zero-frequency limit the stationary instanton action
\begin{equation}\label{eq:150}
    S[Q_s]|_{\omega=0}= \left(s i \frac{\sigma}{2}+\frac{\sigma_{\rm H}}{4}\right) {\rm Str} (Q_s \nabla_1 Q_s \nabla_2 Q_s).
\end{equation}
We first consider $S[Q_+]|_{\omega=0}$. Substituting Eqs.~(\ref{eq:142}), (\ref{eq:143}) and (\ref{eq:233})
into it we find
\begin{eqnarray}
S[Q_+]|_{\omega=0}
&=& \left(\frac{i\sigma}{2}+\frac{\sigma_{\rm H}}{4}\right) {\rm Str} (\Lambda_+ \nabla_1 \Lambda_+ \nabla_2 \Lambda_+) \nonumber\\
&=& \left(i\sigma+\frac{\sigma_{\rm H}}{2}\right) {\rm Str} (\nabla \times (\Lambda R\nabla R^{-1})).
\label{eq:151}
\end{eqnarray}
Then, we apply the Stokes' theorem to Eq.~(\ref{eq:151}). This directly relates the instanton action to the local $U(1)$
gauge symmetry carried by $R_+(\vartheta)$, i.e.,
\begin{eqnarray}
&&S[Q_+]|_{\omega=0}\nonumber\\
  &=& \left(i\sigma+\frac{\sigma_{\rm H}}{2}\right) \oint_{|z|\rightarrow \infty} d{\bf l}\cdot {\rm str}(\Lambda R_+\nabla R_+^{-1}) \nonumber\\
  &=& \left(i\sigma+\frac{\sigma_{\rm H}}{2}\right) \int_0^{2\pi} d\vartheta \lim_{|z|\rightarrow \infty}{\rm str}(\Lambda R_+\partial_\vartheta R_+^{-1}). \label{eq:271}
\end{eqnarray}
Taking the limit of $|z|\rightarrow \infty$ we obtain
\begin{widetext}
\begin{eqnarray}
S[Q_+]|_{\omega=0} &=& -\left(i\sigma+\frac{\sigma_{\rm H}}{2}\right) \int_0^{2\pi} d\vartheta \lim_{|z|\rightarrow \infty} {\rm tr}\left(\sigma_{ar}^3\left(
                                   \begin{array}{cc}
                                     e_1^* & e_0 \\
                                     -e_0 & e_1 \\
                                   \end{array}
                                 \right)_{ar}\partial_\vartheta
                                 \left(
                                   \begin{array}{cc}
                                     e_1 & -e_0 \\
                                     e_0 & e_1^* \\
                                   \end{array}
                                 \right)_{ar}
  \right)\nonumber\\
  &=& -\left(i\sigma+\frac{\sigma_{\rm H}}{2}\right) \int_0^{2\pi} d\vartheta {\rm tr}\left(\sigma_{ar}^3\left(
                                   \begin{array}{cc}
                                     e^{-i\vartheta} & 0 \\
                                     0 & e^{i\vartheta} \\
                                   \end{array}
                                 \right)_{ar}\partial_\vartheta
                                 \left(
                                   \begin{array}{cc}
                                     e^{i\vartheta} & 0 \\
                                     0 & e^{-i\vartheta} \\
                                   \end{array}
                                 \right)_{ar}
  \right)\nonumber\\
  &=&4\pi \sigma -2\pi i\sigma_{\rm H}.
  \label{eq:152}
\end{eqnarray}
Taking the complex conjugate of Eq.~(\ref{eq:152}) we obtain
the stationary action of the anti-instanton.

\section{Eigenvalues and eigenfunctions of $\hat O^{(a)}$}
\label{sec:eigenvalue_eigenfunction}

Because of $[\partial_\vartheta,\hat O^{(a)}]=0$ there is a good quantum number associated with the operator $\partial_\vartheta$,
which is denoted as $M$.
Taking this into account we can assume the general form,
\begin{equation}\label{eq:167}
    \Phi^{(a)}(\eta,\vartheta)=C^{(a)} e^{-iM\vartheta} (1-\eta^2)^{\frac{M}{2}}(1-\eta)^{\frac{a}{2}} \Phi'^{(a)}(\eta),
\end{equation}
for the eigenfunctions, with $C^{(a)}$ being the normalization constant. Substituting it into the eigenfunction equation gives the hypergeometric equation of Gauss,
\begin{eqnarray}
\left((1-\eta^2)\partial_\eta^2+(\beta-\alpha-(\alpha+\beta+2)\eta)\partial_\eta+(E-((1+a)M+M^2))\right)\Phi'^{(a)}(\eta)=0
\label{eq:168}
\end{eqnarray}
with
\begin{equation}\label{eq:169}
    \alpha=M+a,\quad \beta=M.
\end{equation}
On the other hand, the Hilbert space of square integrable functions is spanned by $\Phi^{(a)}$ such that $\Phi'^{(a)}$ is a polynominal.
Equation (\ref{eq:168}) has a polynominal solution only if
\begin{eqnarray}
E-((1+a)M+M^2)=n(n+2M+a+1),\quad n=0,1,2,\cdots
\label{eq:170}
\end{eqnarray}
(cf. Theorem 4.2.2 in Ref.~\onlinecite{Szego39}). Setting $n=J-M$ with $M\leq J\in \mathbb{Z}$ we obtain $E=J(J+a+1)$ and
$\Phi'^{(a)}(\eta)=P_{J-M}^{M+a,M}(\eta)$. As we will see shortly,
the normalizability condition requires $J\geq 0$. Therefore, we prove the eigenvalue spectrum (\ref{eq:159}).

Next, we calculate the normalization constant, which is given by
\begin{eqnarray}
\frac{1}{(C^{(a)})^{2}}&=&\int_{-1}^1d\eta\int_0^{2\pi}d\vartheta (1-\eta)^{M+a}(1+\eta)^M \left|e^{-iM\vartheta}P_{J-M}^{M+a,M}(\eta)\right|^2\nonumber\\
&=&2\pi\int_{-1}^1d\eta (1-\eta)^{M+a}(1+\eta)^M\left(P_{J-M}^{M+a,M}(\eta)\right)^2
\label{eq:171}
\end{eqnarray}
by the definition (\ref{eq:162}). Applying the Rodrigues' formula \cite{Szego39},
\begin{eqnarray}
(1-\eta)^\alpha(1+\eta)^\beta P_n^{\alpha,\beta}(\eta)
=\frac{(-1)^n}{2^n n!}\frac{d^n}{d\eta^n} \left((1-\eta)^{n+\alpha}(1+\eta)^{n+\beta}\right),
\label{eq:172}
\end{eqnarray}
which is valid for arbitrary $\alpha,\beta$, we obtain
\begin{eqnarray}
\frac{1}{(C^{(a)})^{2}}=2\pi\frac{(-1)^{J-M}}{2^{J-M}(J-M)!}
\int_{-1}^1d\eta=\frac{d^{J-M}}{d\eta^{J-M}}\left((1-\eta)^{J+a}(1+\eta)^J\right)P_{J-M}^{M+a,M}(\eta)
\label{eq:173}
\end{eqnarray}
from Eq.~(\ref{eq:171}). Then, we repeatedly perform the integration by parts. Because the polynomial $(1-\eta)^{J+a}(1+\eta)^JP_{J-M}^{M+a,M}(\eta)$
vanishes at $\eta=\pm 1$ (noticing that $J>0$), no boundary contributions arise. As a result,
\begin{eqnarray}
\frac{1}{(C^{(a)})^{2}}=\frac{2\pi}{2^{J-M}(J-M)!}\int_{-1}^1d\eta(1-\eta)^{J+a}(1+\eta)^J \frac{d^{J-M}}{d\eta^{J-M}}P_{J-M}^{M+a,M}(\eta).
\label{eq:174}
\end{eqnarray}
We substitute Eq.~(\ref{eq:161}) into it. Since only the highest order term $\sim \eta^{J-M}$ survives after taking the derivative $(J-M)$ times with respect to $\eta$, we reduce Eq.~(\ref{eq:174}) to
\begin{eqnarray}
\frac{1}{(C^{(a)})^{2}}&=&\frac{2\pi}{2^{2(J-M)}(J-M)!}\frac{\Gamma(2J+a+1)}{\Gamma(J+M+a+1)}\int_{-1}^1d\eta(1-\eta)^{J+a}(1+\eta)^J\nonumber\\
&=&\frac{\pi 2^{2M+a+2}}{2J+a+1}\frac{\Gamma(J+1)\Gamma(J+a+1)}{\Gamma(J-M+1)\Gamma(J+M+a+1)},
\label{eq:175}
\end{eqnarray}
which gives the normalization constant in Eq.~(\ref{eq:160}). From this we see that to ensure that the eigenfunctions are
normalizable it is necessary that $J\geq 0$ and $-J-a\leq M\leq J$.

\section{Instanton manifold as zero modes}
\label{sec:discussions_zero_modes}

From Eq.~(\ref{eq:178}) we obtain
\begin{eqnarray}
    v=-2\left(R({\boldsymbol{c}})(T({\boldsymbol{\xi}})d
    T({\boldsymbol{\xi}})^{-1})R({\boldsymbol{c}})^{-1}+R({\boldsymbol{c}})d
    R({\boldsymbol{c}})^{-1}\right)_{+-},
    \label{eq:179}
\end{eqnarray}
\begin{eqnarray}
    \tilde v=-2\left(R({\boldsymbol{c}})(T({\boldsymbol{\xi}})d
    T({\boldsymbol{\xi}})^{-1})R({\boldsymbol{c}})^{-1}+R({\boldsymbol{c}})d
    R({\boldsymbol{c}})^{-1}\right)_{-+}.\label{eq:180}
\end{eqnarray}
We substitute Eqs.~(\ref{eq:143}) and (\ref{eq:233}) into them. After tedious but straightforward calculations we obtain
\begin{eqnarray}
v_{bb} = -2\sqrt{4\pi} t_{+-}^{bb}= -2\sqrt{4\pi} t_{+-}^{bb}\Phi_{0,0}^{(0)},\label{eq:181}
\end{eqnarray}
\begin{eqnarray}
\tilde v_{bb} = -2\sqrt{4\pi} t_{-+}^{bb}= -2\sqrt{4\pi} t_{-+}^{bb}\Phi_{0,0}^{(0)},\label{eq:182}
\end{eqnarray}
\begin{eqnarray}
v_{bf} = 2\left(t_{++}^{bf}e_0-t_{+-}^{bf}e_1^*\right)=2\sqrt{2\pi} \left(t_{++}^{bf}\Phi_{0,0}^{(1)}-t_{+-}^{bf}\Phi_{0,-1}^{(1)}\right),\label{eq:183}
\end{eqnarray}
\begin{eqnarray}
\tilde v_{bf} = -2\left(t_{-+}^{bf}e_1+t_{--}^{bf}e_0\right)=-2\sqrt{2\pi} \left(t_{-+}^{bf}\Phi_{0,0}^{(1)*}+t_{--}^{bf}\Phi_{0,0}^{(1)}\right),\label{eq:184}
\end{eqnarray}
\begin{eqnarray}
v_{fb} = -2\left(e_1^*t_{+-}^{fb}+e_0 t_{--}^{fb}\right)=-2\sqrt{2\pi} \left(t_{+-}^{fb}\Phi_{0,-1}^{(1)}+t_{--}^{fb}\Phi_{0,0}^{(1)}\right),\label{eq:185}
\end{eqnarray}
\begin{eqnarray}
\tilde v_{fb} = 2\left(e_0t_{++}^{fb}-e_1 t_{-+}^{fb}\right)=2\sqrt{2\pi} \left(t_{++}^{fb}\Phi_{0,0}^{(1)}-t_{-+}^{fb}\Phi_{0,-1}^{(1)*}\right),\label{eq:186}
\end{eqnarray}
\begin{eqnarray}
v_{ff} &=& 2\left(e_0e_1^*\left((t_{++}^{ff}-t_{--}^{ff})+\frac{d\lambda}{\lambda}\right)+e_0^2\left(t_{-+}^{ff}-\frac{dz_0^*}{\lambda}\right)-e_1^{*2}t_{+-}^{ff}
\right)\nonumber\\
&=&2\sqrt{\frac{4\pi}{3}}\left(\frac{1}{\sqrt{2}}\left((t_{++}^{ff}-t_{--}^{ff})+\frac{d\lambda}{\lambda}\right)
\Phi_{0,-1}^{(2)}+\left(t_{-+}^{ff}-\frac{dz_0^*}{\lambda}\right)\Phi_{0,0}^{(2)}-t_{+-}^{ff}\Phi_{0,-2}^{(2)}
\right),\label{eq:187}
\end{eqnarray}
\begin{eqnarray}
\tilde v_{ff} &=& 2\left(e_0e_1\left((t_{++}^{ff}-t_{--}^{ff})-\frac{d\lambda}{\lambda}\right)+e_0^2\left(t_{+-}^{ff}+\frac{dz_0}{\lambda}\right)-e_1^{2}t_{-+}^{ff}
\right)\nonumber\\
&=&2\sqrt{\frac{4\pi}{3}}\left(\frac{1}{\sqrt{2}}\left((t_{++}^{ff}-t_{--}^{ff})-\frac{d\lambda}{\lambda}\right)
\Phi_{0,-1}^{(2)*}+\left(t_{+-}^{ff}+\frac{dz_0}{\lambda}\right)\Phi_{0,0}^{(2)}-t_{-+}^{ff}\Phi_{0,-2}^{(2)*}
\right).\label{eq:188}
\end{eqnarray}
\end{widetext}
Recall $t_{\lambda\lambda'}^{\alpha\alpha'}=(T
d
T^{-1}
)_{\lambda\alpha,\lambda'\alpha'}$. The second equalities in Eqs.~(\ref{eq:181})-(\ref{eq:188}) show that $v_{\alpha\alpha'},\tilde v_{\alpha\alpha'}$ are all expanded by the zero-mode
bases. The expansion coefficients are the coordinate $\boldsymbol{c}$ and the generators of $G/H$. The latter are $t_{\lambda\lambda'}^{\alpha
\alpha'}$ ($\lambda\neq \lambda'$), $t_{\pm\pm}^{\alpha
\alpha'}$ ($\alpha\neq \alpha'$) and $(t_{++}^{ff}-t_{--}^{ff})$.
They generate the first, second, and last factor in Eq.~(\ref{eq:154}), respectively.

\section{Factorization of measure $DR_sDT$}
\label{sec:splitting_measure}

In this appendix we show that the measure $DR_sDT$ can be factorized according to Eq.~(\ref{eq:232}).
Since $z_0,z_0^*$ are independent variables we can make the shift:
\begin{equation}\label{eq:298}
    \frac{dz_0^*}{\lambda}-t_{-+}^{ff}\rightarrow \frac{dz_0^*}{\lambda},\quad
\frac{dz_0}{\lambda}+t_{+-}^{ff}\rightarrow \frac{dz_0}{\lambda}.
\end{equation}
As a result, the elementary length (\ref{eq:200}) becomes
\begin{eqnarray}
ds^2&=&\overline{e_0^2|e_1|^2}\,\,\left(\frac{d\lambda}{\lambda}\right)^2 + \overline{e_0^2|e_1|^2}\,\, d\phi^2\nonumber\\
&&+\overline{e_0^4}\,\,\Big|\frac{dz_0}{\lambda}\Big|^2+\sum_{i=1}^4 ds_i^2,
\label{eq:241}
\end{eqnarray}
where
\begin{eqnarray}
ds_1^2\equiv-\overline{|e_1|^4}\,\, t_{-+}^{ff}t_{+-}^{ff},
  \label{eq:215}
\end{eqnarray}
\begin{eqnarray}
ds_2^2\equiv\overline{ 1 }\,\, t_{-+}^{bb} t_{+-}^{bb},
\label{eq:242}
\end{eqnarray}
\begin{eqnarray}
ds_3^2\equiv-\overline{|e_1|^2}\,\, \left(t_{-+}^{fb}t_{+-}^{bf}-t_{-+}^{bf}t_{+-}^{fb}\right),
\label{eq:243}
\end{eqnarray}
and
\begin{equation}\label{eq:213}
ds_4^2\equiv-\overline{e_0^2}\,\, \left(t_{++}^{fb}t_{++}^{bf}-t_{--}^{bf}t_{--}^{fb}\right).
\end{equation}
We see that $\sum_{i=1}^4 ds_i^2$ involve a set of generators which are not entangled with the
coordinates $\lambda$, $n_{10}$, $n_{20}$ and $\phi$. The first three terms in Eq.~(\ref{eq:241}) induce a measure,
\begin{equation}\label{eq:244}
    \overline{e_0^2|e_1|^2}\,\,\overline{e_0^4}\,\, \frac{d\lambda}{\lambda^3}d\phi dn_{10}dn_{20}.
\end{equation}

Below we find the measure induced by $\sum_{i=1}^4ds_i^2$.
First of all, the generators in the length
element $ds_4^2$ are not entangled with those in $ds_{1,2,3}^2$. So, we separate $ds_4^2$ from $ds_i^2,\, i=1,2,3$. Note that
any $T\in U(1,1|2)$ can be factorized as
\begin{equation}\label{eq:239}
    T=\tilde U\tilde T,\quad \tilde U=h u_0 u,
\end{equation}
where $\tilde T\in \frac{U(1,1|2)}{U(1|1)\times U(1|1)}$,
$h\in H$, $u_0$ is given by Eq.~(\ref{eq:236}), and $u\in (\frac{U(1|1)}{U(1)\times U(1)})^2$
with Eqs.~(\ref{eq:196})-(\ref{eq:218}) as its representation. Substituting Eq.~(\ref{eq:239}) into Eq.~(\ref{eq:240}), we obtain
\begin{eqnarray}
t_{++}^{fb} &=& \left(\tilde Ud
\tilde U^{-1}\right)_{+f,+b}\nonumber\\
&=&(u_1d
u_1^{-1})e^{i(\gamma+\phi/2-\alpha_1)}\nonumber\\
&=&-d\zeta_1e^{i(\gamma+\phi/2-\alpha_1)},\label{eq:207}
\end{eqnarray}
\begin{eqnarray}
t_{++}^{bf} = -d\zeta_1^*e^{-i(\gamma+\phi/2-\alpha_1)},
\label{eq:208}
\end{eqnarray}
\begin{eqnarray}
t_{--}^{fb} = -id\zeta_2^*e^{i(\gamma-\phi/2-\alpha_2)},
\label{eq:209}
\end{eqnarray}
and
\begin{eqnarray}
t_{--}^{bf} = id\zeta_2e^{-i(\gamma-\phi/2-\alpha_2)}.
\label{eq:210}
\end{eqnarray}
Upon inserting them into Eq.~(\ref{eq:213}) we obtain
\begin{eqnarray}
ds_4^2=-\overline{e_0^2}\,\, \left(d\zeta_1d\zeta_1^*+d\zeta_2^*d\zeta_2\right).
\label{eq:211}
\end{eqnarray}
This implies that Eqs.~(\ref{eq:217}) and (\ref{eq:218}) constitute a flat parametrization of $(\frac{U(1|1)}{U(1)\times U(1)})^2$
with
\begin{equation}\label{eq:227}
    \frac{4}{\pi^2} \frac{1}{\overline{e_0^2}^2}d\zeta_1d\zeta_1^*d\zeta_2^*d\zeta_2
\end{equation}
as its measure.

Next, we find the measure induced by $\sum_{i=1}^3ds_i^2$. To this end we choose the coordinates $\{\xi_\mu^c\}$  of $\tilde T$ so that
$(T\partial_{\xi_{\mu}^c}T^{-1})_{-\alpha,+\alpha'}$ are diagonal in the indices $\mu$ and $\alpha,\alpha'$.
Specifically, we can introduce a set of complex numbers and Grassmannians $\chi_{\alpha\alpha'}^q$ ($q=0,1$)
such that
\begin{eqnarray}
t_{-+}^{\alpha\alpha'}\equiv d\chi_{\alpha\alpha'}^q\Delta_{\alpha\alpha'}^q,\quad
t_{+-}^{\alpha'\alpha}\equiv d\chi_{\alpha\alpha'}^q\hat\Delta_{\alpha\alpha'}^q,
\label{eq:214}
\end{eqnarray}
where $\chi_{\alpha\alpha'}^q$ are complex variables (Grassmannians) for $\alpha=\alpha'$ ($\alpha\neq\alpha'$).
Then, we substitute them into $ds_i^2,\, i=1,2,3$. It turns
out that the coordinates in different $ds_i^2$ are not entangled with each other and, therefore, the measure is factorized.
Specifically, substituting Eq.~(\ref{eq:214}) into Eq.~(\ref{eq:215}) gives
\begin{eqnarray}
  ds_1^2 &=&
  -\overline{|e_1|^4}\,\, d\chi_{ff}^q \Delta_{ff}^q \hat \Delta_{ff}^{q'} d\chi_{ff}^{q'}\nonumber\\
  &=&
  -\overline{|e_1|^4}\,\, (d\chi_{ff}^{0}\,\,d\chi_{ff}^{1})
  g_1\left(
       \begin{array}{c}
         d\chi_{ff}^{0} \\
         d\chi_{ff}^{1} \\
       \end{array}
     \right),
  \label{eq:219}
\end{eqnarray}
where $g_1\equiv \{g_{ff}^{qq'}\}$ with
\begin{equation}\label{eq:220}
    g_{\alpha\alpha'}^{qq'}\equiv
    \frac{1}{2}\left(\Delta_{\alpha\alpha'}^q\hat \Delta_{\alpha\alpha'}^{q'}-
    (1-2\delta_{\alpha\alpha'})\Delta_{\alpha\alpha'}^{q'}\hat \Delta_{\alpha\alpha'}^{q}\right).
\end{equation}
Equation (\ref{eq:219}) leads to a measure \cite{note_convention},
\begin{equation}\label{eq:222}
    \overline{|e_1|^4}\,\, \sqrt{{\rm det} g_1}\frac{d\chi_{ff}^0d\chi_{ff}^1}{\pi}.
\end{equation}
Substituting Eq.~(\ref{eq:214}) into Eq.~(\ref{eq:242}) gives
\begin{eqnarray}
ds_2^2
&=&\overline{1}\,\, d\chi_{bb}^q \Delta_{bb}^q \hat \Delta_{bb}^{q'} d\chi_{bb}^{q'}\nonumber\\
&=&(d\chi_{bb}^{0}\,\,d\chi_{bb}^{1})
  g_2\left(
       \begin{array}{c}
         d\chi_{bb}^{0} \\
         d\chi_{bb}^{1} \\
       \end{array}
     \right),
\label{eq:221}
\end{eqnarray}
with $g_2\equiv \{g_{bb}^{qq'}\}$. This induces a measure,
\begin{equation}\label{eq:223}
    \overline{1}\,\, \sqrt{{\rm det} g_2} \frac{d\chi_{bb}^0d\chi_{bb}^1}{\pi}.
\end{equation}
Substituting Eq.~(\ref{eq:214}) into Eq.~(\ref{eq:243}) gives
\begin{widetext}
\begin{eqnarray}\label{eq:224}
    ds_3^2
    &=&-\overline{|e_1|^2}\,\, \left(d\chi_{fb}^q \Delta_{fb}^q\hat \Delta_{fb}^{q'}d\chi_{fb}^{q'}-
    d\chi_{bf}^q \Delta_{bf}^q\hat \Delta_{bf}^{q'}d\chi_{bf}^{q'}\right)\nonumber\\
    &=&-\overline{|e_1|^2}\,\, (d\chi_{fb}^{0}\,\,d\chi_{fb}^{1})
  g_3\left(
       \begin{array}{c}
         d\chi_{fb}^{0} \\
         d\chi_{fb}^{1} \\
       \end{array}
     \right)
     +\overline{|e_1|^2 }\,\, (d\chi_{bf}^{0}\,\,d\chi_{bf}^{1})
  g_4\left(
       \begin{array}{c}
         d\chi_{bf}^{0} \\
         d\chi_{bf}^{1} \\
       \end{array}
     \right),
\end{eqnarray}
with $g_3\equiv \{g_{fb}^{qq'}\}$ and $g_4\equiv \{g_{bf}^{qq'}\}$. This induces a measure,
\begin{equation}\label{eq:225}
    \frac{1}{\overline{|e_1|^2}^2} \frac{1}{\sqrt{{\rm det} (g_3g_4)}} d\chi_{fb}^0d\chi_{fb}^1d\chi_{bf}^0d\chi_{bf}^1.
\end{equation}
Collecting Eqs.~(\ref{eq:241}), (\ref{eq:211}), (\ref{eq:219}), (\ref{eq:221}) and (\ref{eq:224}), we obtain
\begin{eqnarray}
ds^2
&=&\overline{e_0^2|e_1|^2}\,\,\left(\frac{d\lambda}{\lambda}\right)^2 + \overline{e_0^2|e_1|^2}\,\, d\phi^2 +
\overline{e_0^4}\,\, \Big|\frac{dz_0}{\lambda}\Big|^2-\overline{e_0^2}\,\, \left(d\zeta_1d\zeta_1^*+d\zeta_2^*d\zeta_2\right)\nonumber\\
&&-\overline{|e_1|^4}\,\, (d\chi_{ff}^{0}\,\,d\chi_{ff}^{1})
  g_1\left(
       \begin{array}{c}
         d\chi_{ff}^{0} \\
         d\chi_{ff}^{1} \\
       \end{array}
     \right)+\overline{1}\,\,(d\chi_{bb}^{0}\,\,d\chi_{bb}^{1})
  g_2\left(
       \begin{array}{c}
         d\chi_{bb}^{0} \\
         d\chi_{bb}^{1} \\
       \end{array}
     \right)\nonumber\\
     &&-\overline{|e_1|^2}\,\, (d\chi_{fb}^{0}\,\,d\chi_{fb}^{1})
  g_3\left(
       \begin{array}{c}
         d\chi_{fb}^{0} \\
         d\chi_{fb}^{1} \\
       \end{array}
     \right)+\overline{ |e_1|^2 }\,\, (d\chi_{bf}^{0}\,\,d\chi_{bf}^{1})
  g_4\left(
       \begin{array}{c}
         d\chi_{bf}^{0} \\
         d\chi_{bf}^{1} \\
       \end{array}
     \right).
\label{eq:226}
\end{eqnarray}
This elementary length induces a measure,
\begin{equation}\label{eq:228}
    \frac{4}{\pi^2}\frac{\overline{ e_0^2|e_1|^2}\,\, \overline{ e_0^4}\,\, \overline{ |e_1|^4}\,\, \overline{1}}{\left(\overline{e_0^2}\,\,
    \overline{|e_1|^2}\right)^2} \frac{d\lambda}{\lambda^3} d\phi dn_{10}dn_{20}d\zeta_1d\zeta_1^*d\zeta_2^*d\zeta_2 {\cal M}_{Q},
\end{equation}
namely, the product of the measures (\ref{eq:244}), (\ref{eq:227}), (\ref{eq:222}), (\ref{eq:223}) and (\ref{eq:225}), where
\begin{equation}\label{eq:229}
    {\cal M}_{Q}=\sqrt{{\rm det}\left(\frac{g_1g_2}{g_3g_4}\right)}
    \frac{d\chi_{ff}^0d\chi_{ff}^1}{\pi}\frac{d\chi_{bb}^0d\chi_{bb}^1}{\pi}d\chi_{fb}^0d\chi_{fb}^1d\chi_{bf}^0d\chi_{bf}^1
\end{equation}
with ${\rm det}(g_1g_2),{\rm det}(g_3g_4)>0$.

The measure ${\cal M}_{Q}$ has a gauge ($U(1|1)\times U(1|1)$) invariant form. To see this we consider a gauge invariant elementary length on the
coset space $\frac{U(1,1|2)}{U(1|1)\times U(1|1)}$,
\begin{eqnarray}
{\rm str}(dQ)^2&=&{\rm str}(d(T^{-1}\Lambda T)^2)\nonumber\\
&=&-8{\rm str}\left((TdT^{-1})_{+-}(TdT^{-1})_{-+}
\right).
\label{eq:230}
\end{eqnarray}
Then, we parametrize $Q=T^{-1}\Lambda T=
\tilde T^{-1}\Lambda \tilde T$ by the coordinates $\chi_{\alpha\alpha'}^q$. Taking into account Eq.~(\ref{eq:214}), we obtain
\begin{eqnarray}
{\rm str}(dQ)^2&=&8\bigg((d\chi_{ff}^{0}\,\,d\chi_{ff}^{1})
  g_1\left(
       \begin{array}{c}
         d\chi_{ff}^{0} \\
         d\chi_{ff}^{1} \\
       \end{array}
     \right)-(d\chi_{bb}^{0}\,\,d\chi_{bb}^{1})
  g_2\left(
       \begin{array}{c}
         d\chi_{bb}^{0} \\
         d\chi_{bb}^{1} \\
       \end{array}
     \right)\nonumber\\
     &&+(d\chi_{fb}^{0}\,\,d\chi_{fb}^{1})
  g_3\left(
       \begin{array}{c}
         d\chi_{fb}^{0} \\
         d\chi_{fb}^{1} \\
       \end{array}
     \right)-(d\chi_{bf}^{0}\,\,d\chi_{bf}^{1})
  g_4\left(
       \begin{array}{c}
         d\chi_{bf}^{0} \\
         d\chi_{bf}^{1} \\
       \end{array}
     \right)\bigg),
\label{eq:231}
\end{eqnarray}
\end{widetext}
which gives the measure ${\cal M}_{Q}$ exactly. So, Eq.~(\ref{eq:232}) is proven.

\section{Constrained instanton}
\label{sec:constrained_instanton}

The infrared divergence of the second term of Eq.~(\ref{eq:249}) merely informs
that effects of the frequency term ${\eta} {\rm Str}(Q\Lambda)$ in the action
$S[Q]$ are significant when we consider the scale much larger than $\sqrt{\sigma/{\eta}}$. Since the stationary equation
(\ref{eq:135}) and (\ref{eq:136}) or the self-dual equation (\ref{eq:137}) are obtained by setting ${\eta}$ to zero,
they must break down for large $N$. So, exactly speaking, the instanton solution described by Eqs.~(\ref{eq:142}), (\ref{eq:143}) and (\ref{eq:233}) is no longer valid for $\eta>0$.
As we will see below, the solution indeed is dramatically modified for $|N|\gg \sqrt{\sigma/{\eta}}$.

Here we adopt the method of Ref.~\onlinecite{Pruisken05} to cure this infrared divergence. This method involves the concept of the so-called constrained instanton. To start, we note that field configurations constrained in certain way might minimize the action,
and such field configuration of finite
action for nonvanishing ${\eta}$ has to converge to the instanton solution namely Eqs.~(\ref{eq:142}), (\ref{eq:143}) and (\ref{eq:233}) in the limit of ${\eta}=0$.
To achieve these conditions we assume that the field configuration has the same form as Eq.~(\ref{eq:142}) except that $T$ is suppressed due
to the symmetry breaking, i.e.,
\begin{eqnarray}
Q'_s = R_s'^{-1}\Lambda R'_s,
  \label{eq:251}
\end{eqnarray}
and in Eq.~(\ref{eq:233}) the instanton size $\lambda$ has an $N$ dependence:
$\lambda^2\rightarrow \lambda^2 f(x,{\tilde \eta})$ with $x=(|z-z_0|/\lambda)^2$ and ${\tilde \eta}=2{\eta} \lambda^2/\sigma$.
The primes in Eq.~(\ref{eq:251}) are the bookkeeping of this replacement.
The function $f(x,{\tilde \eta})$ is constrained by
$f(x,0)=f(0,{\tilde \eta})=1$ and $f(x\rightarrow \infty,{\tilde \eta}>0)=0$.

Substituting Eq.~(\ref{eq:251}) into the action gives
\begin{eqnarray}
S[Q'_s]|_{\omega\rightarrow\frac{i{\eta}}{2}}&=&2\pi\sigma \int_0^\infty dx \frac{f}{(x+f)^2}\nonumber\\
&&\times\left(
1+\left(1-\frac{x\partial_x f}{f}\right)^2+{\tilde \eta} (x+f)\right).\,\,\,\,\,\,
\label{eq:245}
\end{eqnarray}
The behavior of $f$ optimizing this action is as follows,
\begin{eqnarray}
f(x,{\tilde \eta})=\Bigg\{\begin{array}{c}
                   1+4{\tilde \eta} x,\quad x\ll 1, \\
                   1+\frac{{\tilde \eta} x}{2}\ln \frac{{\tilde \eta} x}{4}, \quad 1\ll x\ll {\tilde \eta}^{-1}, \\
                   \frac{\pi}{2}\sqrt{{\tilde \eta} x} e^{-\sqrt{{\tilde \eta} x}}, \quad x\gg {\tilde \eta}^{-1}.
                 \end{array}
\label{eq:246}
\end{eqnarray}
The last line implies that the constrained instanton exhibits an exponential decay in $|z-z_0|$ for $|z-z_0|$ much exceeding $\sqrt{\sigma/{\eta}}$,
while behaves as the unconstrained instanton for $|z-z_0|\ll\sqrt{\sigma/{\eta}}$.

Substituting Eq.~(\ref{eq:246}) into the action (\ref{eq:245}) we find that the frequency term has a finite action,
\begin{equation}\label{eq:250}
    {\eta} {\rm Str} (Q'_s\Lambda)\sim {\eta}\lambda^2 \ln {\tilde \eta},
\end{equation}
instead of the naive result suffering infrared divergence (cf. Eq.~(\ref{eq:249})). Equation (\ref{eq:250}) can be established in a more sophisticated way
based on the fact that ${\eta}$ is not renormalized \cite{Pruisken05}.
Most importantly, Eq.~(\ref{eq:250}) gives a vanishing result in the limit of ${\eta}\rightarrow 0$. Correspondingly, the action of the constrained instanton converges to that of unconstrained instanton.

\section{Perturbative RG function}
\label{sec:beta_function}

In this appendix we derive $\beta_{{\rm L},p}$ from Eq.~(\ref{eq:119}). For this purpose we use the dimensional
regularization to treat the ultraviolet divergence. Specifically, we
calculate Eq.~(\ref{eq:119}) for $\epsilon=d-2<0$ and finally extend the result to $\epsilon=0$. Applying this regularization, we obtain
\begin{equation}\label{eq:126}
    \tilde \sigma = \sigma \left(1+\frac{1}{\epsilon} \frac{1}{(
    4\pi\sigma \tilde\lambda^\epsilon)^2}\right).
\end{equation}
Note that throughout this appendix the nonperturbative part of $\tilde \sigma$ is excluded.
Define
\begin{equation}\label{eq:127}
    \tilde \sigma\equiv \frac{1}{
    4\pi \tilde\lambda^\epsilon t},\quad \sigma\equiv \frac{1}{
    4\pi \tilde\lambda^\epsilon t} Z_\sigma^{-1}.
\end{equation}
We rewrite Eq.~(\ref{eq:126}) as
\begin{equation}\label{eq:128}
    Z_\sigma=1+\frac{t^2}{\epsilon}+{\cal O}(t^4).
\end{equation}
From the second equation of (\ref{eq:127}) we find
\begin{eqnarray}\label{eq:129}
    \frac{d\ln t}{d\ln \tilde\lambda}&=&-\frac{\epsilon}{1+\frac{d\ln Z_\sigma}{d\ln t}}\nonumber\\
    &\approx&-\epsilon + 2t^2.
\end{eqnarray}
On the other hand, the first equation in
(\ref{eq:127}) gives
\begin{equation}\label{eq:267}
    \frac{d\ln \tilde \sigma}{d\ln \tilde\lambda}
=-\epsilon -\frac{d\ln t}{d\ln \tilde\lambda}.
\end{equation}
Substituting it into Eq.~(\ref{eq:129}) we obtain
the perturbative RG function (\ref{eq:130}).

\section{Details for numerical computation of Eq.~(\ref{eq:S5})}
\label{sec:quantum_critical_points}

To numerically compute Eq.~(\ref{eq:S5}) we rewrite it as
\begin{eqnarray}
\label{eq:18}
  \sigma_{\rm H}^{I} = \sigma_{\rm H}^{I,a}+\sigma_{\rm H}^{I,b},
\end{eqnarray}
where
\begin{eqnarray}
\label{eq:247}
  \sigma_{\rm H}^{I,a} =
    \frac{2\beta}{Kh_e}
    \int\!\!\!\!\int \frac{d\theta_1}{2\pi}\frac{d\theta_2}{2\pi}
    \!\!\int_1^{+\infty}\!\! d\mu \frac{\cos\theta_1\cos\theta_2}{d^2_{\mu}(d^2_{\mu}+
    K^{-2}
    )}
\end{eqnarray}
and
\begin{eqnarray}
\label{eq:248}
  \sigma_{\rm H}^{I,b} =
    \frac{2\beta}{Kh_e}
    \int\!\!\!\!\int \frac{d\theta_1}{2\pi}\frac{d\theta_2}{2\pi}
    \!\!\int_1^{+\infty}\!\! d\mu \frac{\cos2\varphi_{\mu}\cos\theta_1\cos\theta_2}{d^2_{\mu}(d^2_{\mu}+
    K^{-2}
    )}.\,\,\,\,
\end{eqnarray}
For $\sigma_{\rm H}^{I,a}$, carrying out the $\mu$-integral we obtain
\begin{widetext}
\begin{eqnarray}
\label{eq:285}
  \sigma_{\rm H}^{I,a} &=&
    -\frac{2K}{h_e}
    \int\!\!\!\!\int \frac{d\theta_1}{2\pi}\frac{d\theta_2}{2\pi}
    \cos\theta_1\cos\theta_2\nonumber\\
    &&\times\left(\frac{1}{s}
    \arctan\frac{\beta(1-\cos\theta_1-\cos\theta_2)}{s}
    -\frac{1}{\sqrt{s^2+K^{-2}}}\arctan\frac{\beta(1-\cos\theta_1-\cos\theta_2)}{\sqrt{s^2+K^{-2}}}\right),
\end{eqnarray}
with $s^2=s_1^2+s_2^2,\,s_i=\sin\theta_i$. Introducing
\begin{equation}\label{eq:286}
f_{\alpha_{1},\alpha_{2}}(s_1,s_2;\gamma)\equiv
\frac{1}{\sqrt{s^2+\gamma^{2}}}\arctan\frac{\beta\left(1-\alpha_1\sqrt{1-s_1^2}-\alpha_2\sqrt{1-s_2^2}\right)}{\sqrt{s^2+\gamma^2}},
\end{equation}
we rewrite Eq.~(\ref{eq:285}) as
\begin{eqnarray}
\label{eq:19}
\sigma_{\rm H}^{I,a} = -\frac{2K}{(2\pi)^2 h_e}
    \int\!\!\!\!\int_{-1}^1 ds_1ds_2\sum_{\alpha_{1,2}=\pm}\alpha_1\alpha_2\left(f_{\alpha_{1},\alpha_{2}}(s_1,s_2;0)-
    f_{\alpha_{1},\alpha_{2}}(s_1,s_2;K^{-1})\right).
\end{eqnarray}
Passing to the polar coordinate, $(s,\phi)$ with $\tan\phi=s_2/s_1$, we rewrite Eq.~(\ref{eq:19}) as
\begin{eqnarray}
\label{eq:20}
  \sigma_{\rm H}^{I,a} = -\frac{2K}{(2\pi)^2 h_e} \sum_{\alpha_{1,2}=\pm}\alpha_1\alpha_2
    \int_{-\pi}^\pi d\phi\int_0^{f(\phi)} ds s\left(\tilde f_{\alpha_{1},\alpha_{2}}(s,\phi;0)-
    \tilde f_{\alpha_{1},\alpha_{2}}(s,\phi;K^{-1})\right),
\end{eqnarray}
with
\begin{eqnarray}
\tilde f_{\alpha_{1},\alpha_{2}}(s,\phi;\gamma)\equiv
\frac{1}{\sqrt{s^2+\gamma^{2}}}\arctan\frac{\beta\left(1-\alpha_1\sqrt{1-s^2\cos^2\phi}-\alpha_2\sqrt{1-s^2\sin^2\phi}\right)}{\sqrt{s^2+\gamma^2}}
\label{eq:261}
\end{eqnarray}
and
\begin{eqnarray}
f(\phi)=\bigg\{\begin{array}{cc}
                 1/|\cos\phi|,&\phi\in [-\pi/4,\pi/4]\cup[-\pi,-3\pi/4]\cup[3\pi/4,\pi]; \\
                 1/|\sin\phi|,&\phi\in [-3\pi/4,-\pi/4]\cup[\pi/4,3\pi/4].
               \end{array}
\label{eq:262}
\end{eqnarray}
After the integration by parts, we find
\begin{eqnarray}
\label{eq:21}
  \sigma_{\rm H}^{I,a} &=& -\frac{1}{\pi h_e}\left(\arctan 3\beta K-3\arctan \beta K\right)\nonumber\\
  &&  -\frac{2\beta}{(2\pi)^2 h_e K} \sum_{\alpha_{1,2}=\pm}\alpha_1\alpha_2
    \int_{-\pi}^\pi d\phi\int_0^{f(\phi)} ds s\frac{1}{d^2(d^2+K^{-2})}\left(1-\frac{1}{\alpha_1\sqrt{1-s^2\cos^2\phi}}-\frac{1}{\alpha_2\sqrt{1-s^2\sin^2\phi}}\right)\nonumber\\
  &&  -\frac{2\beta}{(2\pi)^2 h_e K} \sum_{\alpha_{1,2}=\pm}\alpha_1\alpha_2
    \int_{-\pi}^\pi d\phi\int_0^{f(\phi)} ds s\frac{1}{s^2(d^2+K^{-2})}\left(\frac{s^2\cos^2\phi}{\alpha_1\sqrt{1-s^2\cos^2\phi}}+\frac{s^2\sin^2\phi}{\alpha_2\sqrt{1-s^2\sin^2\phi}}\right)\nonumber\\
    &=& -\frac{1}{\pi h_e}\left(\arctan 3\beta K-3\arctan \beta K\right)\nonumber\\
  &&  -\frac{2\beta}{(2\pi)^2 h_e K} \sum_{\alpha_{1,2}=\pm}\alpha_1\alpha_2
    \int\!\!\!\!\int_{-1}^1 ds_1ds_2 \frac{1}{d^2(d^2+K^{-2})}\left(1-\frac{1}{\alpha_1\sqrt{1-s_1^2}}-\frac{1}{\alpha_2\sqrt{1-s_2^2}}\right)\nonumber\\
  &&  -\frac{2\beta}{(2\pi)^2 h_e K} \sum_{\alpha_{1,2}=\pm}\alpha_1\alpha_2
    \int\!\!\!\!\int_{-1}^1 ds_1ds_2 \frac{1}{s^2(d^2+K^{-2})}\left(\frac{s_1^2}{\alpha_1\sqrt{1-s_1^2}}+\frac{s_2^2}{\alpha_2\sqrt{1-s_2^2}}\right),
\end{eqnarray}
\end{widetext}

\noindent where in the second equality we have returned to the coordinate $(s_1,s_2)$. Passing to the coordinate $\Theta$, we obtain
\begin{eqnarray}
\label{eq:22}
  \sigma_{\rm H}^{I,a}&=&-\frac{1}{\pi h_e}\left(\arctan 3\beta K-3\arctan \beta K\right)\nonumber\\
  &&+\sigma_{\rm H}^{I,a1}+\sigma_{\rm H}^{I,a2},
\end{eqnarray}
where
\begin{eqnarray}
\label{eq:289}
\sigma_{\rm H}^{I,a1}&=&\frac{2\beta}{h_e K}\int\!\!\!\!\int \frac{d\theta_1}{2\pi}\frac{d\theta_2}{2\pi}\nonumber\\
&&\times
  \frac{\cos\theta_1+\cos\theta_2-\cos\theta_1\cos\theta_2}{d^2(d^2+K^{-2})},
\end{eqnarray}
and
\begin{eqnarray}
\label{eq:290}
\sigma_{\rm H}^{I,a2}&=&-\frac{2\beta}{h_e K} \int\!\!\!\!\int \frac{d\theta_1}{2\pi}\frac{d\theta_2}{2\pi}\nonumber\\
&&\times\frac{\sin^2\theta_1\cos\theta_2+\sin^2\theta_2\cos\theta_1}{(\sin^2\theta_1+\sin^2\theta_2)(d^2+K^{-2})}.
\end{eqnarray}
Therefore, the numerical calculation of $\sigma_{\rm H}^I$ is reduced to three numerical integrals, namely, Eqs.~(\ref{eq:248}), (\ref{eq:289}) and (\ref{eq:290}).

\section{Finite-time effects in simulations}
\label{sec:finite_time_effects}

\begin{figure}[h]
\includegraphics[width=8.6cm]{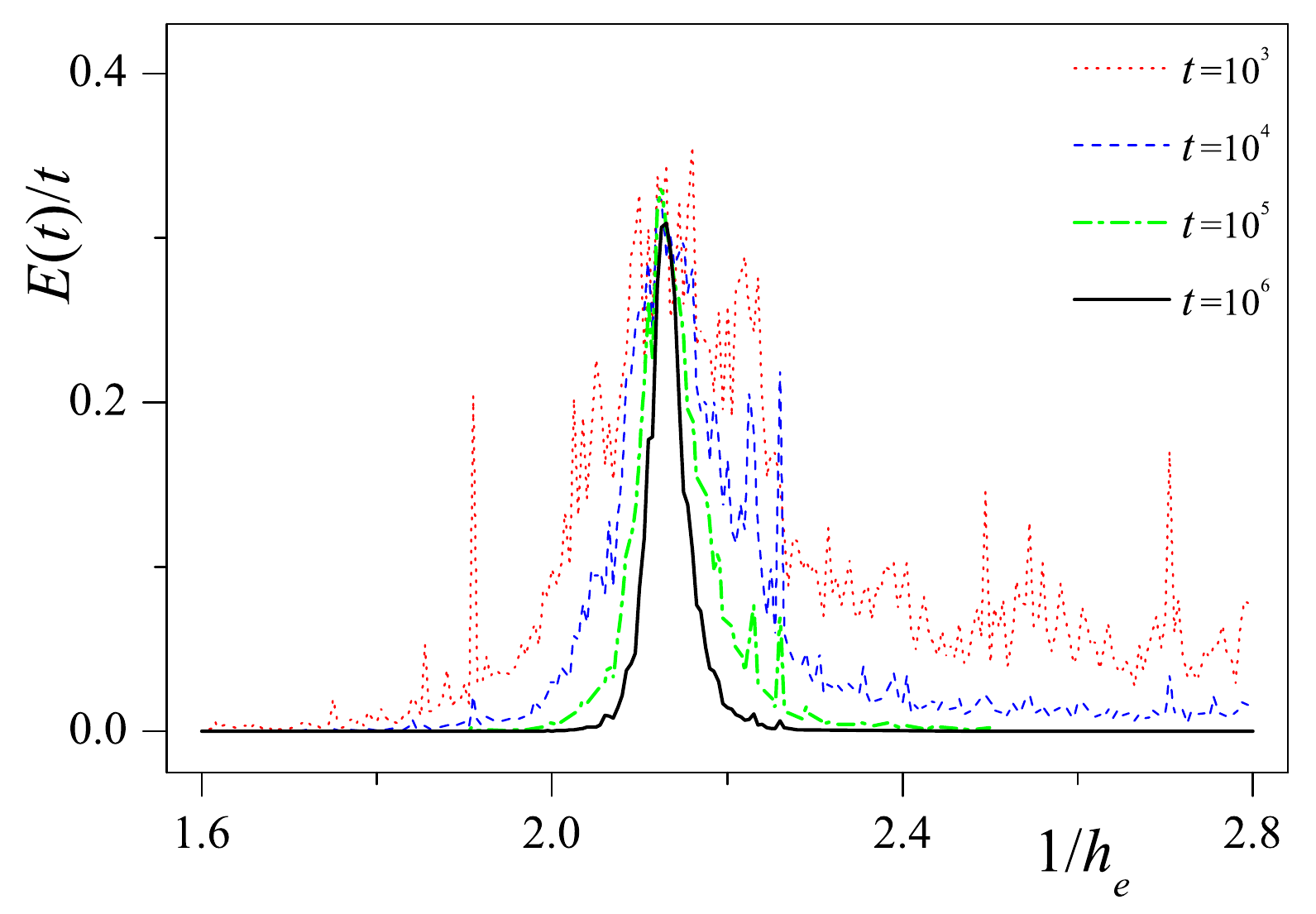}
\caption{Simulation results of $E(t)/t$ at different times. The center of these profiles is located at $h_e^{-1}=2.19$, corresponding to the central peak in Fig.~\ref{fig:3}(c).}
\label{fig:11}
\end{figure}

We find in simulations that the three peaks in Fig.~\ref{fig:3}(c) are feasible already at short times, i.e., $t=10^3$, but relatively broad and exhibit large fluctuations. In Fig.~\ref{fig:11}, we present an example which corresponds to the transition at the critical value of $h_e^{-1}=2.19$. We see that when the simulation time is longer and longer, this peak is narrower and narrower and the fluctuations are gradually suppressed, with the peak center pinned. In addition, we observe that the peak height converges to a finite value. Practically, for $t=10^6$ the peak reaches a stable height, and has a sufficiently sharp profile.

\begin{thebibliography}{}
\bibitem{note_nongeneric_omega} It is likely that there are nongeneric irrational values of $\tilde \omega/(2\pi)$ at which the system behaves essentially the same as the one at rational $\tilde \omega/(2\pi)$.
\bibitem{footnote_1} For certain form of $H_0$,
the translation symmetry in $n_1$ space arises at rational $h_e/(4\pi)$. However, this is a nongeneric case. Therefore, we do not consider it throughout this work.
\bibitem{note_coupling_constant} The (inverse) coupling constant $\sigma$ in the effective field theory (\ref{eq:111}) differs from that in Ref.~\onlinecite{Pruisken10} by a factor of $2$.
\bibitem{note_special_case} From Fig.~\ref{fig:8} we find that in special case of $h_e=1$ two transitions occur when $\mu$ is tuned, and the system is topologically nontrivial ($\sigma_{\rm H}^*=1$) for $\mu$ between two critical values otherwise trivial ($\sigma_{\rm H}^*=0$). This result is consistent with a numerical finding in Ref.~\onlinecite{Beenakker11}.
\bibitem{note_convention} We follow the convention \cite{Efetov97} that the integration over each ordinary number is weighted by $\pi^{-\frac{1}{2}}$.
\end{thebibliography}
\end{document}